\documentclass[aps,prd,11pt,notitlepage,longbibliography,nofootinbib,tightenlines,preprintnumbers,superscriptaddress]{revtex4-1}
\pdfoutput=1

\makeatletter
\renewcommand{\p@subsection}{}
\renewcommand{\p@subsubsection}{}
\makeatother

\usepackage{amsmath,amssymb,amsfonts}
\usepackage{graphicx}
\usepackage{color}
\usepackage{tikz}
\usepackage{dsfont}
\usepackage[pdftex]{hyperref}
\hypersetup{colorlinks=true, linkcolor=darkred, citecolor=blue, linktoc=page}
\definecolor{darkred}{rgb}{0.8,0.1,0.1}

\usepackage{diagbox}

\usepackage{subfigure}

\def\cA{{\cal A}}

\def\cC{{\cal C}}

\def\CC{{\mathds{C}}}
\def\RR{{\mathds{R}}}
\def\ZZ{{\mathds{Z}}}

\def\k{{\mathds{k}}}

\DeclareMathOperator{\Li}{Li}

\makeatletter\def\l@subsubsection#1#2{}%
\makeatother

\def\cN{\mathcal N}
\def\cP{\mathcal P}

\newcommand{\nocontentsline}[3]{}
\newcommand{\tocless}[2]{\bgroup\let\addcontentsline=\nocontentsline#1{#2}\egroup}

\def\Im{\mathop{\rm Im}}
\def\Re{\mathop{\rm Re}}

\begin{document}

\title{Double holography in string theory}

\author{Andreas Karch}
\email{karcha@utexas.edu}
\author{Haoyu Sun}
\email{hkdavidsun@utexas.edu}

\affiliation{University of Texas, Austin, Physics Department, Austin TX 78712, USA}

\author{Christoph F.~Uhlemann}
\email{uhlemann@umich.edu}

\affiliation{Leinweber Center for Theoretical Physics, Department of Physics	\\
	University of Michigan, Ann Arbor, MI 48109-1040, USA}
\preprint{LCTP-22-08}

\begin{abstract}
We develop the notion of double holography in Type IIB string theory realizations of braneworld models. The Type IIB setups are based on the holographic duals of 4d BCFTs comprising 4d $\mathcal N=4$ SYM on a half space coupled to 3d $\mathcal N=4$ SCFTs on the boundary. Based on the concrete BCFTs and their brane construction, we provide microscopic realizations of the intermediate holographic description, obtained by dualizing only the 3d degrees of freedom.
Triggered by recent observations in bottom-up models, we discuss the causal structures in the full BCFT duals and intermediate descriptions. This confirms qualitative features found in the bottom-up models but suggests a refinement of their interpretation.
\end{abstract}

\maketitle
\tableofcontents
\parskip 1mm

\section{Introduction}

Double holography, the notion that the holographic spacetime dual to a boundary CFT (BCFT) has not just two but in fact three equivalent descriptions, has recently found many interesting applications in the context of black hole physics \cite{Almheiri:2019hni,Almheiri:2019psy,Chen:2020uac,Chen:2020hmv,Geng:2020fxl,Geng:2021mic,Uhlemann:2021nhu,Demulder:2022aij}. Double holography was first discovered in  bottom-up models \cite{Karch:2000gx,Karch:2000ct,Takayanagi:2011zk,Fujita:2011fp} where a single Randall-Sundrum brane \cite{Randall:1999vf} serves as ``end of the world" brane that terminates spacetime. While, as they stand, these bottom-up models can not be realized within string theory and have no known holographic dual, they capture crucial aspects of top-down constructions which were developed later based on D3 branes ending on D5 and NS5 branes \cite{DHoker:2007zhm,DHoker:2007hhe,Aharony:2011yc,Assel:2011xz}, as anticipated in
\cite{Karch:2001cw}. Double holography is the statement that besides their description in terms of a $d$-dimensional BCFT or in terms of gravity in the $(d+1)$-dimensional bulk, these setups allow a third, intermediate description: a $d$-dimensional CFT living on the end-of-the-world brane, coupled to $d$-dimensional gravity on the brane and communicating with a $d$-dimensional non-gravitating BCFT via transparent boundary conditions. While the bulk and BCFT descriptions, at least in the top-down models, are completely well-defined theories, the intermediate picture relies on intuition about the holographic interpretation of the (massive) localized graviton one finds on the brane \cite{Karch:2000gx}. Roughly speaking, in the intermediate picture the $d$-dimensional gravity on the brane is the holographic dual of the boundary of the BCFT (often referred to as defect to reserve the term ``boundary'' for the holographic boundary of the $(d+1)$-dimensional bulk spacetime), whereas the full $(d+1)$-dimensional geometry also captures the BCFT degrees of freedom on the $d$-dimensional ambient space.

The reason for the recent resurgence in interest in double holography is the fact that gravity coupled to a bath is the setting for much of the recent progress on the problem of black hole evaporation initiated in \cite{Penington:2019npb,Almheiri:2019psf} and reviewed in \cite{Almheiri:2020cfm}. The basic ideas of how to use gravity coupled to a bath to obtain the Page curve \cite{Penington:2019npb} have first been made concrete in 2d gravity \cite{Almheiri:2019psf}, but all higher dimensional examples constructed so far \cite{Almheiri:2019hni,Almheiri:2019psy,Chen:2020uac,Chen:2020hmv,Geng:2020fxl,Geng:2021mic,Uhlemann:2021nhu,Demulder:2022aij} rely on the intermediate picture interpretation of BCFTs. Furthermore, several important aspects of the problem are highlighted by these braneworld constructions: the graviton in these brane constructions is massive \cite{Karch:2000gx} (see also \cite{Bachas:2018zmb}), which is indeed an unavoidable consequence of the coupling to the external bath \cite{Aharony:2006hz}. It has in fact been argued that this graviton mass is crucial for the Page curve to emerge \cite{Geng:2021hlu}.

Given the importance of this intermediate interpretation of holographic BCFT spacetimes, it is somewhat disconcerting that the precise dictionary is in fact not understood in the bottom-up models. One attempt at such a dictionary has appeared recently \cite{Neuenfeld:2021wbl} for the so-called ``near critical" limit. This is the limit in which the curvature radius on the brane is much larger than the bulk curvature radius and the brane is close to the original conformal boundary. The corresponding BCFT statement is that the number of (local) degrees of freedom on the defect is much larger than the corresponding number in the ambient space. A challenge for the intermediate picture has recently been pointed out in \cite{Omiya:2021olc} where it was shown that a na\"ive implementation of the intermediate picture as a theory living on the brane coupled to a bath is inconsistent with causality: the $(d+1)$-dimensional bulk supports shortcuts when compared to the $d$-dimensional universe of brane and bath (fig.~\ref{fig:braneworld-3d4d}). This was interpreted as signaling non-locality of the intermediate picture.

In this work we give a crisp definition of an intermediate holographic description by making the idea behind it precise: we dualize the degrees of freedom on the defect, but not those on the ambient space. We work with top-down string theory realizations of 4d BCFTs based on D3-branes ending on D5 and NS5 branes.
As concrete examples we take the BCFTs and holographic duals which were used for black hole studies in \cite{Uhlemann:2021nhu}.
In this setting we can isolate the 3d defect degrees of freedom in the brane construction and explicitly construct an intermediate description which dualizes only the 3d degrees of freedom using standard AdS/CFT methods.
Our new ``proper" construction of the intermediate picture is local and manifestly consistent with BCFT causality.

To compare our 10d construction of the intermediate picture to the bottom-up discussions, we will distinguish two notions of intermediate picture:
\begin{itemize}
	\item[$(I_p)$] the ``proper" intermediate picture obtained by geometrizing only the 3d defect d.o.f.
	\item[$(I_n)$] the ``na\"ive" intermediate picture obtained by taking apart the full BCFT dual
\end{itemize}
The na\"ive picture $(I_n)$ is the 10d analog of the intermediate picture advertised in the bottom-up models: it amounts to identifying an end-of-the-world brane region in the 10d solutions and taking a supergravity theory in that region coupled to the 4d ambient CFT as intermediate description. We find that this na\"ive picture $(I_n)$ only provides an approximation to the proper picture $(I_p)$.
Moreover,  we find that analogs of the (apparent) bulk shortcuts of \cite{Omiya:2021olc} exist also in the 10d duals of the full BCFTs, creating the same tension with causality in the na\"ive intermediate picture $(I_n)$.
But the approximate nature of $(I_n)$ obstructs their interpretation as causality violation in the proper local intermediate picture $(I_p)$.
The differences between $(I_n)$ and $(I_p)$ disappear in the near-critical limit, where the shortcuts of \cite{Omiya:2021olc} also disappear; so in this limit the two pictures agree.

This work is organized as follows. In section \ref{sec:stringy-braneworld} we review the brane constructions of the BCFTs and their holographical duals. In section \ref{sec:intermediate} we present our proposal for the proper intermediate description.
The large-$N$ limit is discussed in section \ref{sec:S-duality}. In section \ref{sec:shortcuts} we study geodesics in the BCFT dual and document the appearance of shortcuts compared to the na\"ive intermediate picture. Implications for the proper intermediate picture are discussed in section \ref{sec:causal}. We discuss our interpretation of these findings in section \ref{sec:discussion}. Some technical details can be found in an appendix.

\section{String theory duals for BCFTs}\label{sec:stringy-braneworld}

\begin{figure}
	\centering
	\begin{tikzpicture}[y={(0cm,1cm)}, x={(0.707cm,0.707cm)}, z={(1cm,0cm)}, scale=1.4]
		\draw[gray,fill=gray!100,rotate around={-45:(0,0,1.8)}] (0,0,1.8) ellipse (1.8pt and 3.5pt);
		\draw[gray,fill=gray!100] (0,0,0) circle (1.5pt);
		
		\foreach \i in {-0.05,0,0.05}{ \draw[thick] (0,-1,\i) -- (0,1,\i);}

		\foreach \i in {-0.075,-0.025,0.025,0.075}{ \draw (-1.1,\i,1.8) -- (1.1,\i,1.8);}
		
		\foreach \i in {-0.045,-0.015,0.015,0.045}{ \draw (0,1.4*\i,0) -- (0,1.4*\i,1.8+\i);}
		\foreach \i in  {-0.075,-0.045,-0.015,0.015,0.045,0.075}{ \draw (0,1.4*\i,1.8+\i) -- (0,1.4*\i,4);}
		
		\node at (-0.18,-0.18,3.4) {\scriptsize $2N_5 K$ D3};
		\node at (1.0,0.3,1.8) {\scriptsize $N_5$ D5};
		\node at (0,-1.25) {\footnotesize $N_5$ NS5};
		\node at (0.18,0.18,0.8) {{\scriptsize $N_5 K+\tfrac{N_5^2}{2}$ D3}};
	\end{tikzpicture}
	\caption{Semi-infinite D3-branes ending on a combination of $N_5$ D5-branes and $N_5$ NS5-branes. The D3-branes extend along the (0123) directions, the D5-branes along (012456) and the NS5-branes along (012789).
	All NS5-branes have the same net number of D3-branes ending on them, and likewise for the D5-branes. \label{fig:brane-D5NS5-D3}}
\end{figure}
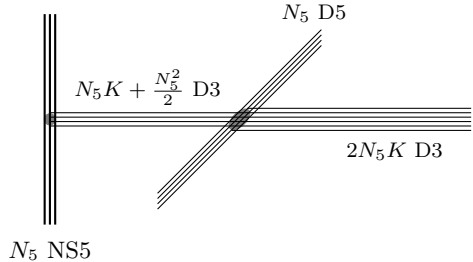

BCFTs based on 4d $\cN=4$ SYM can be engineered in Type IIB string theory by terminating D3-branes on combinations of D5 and NS5 branes \cite{Hanany:1996ie,Gaiotto:2008sa,Gaiotto:2008ak}.
We will focus for simplicity on a setup which was engineered in \cite{Uhlemann:2021nhu} to provide a minimal string theory version of the bottom-up braneworld models.
The brane configuration is shown in fig.~\ref{fig:brane-D5NS5-D3}.
It is characterized by two integers $N_5$ and $K$.
The $2N_5K$ semi-infinite D3-branes realize 4d $\cN=4$ SYM on a half space. At the boundary the 4d $\cN=4$ SYM theory is coupled to a 3d quiver SCFT, which in the string theory construction is realized by D3-brane segments suspended between the 5-branes.

\subsection{BCFTs}

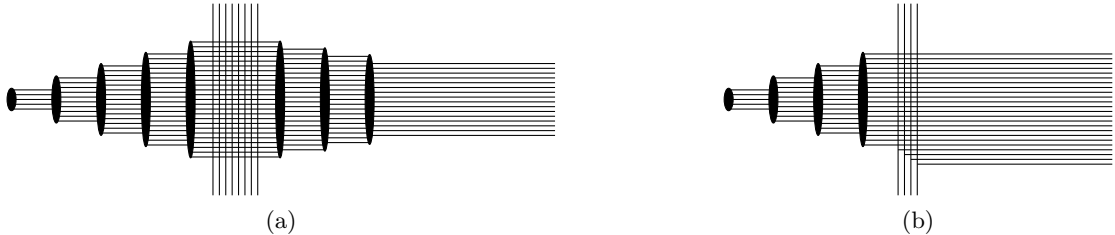
\begin{figure}
	\subfigure[][]{\label{fig:BCFT-brane-1}
		\begin{tikzpicture}[scale=0.85]
			\pgfmathsetmacro{\s}{0.7}
			
			\draw[fill=black] (0,0)  ellipse (2pt and 5pt);	
			\draw[fill=black] (\s,0)  ellipse (2pt and 10.5pt);	
			\draw[fill=black] (2*\s,0)  ellipse (2pt and 16pt);	
			\draw[fill=black] (3*\s,0)  ellipse (2pt and 21pt);	
			\draw[fill=black] (4*\s,0)  ellipse (2pt and 26pt);	
			\draw[fill=black] (6*\s,0)  ellipse (2pt and 26pt);
			\draw[fill=black] (7*\s,0)  ellipse (2pt and 23pt);
			\draw[fill=black] (8*\s,0)  ellipse (2pt and 20pt);
			
			\foreach \i in {-3.5,...,3.5} \draw (5*\s+0.1*\i,-1.5) -- +(0,3);
			
			\foreach \i in {-9,...,9} \draw (7*\s,0.075*\i) -- +(\s,0);
			\foreach \i in {-10.5,...,10.5} \draw (6*\s,0.075*\i) -- +(\s,0);
			\foreach \i in {-12,...,12} \draw (4*\s,0.075*\i) -- +(2*\s,0);
			\foreach \i in {-9.5,...,9.5} \draw (3*\s,0.075*\i) -- +(\s,0);
			\foreach \i in {-7,...,7} \draw (2*\s,0.075*\i) -- +(\s,0);
			\foreach \i in {-4.5,...,4.5} \draw (\s,0.075*\i) -- +(\s,0);
			\foreach \i in {-2,...,2} \draw (0,0.075*\i) -- +(\s,0);
			
			\foreach \i in {-7.5,...,7.5} \draw (8*\s,0.075*\i) -- +(8.5-8*\s,0);
		\end{tikzpicture}
	}\hskip 20mm
	\subfigure[][]{\label{fig:BCFT-brane-2}
	\begin{tikzpicture}[scale=0.85]
		\pgfmathsetmacro{\s}{0.7}
		
		\draw[fill=black] (0,0)  ellipse (2pt and 5pt);	
		\draw[fill=black] (\s,0)  ellipse (2pt and 10.5pt);	
		\draw[fill=black] (2*\s,0)  ellipse (2pt and 16pt);	
		\draw[fill=black] (3*\s,0)  ellipse (2pt and 21pt);	
		
		\foreach \i in {-7,...,7} \draw (2*\s,0.075*\i) -- +(\s,0);
		\foreach \i in {-4.5,...,4.5} \draw (\s,0.075*\i) -- +(\s,0);
		\foreach \i in {-2,...,2} \draw (0,0.075*\i) -- +(\s,0);
		
		\foreach \i in {-9.5,...,9.5} \draw (3*\s,0.075*\i) -- (6,0.075*\i);
		
		\foreach \i in {-1.5,...,1.5}{
			\draw (4*\s-0.1*\i,{0.075*(\i-12)}) -- (6,{0.075*(\i-12)});
			\draw (4*\s+0.1*\i,-1.5) -- +(0,3);
		}
		
	\end{tikzpicture}
	}
\caption{Left: BCFT quiver (\ref{eq:D5NS5K-quiver-2-rep}) for $N_5=8$, $K=1$, with NS5-branes as filled ellipses, D3-branes as horizontal lines, and D5-branes as vertical lines. Right: BCFT quiver (\ref{eq:D5NS5K-quiver-1-rep}) for $N_5=4$, $K=3$.\label{fig:BCFT-brane}}
\end{figure}

The form of the field theories associated with the brane configuration in fig.~\ref{fig:brane-D5NS5-D3} depends on whether $N_5>2K$ or $N_5<2K$.
For $N_5>2K$ the net number of D3-branes ending on the D5-branes from the right in fig.~\ref{fig:brane-D5NS5-D3} is negative, while for $N_5<2K$ it is positive.
To make the field theory manifest in the brane construction for $N_5>2K$, one separates the NS5-branes and uses Hanany-Witten transitions \cite{Hanany:1996ie} to bring the D5-branes to a location where they have no D3-branes ending on them.
The resulting form of the brane configuration is illustrated in fig.~\ref{fig:BCFT-brane-1}. The D3-brane segments suspended between the NS5-branes realize a 3d quiver gauge theory, in which the D5-branes represent 3d flavors. This 3d gauge theory is a UV description for the 3d SCFT to which 4d $\cN=4$ SYM is coupled at the boundary of the BCFT geometry.
The combined 3d/4d quiver gauge theory for $N_5>2K$ is
\begin{align}\label{eq:D5NS5K-quiver-2}
	U(R)-U(2R)-\ldots - &U(R^2) - U(R^2-S)-\ldots - U(2N_5K+S) - \widehat{U(2N_5K)}
	\nonumber\\
	&\ \ \ \vert\\
	\nonumber & \ [N_5]	
\end{align}
where
\begin{align}
	R&=\frac{N_5}{2}+K~, & S&=\frac{N_5}{2}-K~.
\end{align}
The node with a hat in (\ref{eq:D5NS5K-quiver-2}) represents 4d $\cN=4$ SYM on a half space. The remaining nodes are the 3d quiver gauge theory description for the 3d SCFT on the defect.
At the boundary of the 4d half space the 4d $\cN=4$ SYM fields are coupled to the 3d quiver.
Along the first ellipsis in (\ref{eq:D5NS5K-quiver-2}) the rank increases in steps of $R$ from left to right, along the second it decreases in steps of $S$.

For $N_5<2K$, some of the semi-infinite D3-branes end on the D5-branes, as illustrated in fig.~\ref{fig:BCFT-brane-2}.
The combined 3d/4d quiver gauge theory for $N_5<2K$ is
\begin{align}\label{eq:D5NS5K-quiver-1}
	U(R)-U(2R)-\ldots - U((N_5-1)R) - \widehat{U(2N_5K)}
\end{align}
In this case, at the boundary of the 4d half space only a subset of the 4d $\cN=4$ SYM fields couple to the 3d quiver, while the others satisfy Nahm pole boundary conditions imposed by the D5-branes (more details can be found in \cite{Gaiotto:2008ak} or the review in  \cite[sec.~2.2]{Raamsdonk:2020tin}).

Generally speaking, the 3d quiver ``grows" relative to the 4d ambient CFT with increasing $N_5/K$ -- the number of gauge nodes increases with $N_5$ and so do their ranks.
There is a qualitative transition at $N_5=2K$:
For the small-$N_5$ quivers (\ref{eq:D5NS5K-quiver-1}) the ranks of the 3d gauge nodes simply decrease, starting from a number which is smaller than the rank of the 4d $\cN=4$ SYM node on the right end towards zero on the left end of the 3d quiver. For the large-$N_5$ quivers (\ref{eq:D5NS5K-quiver-2}), on the other hand, the rank of the 3d gauge nodes first increases compared to the rank of the 4d gauge node from right to left, up to the node which has flavors attached, and then decreases to zero.

\subsection{Supergravity duals}\label{sec:sugra-duals}

The supergravity solutions associated with brane configurations involving D3 branes ending on and suspended between D5 and NS5 branes were constructed, based on the general class of solutions obtained in \cite{DHoker:2007zhm,DHoker:2007hhe}, in \cite{Aharony:2011yc,Assel:2011xz}.
They involve a non-trivial metric, dilaton, as well as 3- and 5-form fields. The geometry is a warped product of AdS$_4$ and two 2-spheres, $S_1^2$ and $S_2^2$, over a Riemann surface $\Sigma$.
The metric in Einstein frame takes the form
\begin{align}\label{eq:metric-gen}
	ds^2&=f_4^2 ds^2_{AdS_4}+f_1^2 ds^2_{S_1^2}+f_2^2 ds^2_{S_2^2}+4\rho^2 |dz|^2~,
\end{align}
where, for the solutions of interest here, $z$ can be taken as a complex coordinate on the strip
\begin{align}
	\Sigma&=\left\lbrace z\in\CC\,\vert\, 0\leq\Im(z)\leq\frac{\pi}{2}\right\rbrace~.
\end{align}
The solutions are specified by a pair of harmonic functions $h_1$, $h_2$ on $\Sigma$.
The metric functions are
\begin{align}
	f_4^8&=16\frac{N_1N_2}{W^2}~, & f_1^8&=16h_1^8\frac{N_2 W^2}{N_1^3}~, & f_2^8&=16 h_2^8 \frac{N_1 W^2}{N_2^3}~,
	&
	\rho^8&=\frac{N_1N_2W^2}{h_1^4h_2^4}~,
\end{align}
where
\begin{align}
	W&=\partial\bar\partial (h_1 h_2)~, & N_i &=2h_1 h_2 |\partial h_i|^2 -h_i^2 W~.
\end{align}
The axio-dilaton is given by $\tau=\chi+i e^{-2\phi}=i\sqrt{N_1/N_2}$ (using the dilaton normalization of \cite{DHoker:2007zhm,DHoker:2007hhe}). We will not need the expressions for the 3- and 5-form fields here; they can be found in \cite{DHoker:2007zhm,DHoker:2007hhe,Aharony:2011yc,Assel:2011xz}.

The concrete solutions associated with the brane configuration in fig.~\ref{fig:brane-D5NS5-D3} are given by the harmonic functions
\begin{align}\label{eq:h1h2-BCFT}
	h_1&=-\frac{i\pi\alpha^\prime}{4}K e^z-\frac{\alpha^\prime}{4}N_5\ln\tanh\left(\frac{i\pi}{4}-\frac{z}{2}\right)+\rm{c.c.}
	\nonumber\\
	h_2&=\frac{\pi \alpha^\prime}{4} K e^z-\frac{\alpha^\prime}{4}N_5\ln\tanh\left(\frac{z}{2}\right)+\rm{c.c.}
\end{align}
The NS5-branes are represented as poles in $\partial h_2$ at $z=0$, while the D5-branes correspond to the poles in $\partial h_1$ at $z=\frac{i\pi}{2}$.
For $\Re(z)\rightarrow\infty$ the geometry becomes AdS$_5\times$S$^5$; this is where the semi-infinite D3-branes emerge. On the other end of the strip, at $\Re(z)\rightarrow -\infty$, the geometry closes off smoothly.
For a qualitative picture see fig.~\ref{fig:braneworld}, whereas plots of the metric functions are shown in fig.~\ref{fig:metric-plots}.
The 10d solutions specified by this pair of harmonic functions will serve as representative top-down 10d versions of the bottom-up braneworld models.

For the solutions defined by the harmonic functions (\ref{eq:h1h2-BCFT}) the asymptotic value of the dilaton in the AdS$_5\times$S$^5$ region is $\lim_{\Re(z)\rightarrow\infty}e^\phi=1$. This sets the gauge coupling of the 4d $\mathcal N=4$ SYM node to the self-dual value. The asymptotic value of the dilaton can be adjusted using the $SL(2,\RR)$ transformations of type IIB supergravity.
This will be discussed in sec.~\ref{sec:S-duality}.

\subsection{Connection to braneworlds}\label{sec:braneworld-connection}

Qualitatively, the 10d solutions can be described in terms which connect them to bottom-up braneworld models: the part of the 10d geometry around the 5-brane sources can be interpreted as resolved 10d version of the end-of-the-world (ETW) brane, while the AdS$_5\times$S$^5$ region at $\Re(z)\rightarrow\infty$ corresponds to the remaining ``bulk" AdS$_5$.
One would then expect the former to correspond in some sense to 3d degrees of freedom and the latter to 4d degrees of freedom.

\begin{figure}
	\centering
	\begin{tikzpicture}
		\node at (4,0){\includegraphics[width=0.28\linewidth]{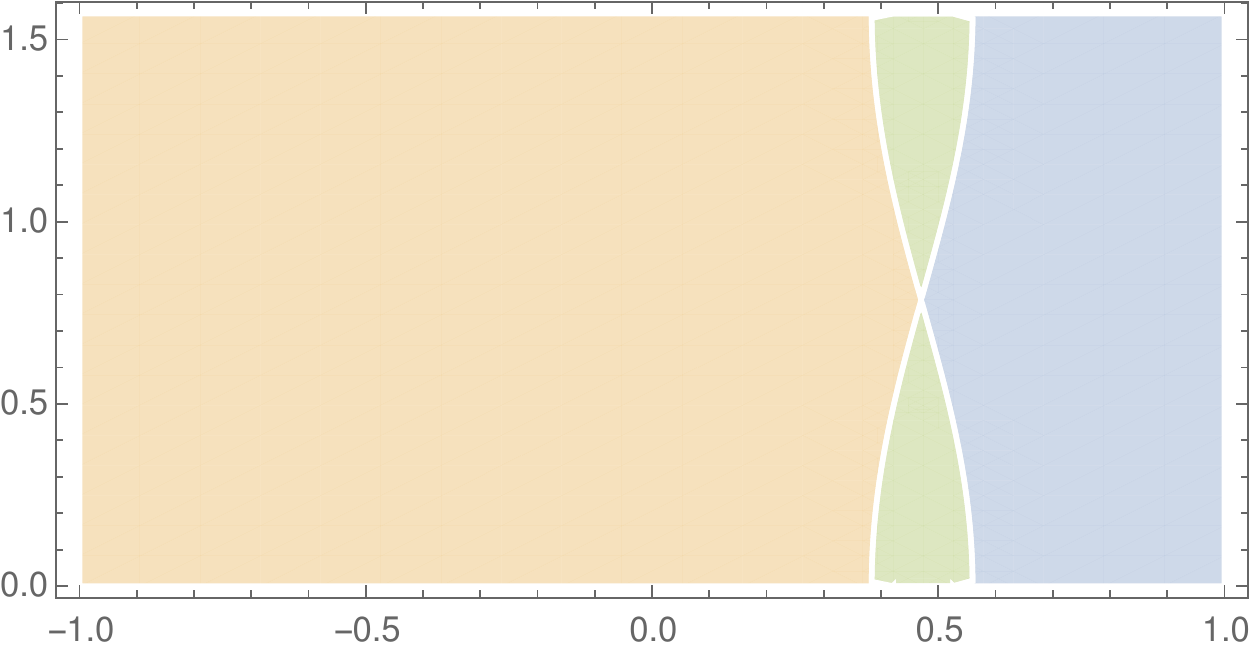}};
		\node at (5.7,0.1) {\small 4d};
		\node at (3.2,0.1) {\small 3d};
		\node at (2.4,0.8) {\small {\boldmath{$\Sigma$}}};
		
		\draw[very thick] (4+0.1,-1.1+0.1) -- (4+0.1,-1.1-0.1) node [anchor=north,yshift=0.5mm] {\small NS5};
		\draw[very thick] (4+0.1,1.25-0.1) -- (4+0.1,1.25+0.1) node [anchor=south,yshift=-0.5mm] {\small D5};
		\node at (6.6,0.1) {\small D3};		
	\end{tikzpicture}
\hfill
	\begin{tikzpicture}
		\node at (4,0){\includegraphics[width=0.28\linewidth]{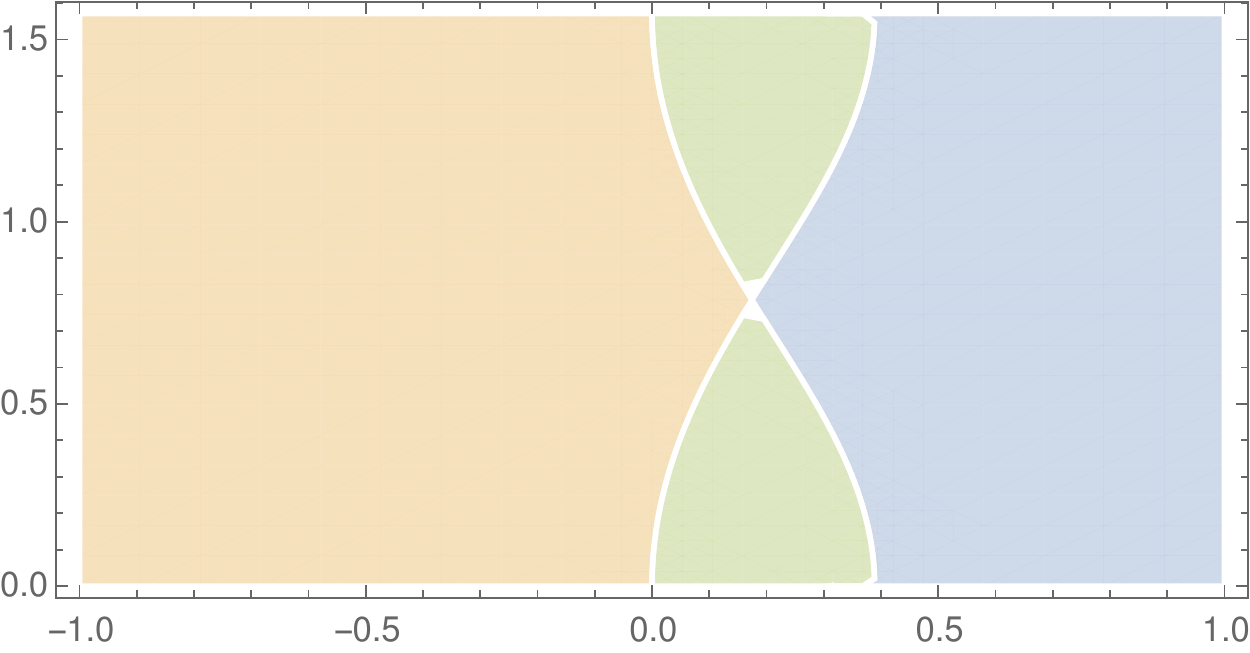}};
		\node at (5.5,0.1) {\small 4d};
		\node at (3.2,0.1) {\small 3d};
		\node at (2.4,0.8) {\small {\boldmath{$\Sigma$}}};
		
		\draw[very thick] (4+0.1,-1.1+0.1) -- (4+0.1,-1.1-0.1) node [anchor=north,yshift=0.5mm] {\small NS5};
		\draw[very thick] (4+0.1,1.25-0.1) -- (4+0.1,1.25+0.1) node [anchor=south,yshift=-0.5mm] {\small D5};
		\node at (6.6,0.1) {\small D3};		
	\end{tikzpicture}	
\hfill
	\begin{tikzpicture}
		\node at (4,0){\includegraphics[width=0.28\linewidth]{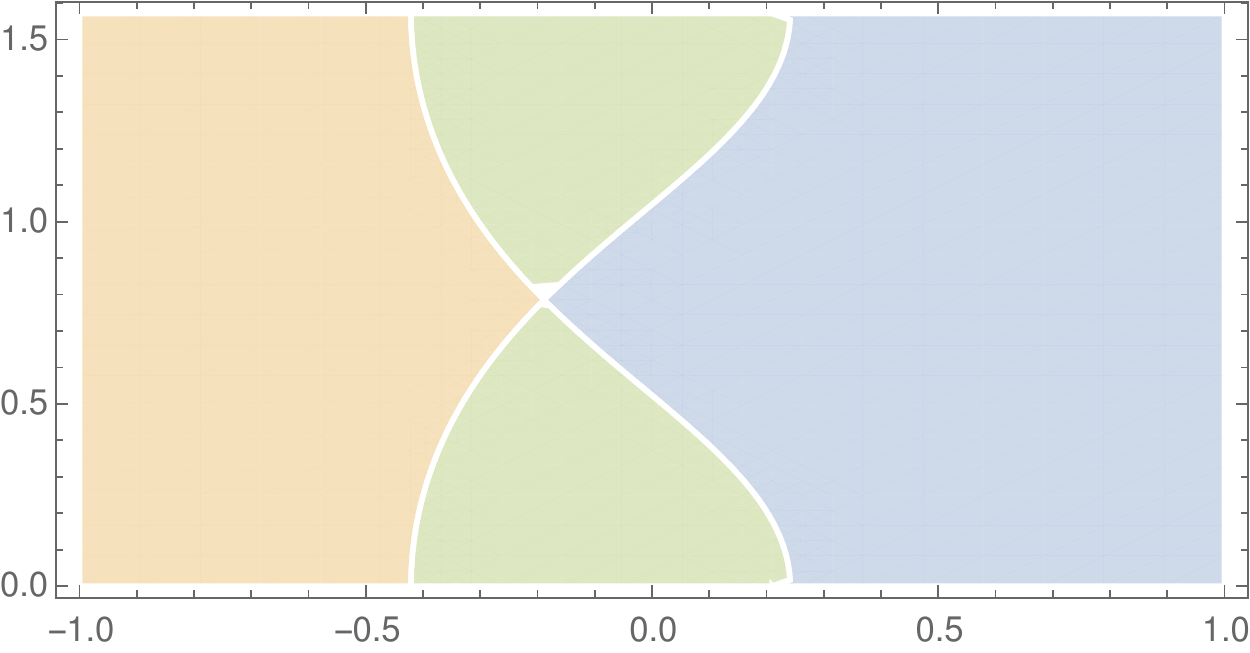}};
		\node at (5.2,0.1) {\small 4d};
		\node at (2.7,0.1) {\small 3d};
		\node at (2.4,0.8) {\small {\boldmath{$\Sigma$}}};
		
		\draw[very thick] (4+0.1,-1.1+0.1) -- (4+0.1,-1.1-0.1) node [anchor=north,yshift=0.5mm] {\small NS5};
		\draw[very thick] (4+0.1,1.25-0.1) -- (4+0.1,1.25+0.1) node [anchor=south,yshift=-0.5mm] {\small D5};
		\node at (6.6,0.1) {\small D3};		
	\end{tikzpicture}	
	\caption{Regions on the strip $\Sigma$ for the BCFT duals (\ref{eq:h1h2-BCFT}) according to which type of loop operators they host, from left to right for $N_5/K\in \lbrace 4,2,1\rbrace$. The horizontal axis is $\tanh(\Re(z))$, the vertical axis $\Im(z)$. The 3d part hosts Wilson and vortex loop operators on the defect, the 4d part hosts neither, while the transition regions host either Wilson or vortex loops on the defect but not both (see also fig.~\ref{fig:BCFT-coords}).\label{fig:braneworld}}
\end{figure}
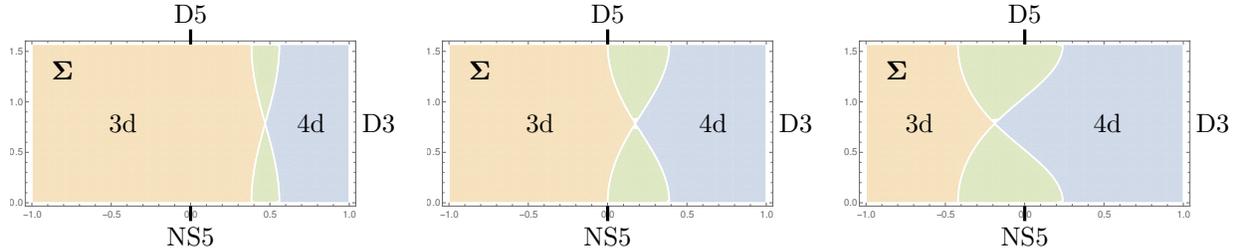

A concrete connection between regions in the 10d geometry and 3d/4d degrees of freedom was made in \cite{Coccia:2021lpp}, by studying $\frac{1}{2}$-BPS Wilson loops in antisymmetric representations of individual 3d gauge nodes (and their mirror-dual vortex loops), identifying their holographic representation as probe D5$^\prime$ (and NS5$^\prime$) branes in the 10d solutions, and matching the expectation values to supersymmetric localization calculations.
In this way, slices of the strip $\Sigma$ can be linked to individual 3d gauge nodes.
The resulting identification of 3d and 4d parts in the 10d solutions (\ref{eq:h1h2-BCFT}) is shown in fig.~\ref{fig:braneworld}.
The part marked as 3d hosts Wilson and vortex loop operators associated with the 3d defect part of the quiver, the part marked as 4d hosts neither type of operator, while the transition regions host either Wilson or vortex loops on the defect but not both.

For $N_5\gg K$ the number of 3d degrees of freedom is large, corresponding to an ETW brane close to the boundary (sharp brane angle). In that limit the 3d quiver (\ref{eq:D5NS5K-quiver-2}) is long and the maximal rank of the 3d nodes is large compared to that of the 4d node. For $K\gg N_5$, on the other hand, the 3d quiver is short and the ranks of the 3d gauge nodes in (\ref{eq:D5NS5K-quiver-1}) just decrease compared to that of the 4d node. This corresponds to an ETW brane at a more blunt angle.\footnote{%
	For any $N_5> 0$ the solutions (\ref{eq:h1h2-BCFT}) close off smoothly at $\Re(z)= -\infty$, but for $N_5=0$ they degenerate --  the D3-branes in fig.~\ref{fig:brane-D5NS5-D3} can not just end in smoke. We note that the $\ZZ_2$ orbifold of AdS$_5\times$S$^5$ which gives a particular 10d realization of the braneworld model with tensionless brane is also included in the general AdS$_4\times$S$^2\times$S$^2\times\Sigma$ solutions. Compared to the solutions here it does not close off smoothly by internal cycles collapsing.}
These features are represented in fig.~\ref{fig:braneworld} as a large /small 3d region for large/small $N_5/K$.
This suggests that the brane angle in the bottom-up models can be seen as effective description for the ratio $N_5/K$ in 10d.

With a qualitative connection between the 10d BCFT duals and the brane world models at our disposal, we can formulate a 10d version of the ``na\"ive" intermediate picture $(I_n)$ advertised in the bottom-up models: a gravitational theory in the ETW-brane region of the 10d BCFT dual should be coupled at the boundary of the AdS$_4$ fibers to 4d $\mathcal N=4$ SYM on a half space.
As ETW-brane region one might take the region on $\Sigma$ hosting Wilson loop operators on the defect, as discussed above, or one may consider other identifications. The discussions to come will apply for more general identifications of ETW-brane regions in the BCFT dual.
For any such identification one may wonder what boundary conditions should be imposed at the boundary of the ETW-brane region on $\Sigma$. This question already suggests that this picture might at best be approximate.
We will not dwell on it now and instead discuss a proper intermediate picture $(I_p)$ next.

\section{Intermediate description in 10d}\label{sec:intermediate}

In this section a proper intermediate holographic description will be made concrete in the full 10d string theory setting.
The strategy will be as follows: The BCFT contains 3d and 4d degrees of freedom. The full BCFT dual, discussed in sec.~\ref{sec:stringy-braneworld}, geometrizes all field theory degrees of freedom. The intermediate description, on the other hand, will be constructed so that it only geometrizes the 3d degrees of freedom; they are replaced by a 10d AdS$_4$ supergravity dual which is coupled at the boundary of AdS$_4$ to the remaining 4d $\mathcal N=4$ SYM d.o.f.\ on the half space.

To make the proposal precise, we start with the quiver description of the BCFT in (\ref{eq:D5NS5K-quiver-2}). That is, we first assume $N_5>2K$. We repeat the combined 3d/4d quiver for convenience:
\begin{align}\label{eq:D5NS5K-quiver-2-rep}
	U(R)-U(2R)-\ldots - &U(R^2) - U(R^2-S)-\ldots - U(2N_5K+S) - \widehat{U(2N_5K)}
	\nonumber\\
	&\ \ \ \vert\\
	\nonumber & \ [N_5]	
\end{align}
and recall that $R=K+N_5/2$ and $S=N_5/2-K$.
The 4d degrees of freedom are represented by the hatted gauge node at the right end, the rest of the quiver represents the 3d boundary degrees of freedom.
The brane configuration for $N_5=8$, $K=1$, so that $S=3$, $R=5$, is shown in fig.~\ref{fig:BCFT-brane-1}.

\begin{figure}
\subfigure[][]{\label{fig:3dmax-brane}
	\begin{tikzpicture}[scale=0.85]
	\pgfmathsetmacro{\s}{0.7}
	
	\draw[fill=black] (0,0)  ellipse (2pt and 5pt);	
	\draw[fill=black] (\s,0)  ellipse (2pt and 10.5pt);	
	\draw[fill=black] (2*\s,0)  ellipse (2pt and 16pt);	
	\draw[fill=black] (3*\s,0)  ellipse (2pt and 21pt);	
	\draw[fill=black] (4*\s,0)  ellipse (2pt and 26pt);	
	\draw[fill=black] (6*\s,0)  ellipse (2pt and 26pt);
	\draw[fill=black] (7*\s,0)  ellipse (2pt and 23pt);
	\draw[fill=black] (8*\s,0)  ellipse (2pt and 20pt);
	
	\foreach \i in {-3.5,...,3.5} \draw (5*\s+0.1*\i,-1.5) -- +(0,3);
	
	\foreach \i in {-9,...,9} \draw (7*\s,0.075*\i) -- +(\s,0);
	\foreach \i in {-10.5,...,10.5} \draw (6*\s,0.075*\i) -- +(\s,0);
	\foreach \i in {-12,...,12} \draw (4*\s,0.075*\i) -- +(2*\s,0);
	\foreach \i in {-9.5,...,9.5} \draw (3*\s,0.075*\i) -- +(\s,0);
	\foreach \i in {-7,...,7} \draw (2*\s,0.075*\i) -- +(\s,0);
	\foreach \i in {-4.5,...,4.5} \draw (\s,0.075*\i) -- +(\s,0);
	\foreach \i in {-2,...,2} \draw (0,0.075*\i) -- +(\s,0);
	
	\foreach \i in {-7.5,...,7.5}{
		\draw (8*\s,0.075*\i) -- +(6.5-8*\s-0.075*\i,0);
		\draw (6.5+0.075*\i,-1.5) -- +(0,3);
	}
	\end{tikzpicture}
}\hskip 30mm
	\subfigure[][]{\label{fig:3dmax-brane-2}
	\begin{tikzpicture}[scale=0.85]
		\pgfmathsetmacro{\s}{0.7}
		
		\draw[fill=black] (0,0)  ellipse (2pt and 5pt);	
		\draw[fill=black] (\s,0)  ellipse (2pt and 10.5pt);	
		\draw[fill=black] (2*\s,0)  ellipse (2pt and 16pt);	
		\draw[fill=black] (3*\s,0)  ellipse (2pt and 21pt);	
		
		\foreach \i in {-7,...,7} \draw (2*\s,0.075*\i) -- +(\s,0);
		\foreach \i in {-4.5,...,4.5} \draw (\s,0.075*\i) -- +(\s,0);
		\foreach \i in {-2,...,2} \draw (0,0.075*\i) -- +(\s,0);
		
		\foreach \i in {-9.5,...,9.5}{
			\draw (5.5*\s-0.1*\i,{0.075*\i}) -- (3*\s,{0.075*\i});
			\draw (5.5*\s+0.1*\i,-1.5) -- +(0,3);
		}
	\end{tikzpicture}
}

	\caption{Left: 3d quiver (\ref{eq:3d-intermediate-quiver}) for $N_5=8$, $K=1$. Right: 3d quiver (\ref{eq:3d-intermediate-quiver-2}) for $N_5=4$, $K=3$.
	As in fig.~\ref{fig:BCFT-brane}, NS5-branes are shown as ellipses, D3-branes as horizontal lines, and D5-branes as vertical lines.}
\end{figure}
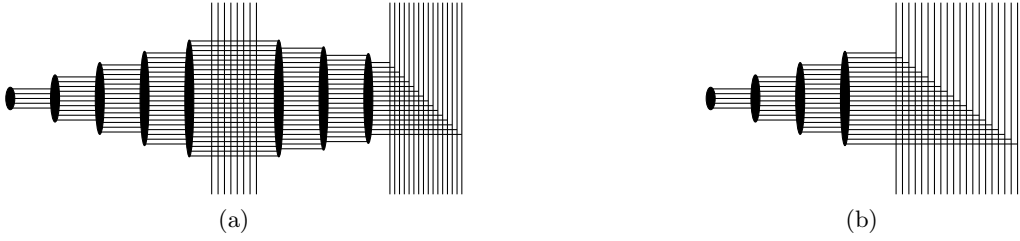

To obtain an intermediate holographic description one has to choose a set of 3d degrees of freedom to geometrize. These 3d d.o.f.\ will be taken out of the quiver and be replaced by their gravity dual, which is then coupled to the remaining ungeometrized degrees of freedom.
There is some freedom in which degrees of freedom to dualize.
We make a maximal choice here and geometrize all 3d degrees of freedom. To this end, we terminate the 3d part of the BCFT quiver in the brane construction of fig.~\ref{fig:BCFT-brane-1} at the 4d gauge node, by inserting D5-branes as in fig.~\ref{fig:3dmax-brane}. This leaves the 3d quiver
\begin{align}\label{eq:3d-intermediate-quiver}
	U(R)-U(2R)-\ldots - &U(R^2) - U(R^2-S)-\ldots - U(2N_5K+S) - [2N_5 K]
	\nonumber\\
	&\ \ \ \vert\\
	\nonumber & \ [N_5]	
\end{align}
On the right end the 4d $U(2N_5K)$ $\cN=4$ SYM node has been replaced by 3d flavors.
This quiver has an $SU(2N_5K)$ flavor symmetry associated with the flavors at the right end, and the coupling to the remaining 4d degrees of freedom amounts to using the 4d $\cN=4$ SYM fields to gauge this flavor symmetry.
The 3d quiver (\ref{eq:3d-intermediate-quiver}) has all nodes balanced and is `good' in the classification of \cite{Gaiotto:2008sa,Gaiotto:2008ak}. This means the 3d gauge theory flows to a well-behaved 3d SCFT in the IR, and this IR fixed point SCFT has a holographic dual.
The brane construction is shown in fig.~\ref{fig:3dmax-brane}. The BCFT brane configuration in fig.~\ref{fig:BCFT-brane-1} can be recovered by moving the D5-branes at the right end off to infinity and then removing them, to reintroduce semi-infinite D3-branes.

The full BCFT (\ref{fig:brane-D5NS5-D3}), (\ref{eq:D5NS5K-quiver-2-rep}) is S-dual to itself: The S-dual brane configuration is obtained by exchanging D5 and NS5 branes in fig.~\ref{fig:brane-D5NS5-D3}. It can be brought back to the original form using Hanany-Witten transitions. The 3d SCFT (\ref{eq:3d-intermediate-quiver}), on the other hand, is not S-dual to itself, since the brane construction involves different numbers of D5-branes and NS5-branes.

The supergravity dual for the 3d SCFT arising from the quiver in (\ref{eq:3d-intermediate-quiver}) is of the general AdS$_4\times$S$^2\times$S$^2\times\Sigma$ form discussed in sec.~\ref{sec:sugra-duals}. But it is a genuine AdS$_4$ solution  -- instead of blowing up to AdS$_5\times$S$^5$ at the right end of the strip, $\Re(z)\rightarrow \infty$, the geometry closes off smoothly.
Concretely, the supergravity dual for the quiver SCFT in (\ref{eq:3d-intermediate-quiver}) is given by
\begin{align}\label{eq:h1h2-intermediate-max}
	h_1&=-\frac{\alpha^\prime}{4}N_5\ln\tanh\left(\frac{i\pi}{4}-\frac{z-\delta_1}{2}\right)-\frac{\alpha^\prime}{4}2N_5K\ln\tanh\left(\frac{i\pi}{4}-\frac{z-\delta_2}{2}\right)+\rm{c.c.}
	\nonumber\\
	h_2&=-\frac{\alpha^\prime}{4}N_5\ln\tanh\left(\frac{z}{2}\right)+\rm{c.c.}
\end{align}
with
\begin{align}
	\delta_1&=-\ln \tan \frac{\pi S}{2N_5}~,
	&
	\delta_2&=-\ln \tan \frac{\pi}{2N_5}~.
\end{align}
To arrive at this form of the supergravity dual, one brings all D5-branes in fig.~\ref{fig:3dmax-brane} to the right side. This results in a brane configuration with $N_5$ NS5 branes with $R$ D3-branes ending on each, as well as $N_5$ D5-branes with $S$ D3-branes ending on each and $2N_5K$ D5-branes with one D3-brane ending on each.
The numbers of 5-branes are then reflected in the residues at the poles of $\partial h_{1/2}$  in (\ref{eq:h1h2-intermediate-max}), and the numbers of attached D3-branes in the locations of the poles \cite{Assel:2011xz}.
The solutions are a special case of the solutions discussed in \cite[sec.~V.A]{Coccia:2020wtk}.
Compared to the BCFT dual (\ref{eq:h1h2-BCFT}), there are no exponentially growing terms in (\ref{eq:h1h2-intermediate-max}); $h_{1/2}$ both vanish at $\Re(z)\rightarrow\infty$. The geometry closes off smoothly at both ends of the strip, $\Re(z)\rightarrow\pm\infty$.
For $N_5\rightarrow\infty$ the D5 pole at $z=\delta_2$ would move to large $z$, which would be similar to the $T[SU(N)]$ solutions discussed in  \cite{Assel:2012cp}. The actual large-$N$ limit will be discussed in sec.~\ref{sec:S-duality} and will lead to fixed 5-brane sources.

The proposal for the intermediate description now is very natural: the AdS$_4$ gravity dual (\ref{eq:h1h2-intermediate-max}) for the 3d SCFT in (\ref{eq:3d-intermediate-quiver}) should be coupled to the remaining 4d $\mathcal N=4$ SYM degrees of freedom, in a way which implements the gauging of the global $SU(2N_5K)$ flavor symmetry of the 3d SCFT that is realized by the coupling to the 4d fields in the brane construction.\footnote{Following \cite{Gaiotto:2008ak} the $U(1)$ sector of the 4d $U(2N_5K)$ $\mathcal N=4$ SYM fields satisfies Neumann boundary conditions.}
The statement of AdS/CFT for the solution (\ref{eq:h1h2-intermediate-max}) is that the partition function of the 3d SCFT (\ref{eq:3d-intermediate-quiver}), as functional of sources for gauge-invariant operators, is identical to the partition function of Type IIB string theory on the background specified by (\ref{eq:h1h2-intermediate-max}), as functional of the boundary conditions for excitations,
\begin{align}\label{eq:AdSCFT}
	\mathcal Z_\text{3d SCFT (\ref{eq:3d-intermediate-quiver})}[A_\mu^{(3d)},\ldots]&=
	\mathcal Z_\text{Type IIB on (\ref{eq:h1h2-intermediate-max})}[A_\mu^{(3d)},\ldots]~.
\end{align}
Here we have highlighted the source for the conserved current $J^\mu$ corresponding to the $SU(2N_5K)$ flavor symmetry associated with the $2N_5K$ flavors at the right end of the quiver (\ref{eq:3d-intermediate-quiver}) and the corresponding boundary condition in the bulk partition function.
In field theory terms, gauging the flavor symmetry of the 3d SCFT using 4d $\mathcal N=4$ SYM on a half space may be understood schematically as integrating over the (so far non-dynamical) source $A_\mu^{(3d)}$, which should be understood as boundary value of the 4d gauge field $A_\mu^{(4d)}$, weighted by the action of the ambient 4d $\mathcal N=4$ SYM fields including $A_\mu^{(4d)}$.
We obtain
\begin{align}\label{eq:intermediate-dual}
	\mathcal Z_\text{BCFT}&=\int \mathcal D A^{(4d)}_{\mu}\ldots\, Z_\text{3d SCFT (\ref{eq:3d-intermediate-quiver})}[A_\mu^{(3d)},\ldots ]\,e^{S_{\text{4d $\mathcal N=4$ SYM}}[A_\mu^{(4d)},\ldots]}
	\nonumber\\
	&=\int \mathcal D A^{(4d)}_{\mu}\ldots\, \mathcal Z_\text{Type IIB on (\ref{eq:h1h2-intermediate-max})}[A_\mu^{(3d)},\ldots ]\,e^{S_{\text{4d $\mathcal N=4$ SYM}}[A_\mu^{(4d)},\ldots]}
\end{align}
The first line is the gauging in field theory.
The second line follows from the AdS/CFT identification (\ref{eq:AdSCFT}), and implements the coupling of the non-gravitational 4d $\mathcal N=4$ SYM degrees of freedom to the 4d gravity dual of the 3d degrees of freedom.
The second line of (\ref{eq:intermediate-dual}) is our proposal for the intermediate holographic description for the BCFT in (\ref{eq:D5NS5K-quiver-2-rep}).

In the bottom-up description of double holography, the intermediate picture contains as crucial part a CFT living on the brane, coupled to the localized graviton \cite{Karch:2000ct}. This CFT is expected to be the same as the CFT defining the non-gravitational bath. A natural question is whether and where this CFT appears in the top-down construction. We note that, by the standard rules of AdS/CFT, gauge symmetries in the gravitational description map to global symmetries of the dual CFT. We can use that to find the gauge theory corresponding to the intermediate picture CFT. By construction, the 3d SCFT we engineered in (\ref{eq:3d-intermediate-quiver}) has an $SU(2 N_5 K)$ global symmetry. Thus, the intermediate picture gravity theory (\ref{eq:h1h2-intermediate-max}) contains the same maximally supersymmetric $SU(2 N_5 K)$ SYM theory that appears as part of the non-gravitating bath. It arises from open string fields -- the corresponding gauge bosons live on the D5-brane source at $z=\delta_2+\frac{i\pi}{2}$ in the bulk, where the supergravity approximation strictly speaking breaks down. The full global symmetry of the 3d SCFT actually is $SU(N_5) \times SU(N_5) \times SU(2 N_5 K)$, with the holographic representation for one $SU(N_5)$ arising from open string modes on the $N_5$ D5 branes and the other from their mirror-dual analogs for NS5 branes (such modes were also discussed in \cite{Bachas:2017wva}).
The bottom-up model misses these extra $SU(N_5)$ factors. This is reasonable: they depend on the detailed nature of the 3d degrees of freedom and this type of information should not be expected to be captured in a bottom-up model.

We now turn to the other regime, $N_5<2K$, with the combined 3d/4d quiver given in (\ref{eq:D5NS5K-quiver-1}). We again repeat the quiver for convenience,
\begin{align}\label{eq:D5NS5K-quiver-1-rep}
	U(R)-U(2R)-\ldots - U((N_5-1)R) - \widehat{U(2N_5K)}
\end{align}
The brane configuration for $N_5=4$ and $K=3$ is shown in fig.~\ref{fig:BCFT-brane-2}.
Compared to fig.~\ref{fig:BCFT-brane-1}, the D5-branes now impose Nahm pole boundary conditions on part of the 4d $\cN=4$ SYM fields instead of being part of the 3d quiver. To obtain the intermediate holographic description we again take the 3d degrees of freedom out of the BCFT quiver (\ref{eq:D5NS5K-quiver-1-rep}), terminate the quiver with 3d flavors instead of a 4d node and then dualize the 3d quiver SCFT separately.
The 3d quiver obtained by introducing a minimal number of flavors is
\begin{align}\label{eq:3d-intermediate-quiver-2}
	U(R)-U(2R)-\ldots - U((N_5-1)R) - [N_5R]
\end{align}
It again has all nodes balanced.
The full BCFT can now be recovered in two steps. One first gauges the flavor symmetry associated with the $N_5R$ flavors at the right end using the boundary values of 4d $\cN=4$ SYM fields on a half space. In the brane setup this amounts to replacing the $N_5R$ D5-branes with semi-infinite D3-branes.
One then brings in additional $N_5(K-N_5/2)$ semi-infinite D3-branes which terminate on $N_5$ D5-branes, to get back to the full quiver in (\ref{eq:D5NS5K-quiver-1-rep}).

The proposal for the intermediate description is that the holographic dual for the 3d quiver in (\ref{eq:3d-intermediate-quiver-2}) is coupled at the boundary of AdS$_4$ to the remaining 4d degrees of freedom, in analogy to the discussion for $N_5>2K$.
The dual for the 3d quiver SCFT in (\ref{eq:3d-intermediate-quiver-2}) is defined by the harmonic functions
\begin{align}\label{eq:h1h2-intermediate-max-2}
	h_1&=-\frac{\alpha^\prime}{4}N_5R\,\ln\tanh\left(\frac{i\pi}{4}-\frac{z-\delta_1}{2}\right)+\rm{c.c.}
	&
	\delta_1&=-\ln \tan \frac{\pi}{2N_5}
	\nonumber\\
	h_2&=-\frac{\alpha^\prime}{4}N_5\ln\tanh\left(\frac{z}{2}\right)+\rm{c.c.}
\end{align}
It has a pole in $\partial h_2$ at $z=0$ which represents $N_5$ NS5-branes and a pole at $z=\delta_1+\frac{i\pi}{2}$ in $\partial h_1$ which represents $N_5R$ D5-branes.

The global symmetry of the 3d SCFT (\ref{eq:3d-intermediate-quiver-2})  is $SU(N_5)\times SU(N_5R)$, and the holographic dual again contains the appropriate 4d $\mathcal N=4$ vector multiplets.
Comparing to the bottom-up models, one may now identify the $SU(N_5R)$ part as the 10d version of the CFT on the end-of-the-world brane. This theory now has a smaller gauge group than the non-gravitating bath CFT, which still is $U(2N_5K)$. The reason is that in the brane construction some of the semi-infinite D3-branes just terminate on D5-branes, and do not lead to flavors in the 3d quiver.
That such details are not captured automatically by the bottom-up models highlights their nature as effective descriptions -- one can adapt them by hand to connect to specific 10d brane constructions.

The decomposition of the full BCFT quivers (\ref{eq:D5NS5K-quiver-2-rep}) and (\ref{eq:D5NS5K-quiver-1-rep}) into the 3d parts (\ref{eq:3d-intermediate-quiver}) and (\ref{eq:3d-intermediate-quiver-2}) and the remaining degrees of freedom is not unique. One could choose other 3d sub-quivers of (\ref{eq:D5NS5K-quiver-2-rep}) and (\ref{eq:D5NS5K-quiver-1-rep}) which are `good' in the sense of \cite{Gaiotto:2008sa,Gaiotto:2008ak} and geometrize those to obtain alternative intermediate descriptions. The choices (\ref{eq:3d-intermediate-quiver}), (\ref{eq:3d-intermediate-quiver-2}) are maximal in the sense that they use the entire 3d parts of the BCFT quivers, and we will focus on those in the following.

\subsection{Comparing geometries}\label{sec:compare-geometry}

\begin{figure}
	\includegraphics[width=0.242\linewidth]{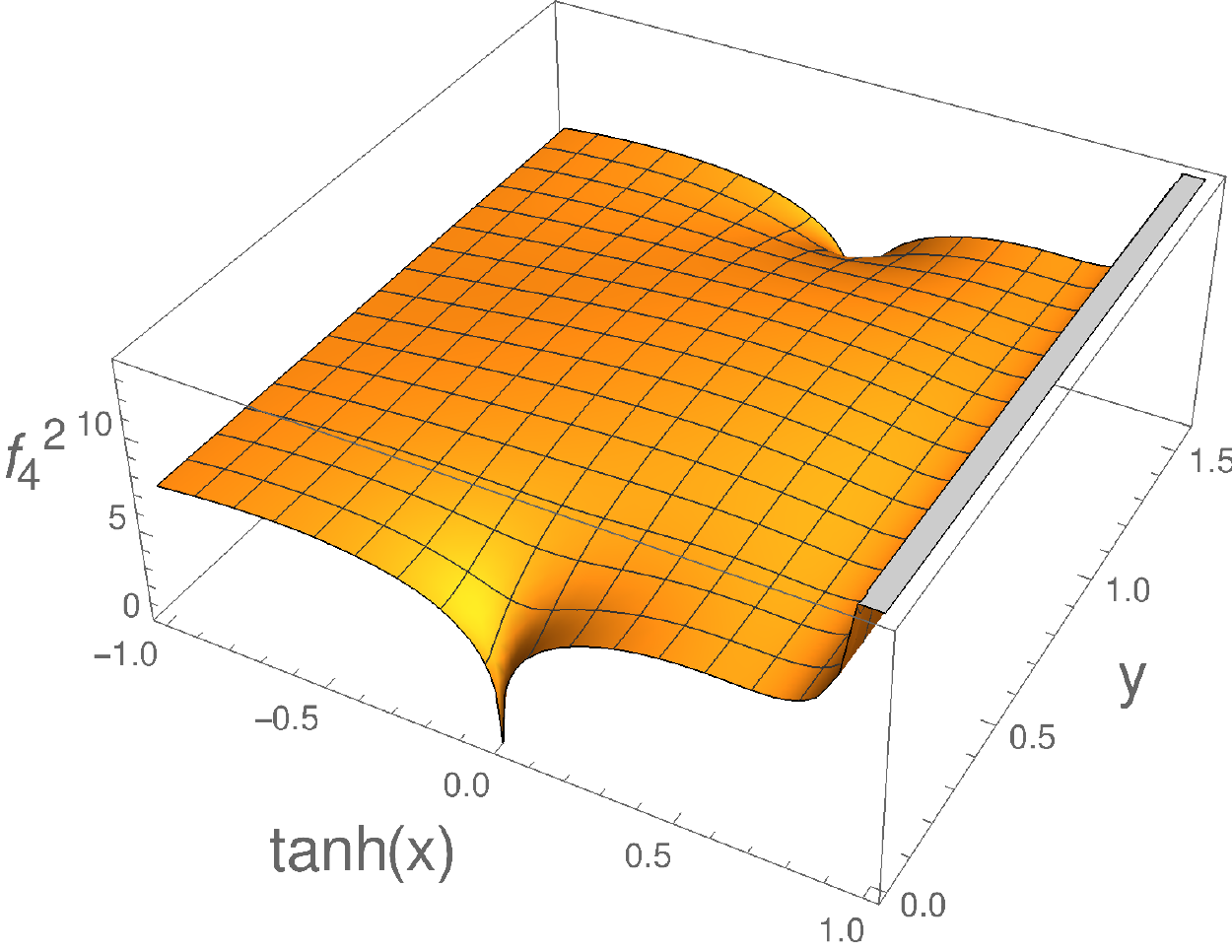}
	\includegraphics[width=0.242\linewidth]{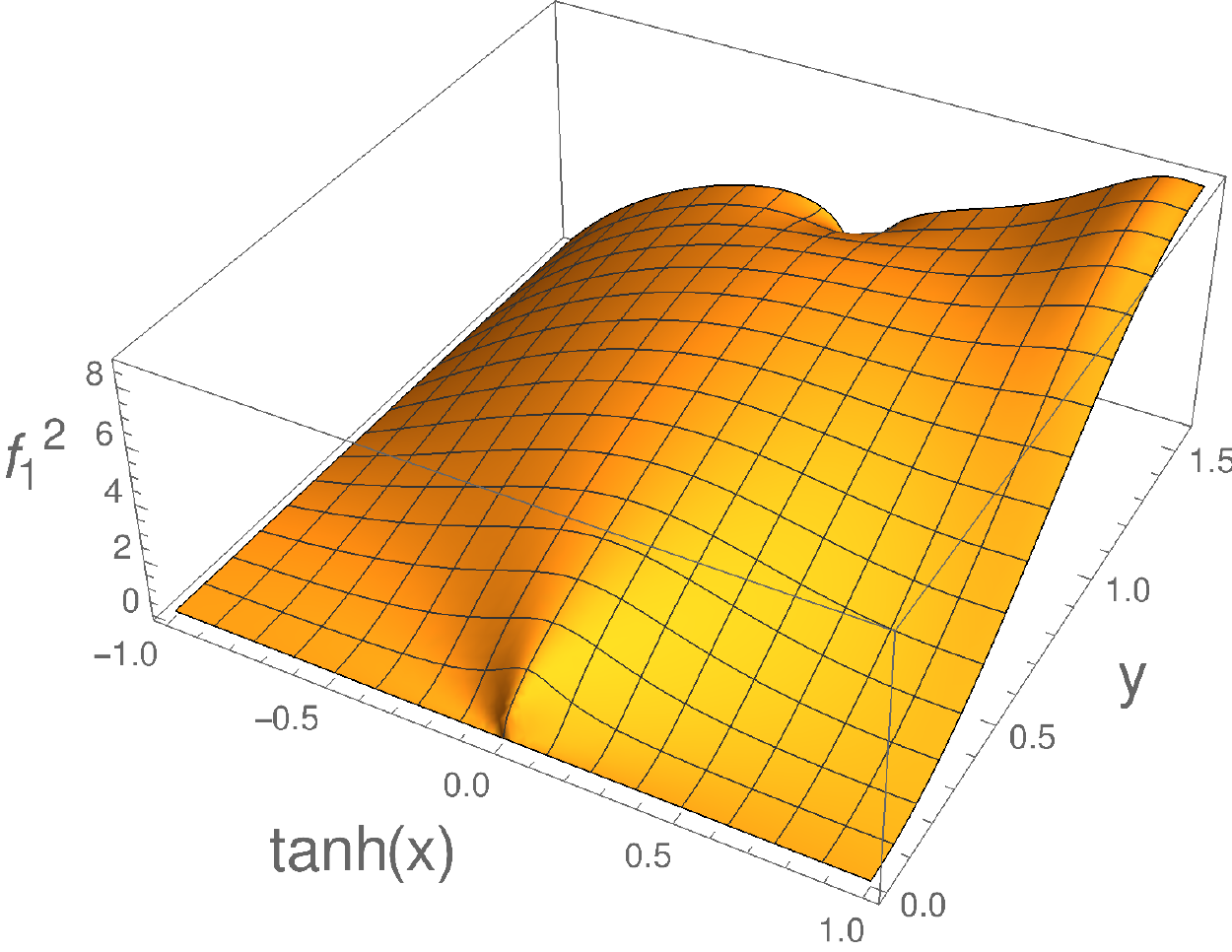}
	\includegraphics[width=0.242\linewidth]{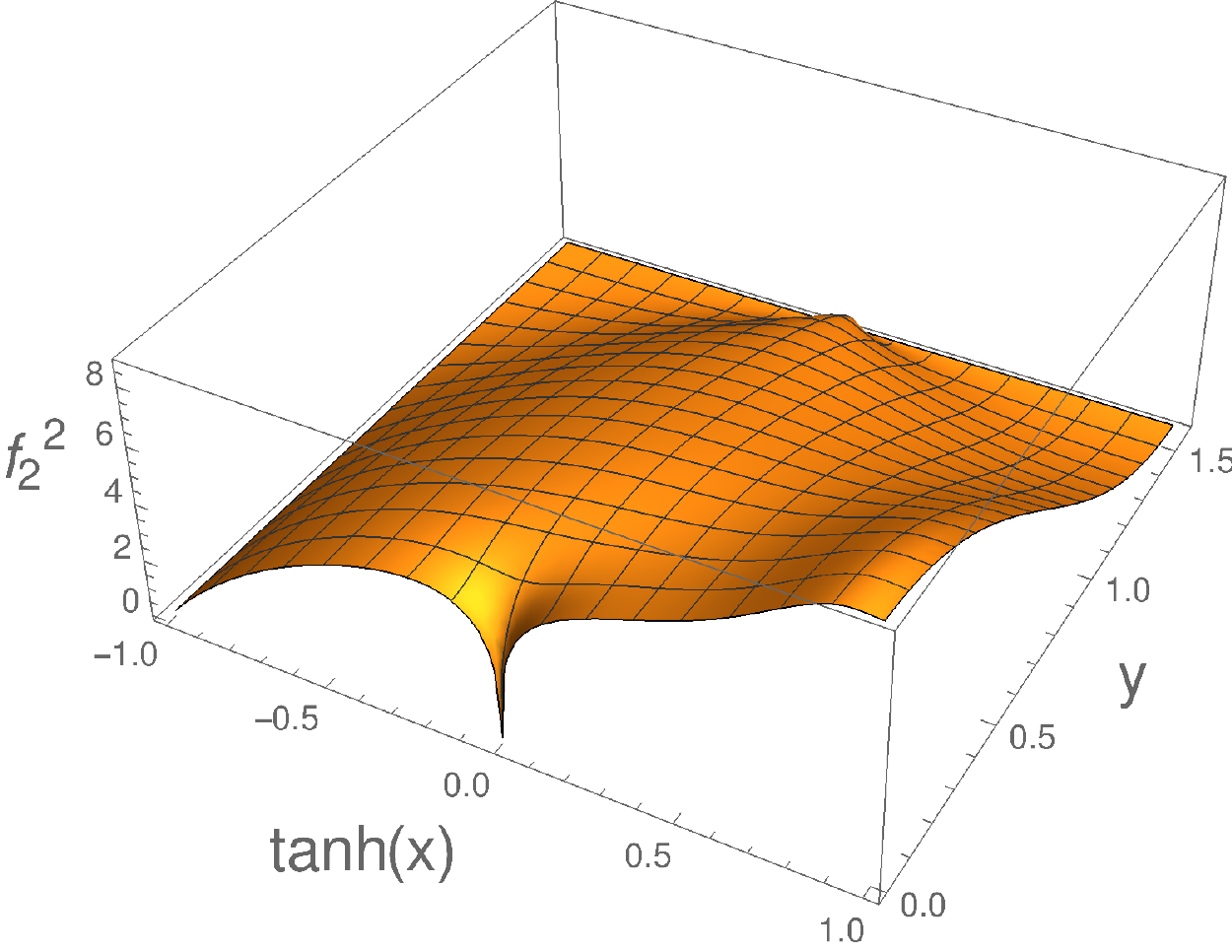}
	\includegraphics[width=0.242\linewidth]{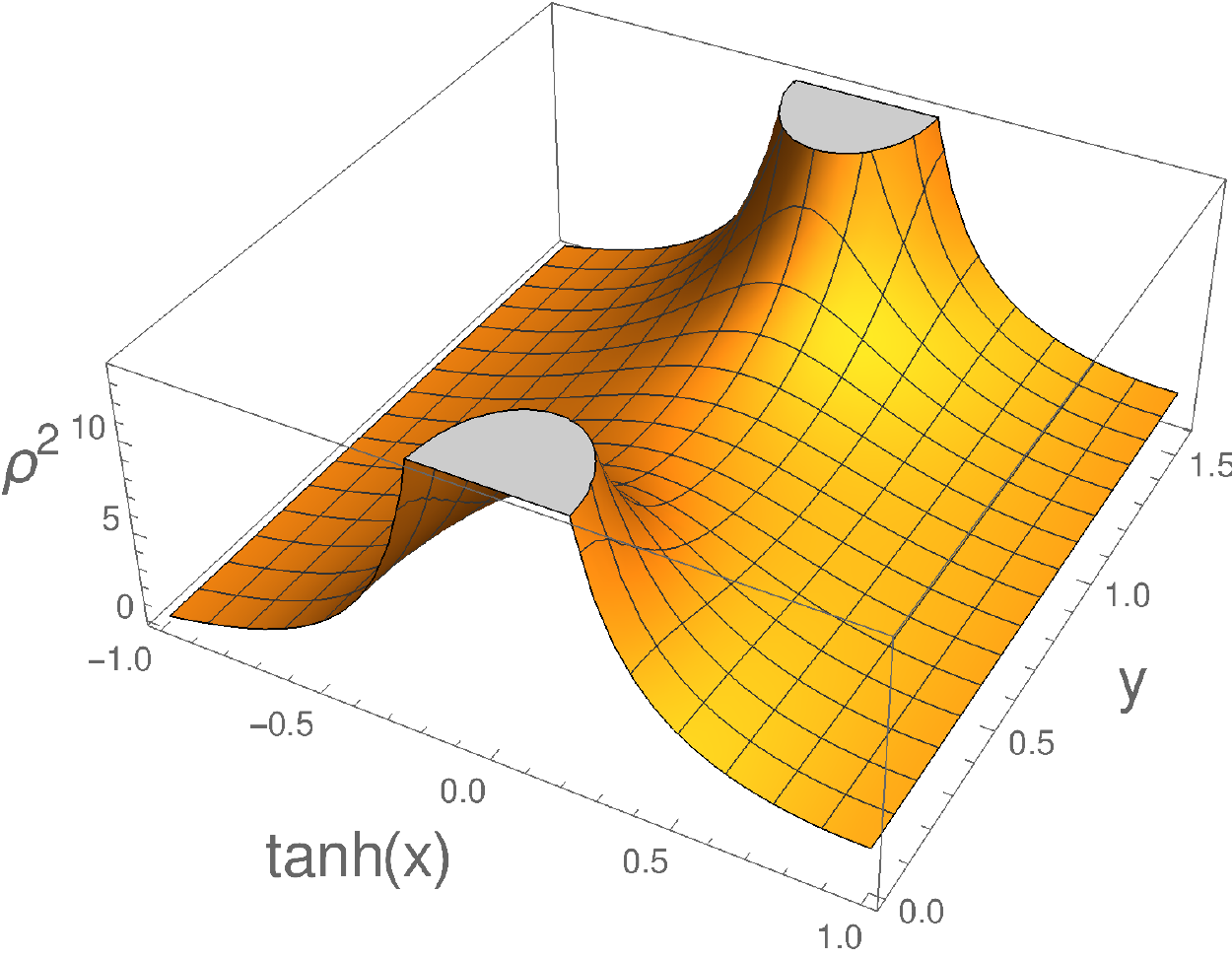}
	\\
	\includegraphics[width=0.242\linewidth]{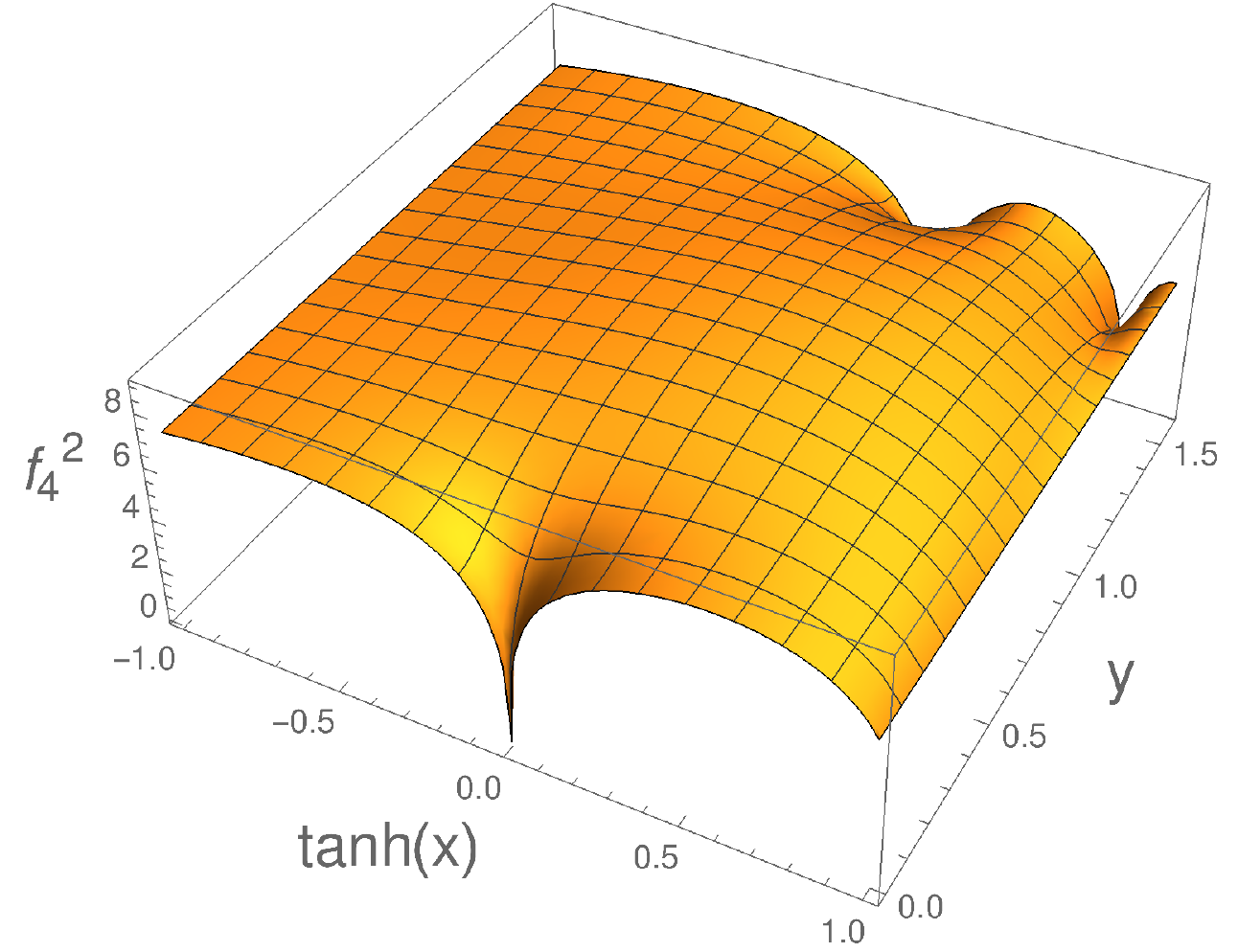}
	\includegraphics[width=0.242\linewidth]{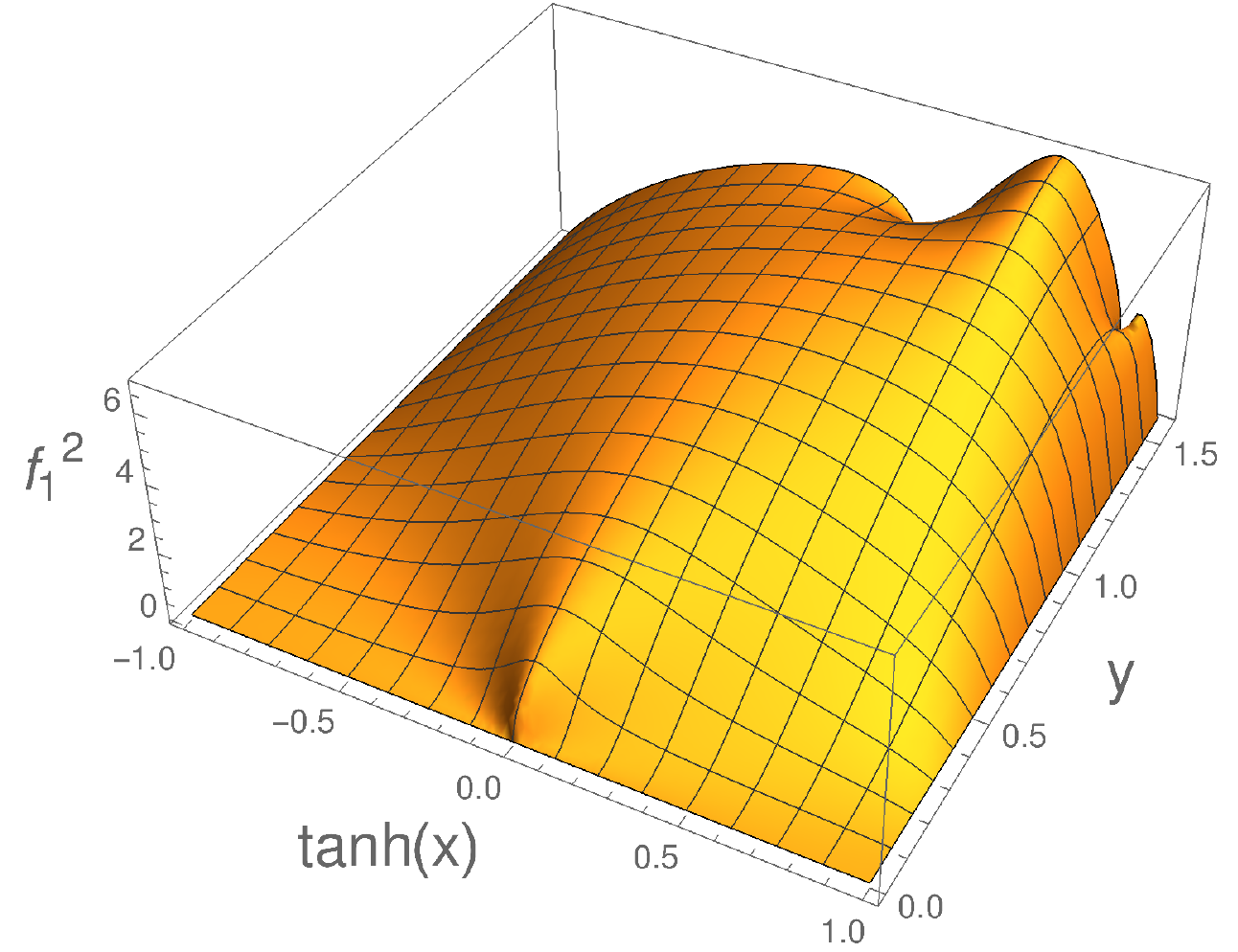}
	\includegraphics[width=0.242\linewidth]{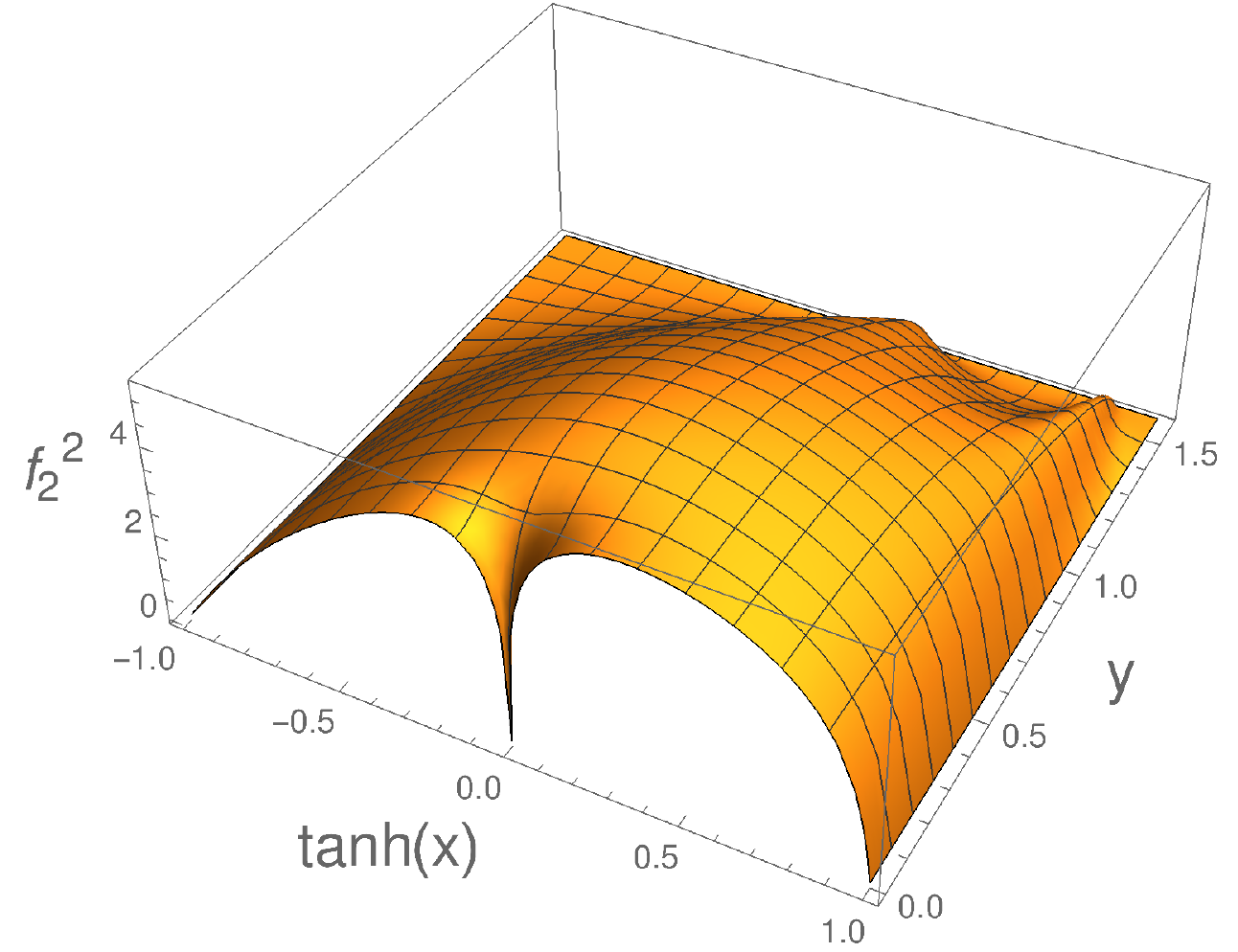}
	\includegraphics[width=0.242\linewidth]{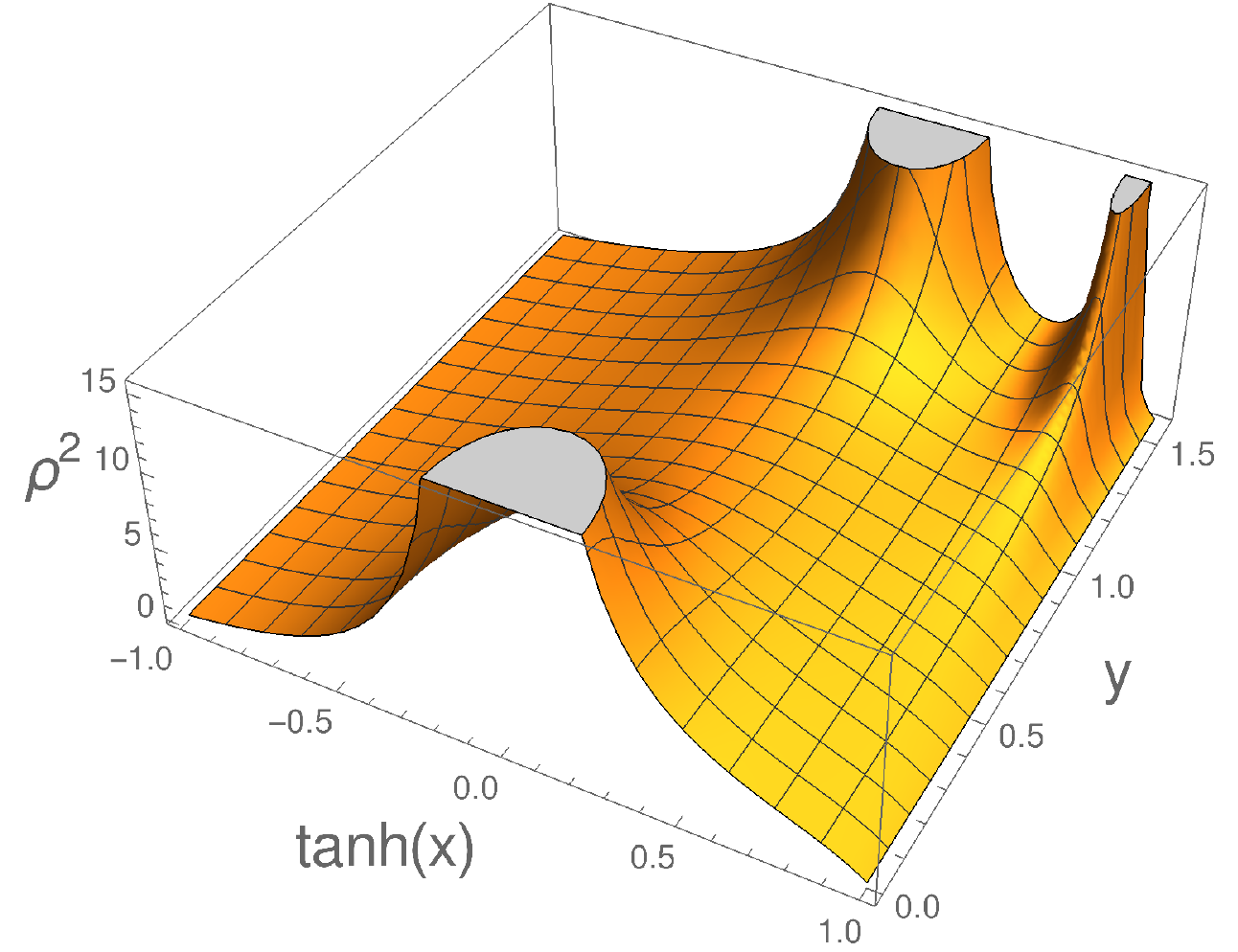}
	
	\caption{Top row: from left to right the Einstein-frame metric functions $f_4^2$, $f_1^2$, $f_2^2$, $\rho^2$ in the ansatz (\ref{eq:metric-gen}) with $z=x+iy$ for the BCFT dual (\ref{eq:h1h2-BCFT}) with $N_5/K=12$. Bottom row: same functions for the 3d dual (\ref{eq:h1h2-intermediate-max}).\label{fig:metric-plots}}
\end{figure}

A natural question is how the geometric features compare between the 10d AdS$_4$ solution dual to the 3d defect SCFT in the intermediate description on the one hand and the full BCFT dual on the other.
This will lead to a clear distinction between the proper intermediate picture $(I_p)$ and the na\"ive intermediate picture $(I_n)$.
Sample plots for the metric functions are shown in fig.~\ref{fig:metric-plots}.

For both solutions the metric functions $f_{1/2}^2$ and $\rho^2$ vanish at $\Re(z)\rightarrow-\infty$, making $\Re(z)=-\infty$ a regular point in the internal space.
In both cases $f_1^2$ vanishes at $\Im(z)=0$ and $f_2^2$ vanishes at $\Im(z)=\frac{\pi}{2}$, closing off the internal space at these boundaries of the strip $\Sigma$.
The qualitative difference is at $\Re(z)\rightarrow+\infty$. In the BCFT solution $f_{1/2}^2$ are finite but non-zero in that limit, while $f_4^2$ blows up. The geometry becomes AdS$_5\times$S$^5$. In the 3d dual, on the other hand, $f_{1/2}^2$ and $\rho^2$ vanish and $f_4^2$ is finite at $\Re(z)\rightarrow\infty$, making it  a regular point in the internal space.

The external branes in the associated brane configurations, which appear explicitly as sources in the supergravity solutions, also differ, and this is reflected in the metric functions.
In the BCFT dual we have $N_5$ NS5-branes at $z=0$ and $N_5$ D5-branes at $z=\frac{i\pi}{2}$, as well as $2N_5K$ D3-branes emerging at $\Re(z)\rightarrow\infty$. The 5-brane sources can be seen in fig.~\ref{fig:metric-plots} as the points where $\rho^2$ blows up while $f_{1/2}^2$ and $f_4^2$ vanish. The 5-brane sources are at infinite proper distance from other points on $\Sigma$, due to the growing behavior of $\rho^2$.
For the 3d dual we have $N_5$ NS5-branes at $z=0$ and $N_5$ D5-branes at $z=\frac{i\pi}{2} + \delta_1$, as well as $2N_5K$ D5-branes at $z=\frac{i\pi}{2}+\delta_2$.
So the $2N_5K$ D3-branes have been replaced by $2N_5K$ D5-branes, combined with a  relocation of the original $N_5$ D5-branes.
This changes the geometry of the internal space so that it closes off smoothly on all boundaries of $\Sigma$.

Clearly the 3d dual is not obtained by taking only a part of $\Sigma$ from the full BCFT dual, such as e.g.\ the 3d region in in fig.~\ref{fig:braneworld}.
The metric functions do not vanish in the interior of $\Sigma$, so slicing up the BCFT dual and taking it apart along any dividing line through the interior of $\Sigma$ would lead to a geometry which is not geodesically complete, and instead has an actual boundary at the dividing line. This would be different from our proposal for the actual 3d dual.

\begin{figure}
	\includegraphics[width=0.37\linewidth]{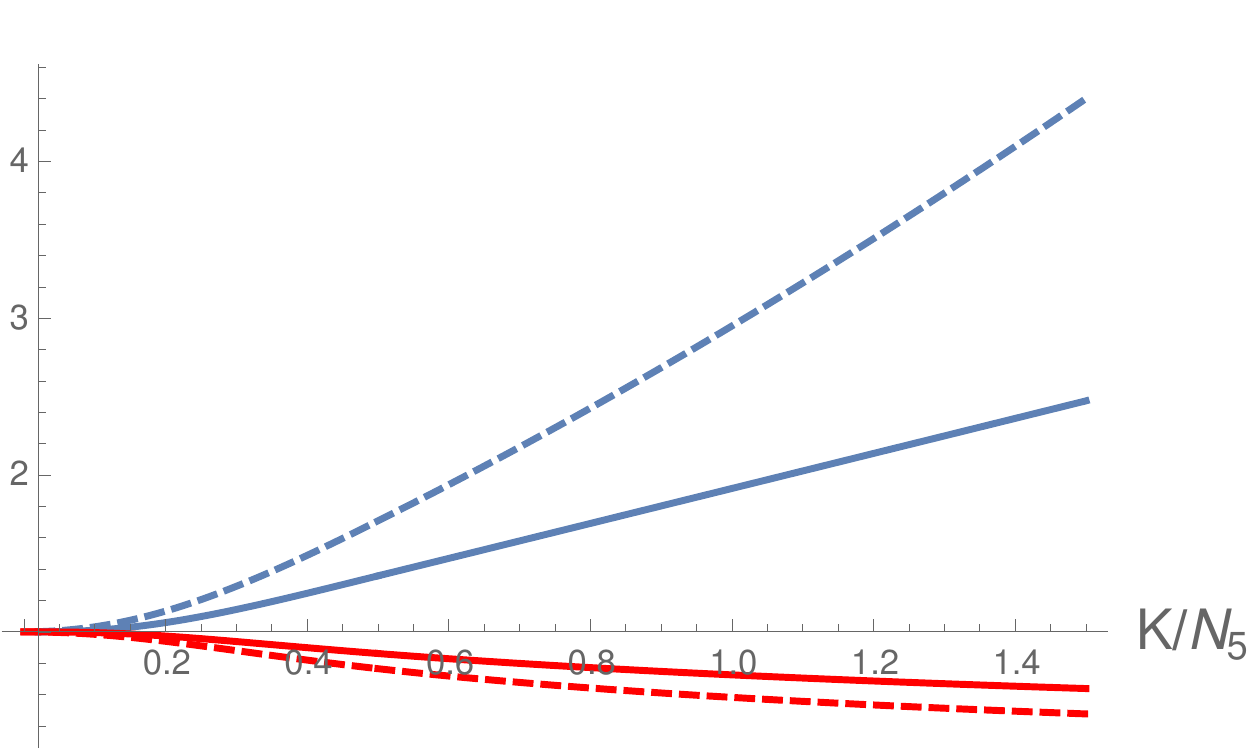}
	\caption{Ratio between geometric quantities at the left end of $\Sigma$, $\Re(z)= -\infty$, in the 3d defect duals (\ref{eq:h1h2-intermediate-max}), (\ref{eq:h1h2-intermediate-max-2}) and the BCFT dual (\ref{eq:h1h2-BCFT}).
	The blue (red) curves show the ratio of the Ricci scalars (AdS$_4$ radii $f_4$). The solid (dashed) curves are for Einstein (string) frame.\label{fig:f4ratio}}
\end{figure}

Moreover, a feature which is only implicit in fig.~\ref{fig:metric-plots} is that the geometry of the 3d dual generally differs from the full BCFT dual even deep in the 3d region, at the left end of the strip.
A comparison between the AdS$_4$ curvature radii and Ricci scalars of the two solutions at the point where both two-spheres collapse, $\Re(z)\rightarrow -\infty$, is shown in fig.~\ref{fig:f4ratio}. The respective ratios approach one for $K/N_5\rightarrow 0$, but are otherwise different.
The solutions thus become identical even at the left end of the strip only for $K/N_5\rightarrow 0$.
The 3d dual of the proper intermediate description can thus also not be obtained by taking a part of $\Sigma$ from the full BCFT dual and only modifying the geometry in a vicinity of the resulting boundary.
The plots show that this holds in Einstein and in string frame.

\subsection{Identifying points on $\Sigma$}\label{sec:identify-Sigma}

We realized an intermediate holographic description which properly geometrizes the 3d degrees of freedom, and found that the geometry in general differs non-trivially between the 3d dual and the full BCFT dual.
To some extent one can still identify points on $\Sigma$ between the two solutions, by characterizing them in terms of CFT data.
We will illustrate this for Wilson loops, based on the results of \cite{Coccia:2021lpp}.
We define a coordinate system on the brane diagram and discuss how it can be imported onto the strip $\Sigma$. By applying this procedure in the full BCFT dual and in the 3d dual, we obtain an identification between points on $\Sigma$ in the 3d dual and in a subset of the BCFT dual.

\begin{figure}
	\centering
	\begin{tikzpicture}[scale=2.5]	
		\node at (0.31,0) {\ldots};
		
		\foreach \i in {-0.08,0,0.08} \draw[thick] (0.5,\i) -- (1,\i);
		\foreach \i in {-0.18,-0.06,0.06,0.18} \draw[thick] (1,\i) -- (2,\i);
		\foreach \i in {-0.04,0.04} \draw[thick] (2,\i) -- (2.5,\i);
		
		\node at (2.7,0) {\ldots};
		
		\foreach \i in {1,2} \draw[fill=black] (\i,0) ellipse (1pt and 6pt);
		
		\fill [red] (1.5,0) circle (1.3pt);
		
		\node at (1,-0.33) {\footnotesize NS5};
		\node at (2,-0.33) {\footnotesize NS5};       			
		\node at (1.5,-0.3) {\footnotesize D3 };
		\node at (1.65,0) {\footnotesize D5$^\prime$};
	\end{tikzpicture}
	\hskip 10mm
	\begin{tikzpicture}[scale=2.5]
		
		\node at (0.31,0) {\ldots};
		
		\foreach \i in {-0.08,0,0.08} \draw[thick] (0.5,\i) -- (1,\i);
		\foreach \i in {-0.18,-0.06,0.06,0.18} \draw[thick] (1,\i) -- (2,\i);
		\foreach \i in {-0.04,0.04} \draw[thick] (2,\i) -- (2.5,\i);
		
		\node at (2.7,0) {\ldots};

		\foreach \i in {1,2} \draw[fill=black] (\i,0) ellipse (1pt and 6pt);
		
		\fill [red] (1.5,0.6) circle (1.3pt);
		
		\draw[thick,blue] (1.48,0.558) -- (1.48, 0.18);
		\draw[thick,blue] (1.52,0.558) -- (1.52, 0.055);
		
		\node at (1,-0.33) {\footnotesize NS5};
		\node at (2,-0.33) {\footnotesize NS5};       			
		\node at (1.5,-0.3) {\footnotesize D3 };
		\node at (1.68,0.6) {\footnotesize D5$^\prime$};
		\node at (1.63,0.35) {\footnotesize F1};
	\end{tikzpicture}
	\caption{D5$^\prime$-brane realization of a Wilson loop in the antisymmetric representation  of rank $k$ at the $t^{\rm th}$ gauge node.
		The D5$^\prime$ extends along the (045678) directions.	
		Moving the D5$^\prime$ out of the D3-brane stack across $k$ D3-branes creates $k$ fundamental strings stretched between the D3-branes and the D5$^\prime$-brane.\label{fig:Wilson}}
\end{figure}
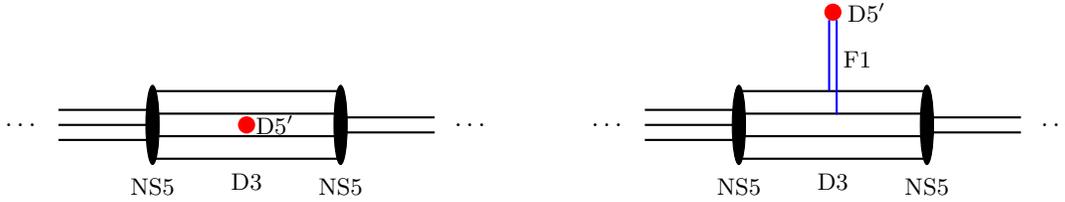

\begin{figure}
	\subfigure[][]{\label{fig:BCFT-coords}		
		\begin{tikzpicture}			
			\node at (0,0){
				\includegraphics[width=0.35\linewidth]{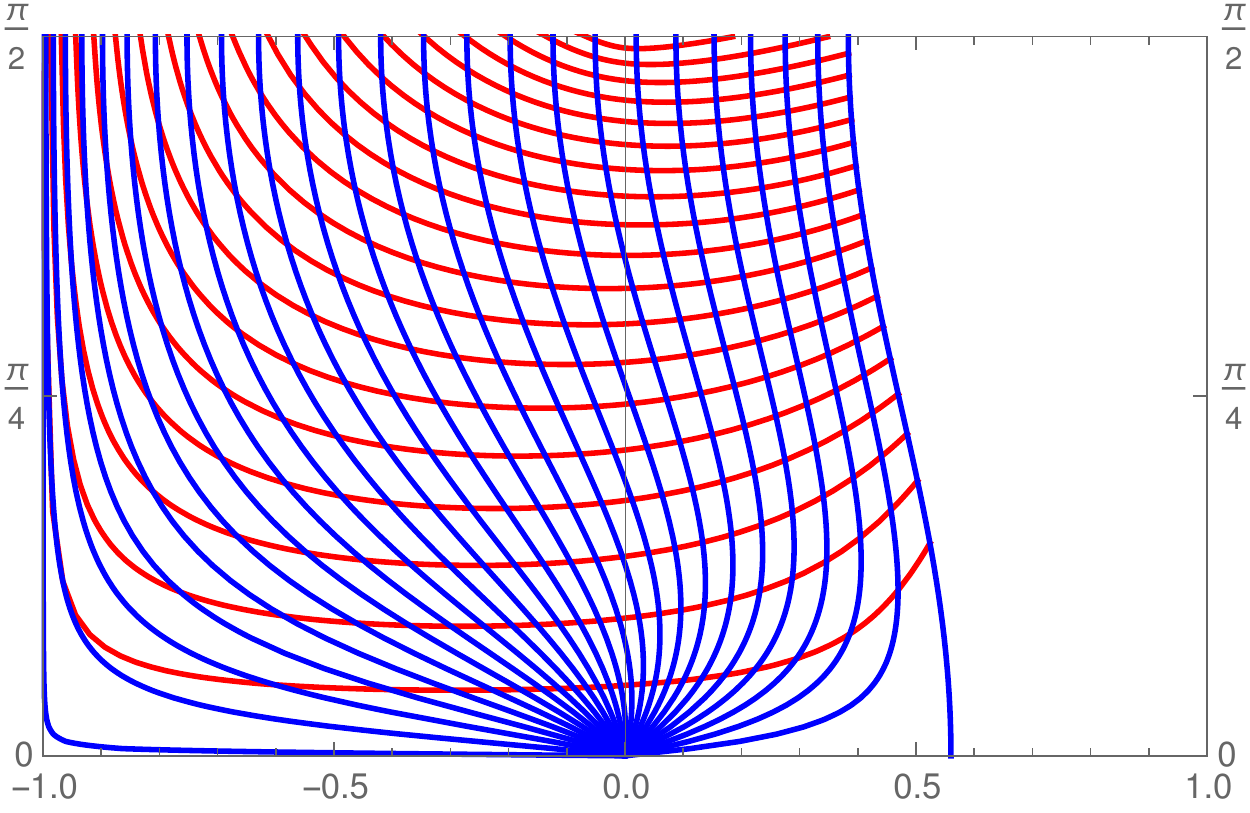}};
			\draw[very thick] (0,-1.55) -- +(0,-0.25) node [anchor=north] {\small NS5};
			\draw[very thick] (0,1.65) -- +(0,0.2) node [anchor=south] {\small D5};
			\node at (3.2,0.05) {\small D3};
			
		\end{tikzpicture}
	}
	\hskip 10mm
	\subfigure[][]{\label{fig:3d-coords}
		\begin{tikzpicture}
			\node at (-8,0) {\includegraphics[width=0.35\linewidth]{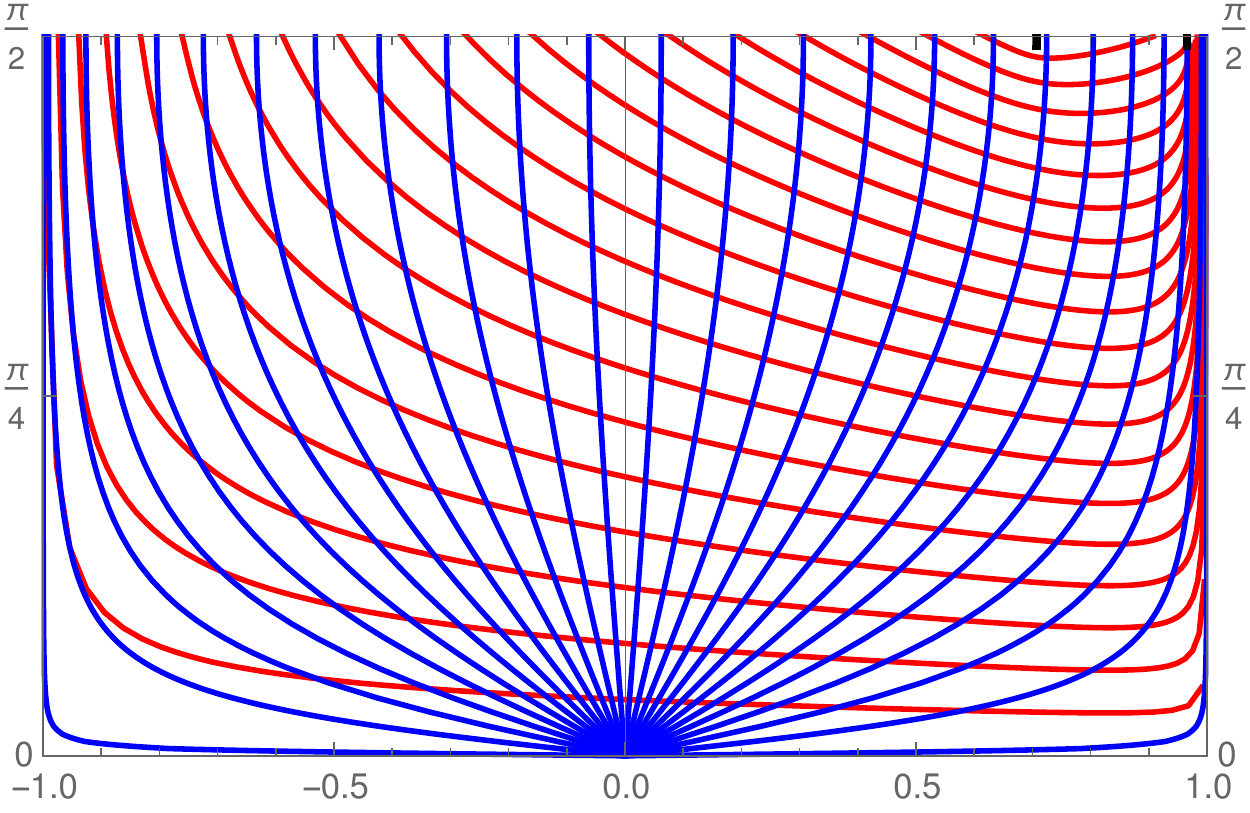}};
			\draw[very thick] (-8,-1.55) -- +(0,-0.25) node [anchor=north] {\small NS5};
			\draw[very thick] (-5.42,1.65) -- +(0,0.2) node [anchor=south] {\small D5};
			\draw[very thick] (-6.1,1.65) -- +(0,0.2) node [anchor=south] {\small D5};			
		\end{tikzpicture}
	}
	\caption{Left: `Wilson loop coordinates' on (part of) $\Sigma$ for the BCFT dual (\ref{eq:h1h2-BCFT}) with $N_5=4K$. The D5$^\prime$-branes extend along the blue curves along which $h_2^D$ is constant. The red curves trace the cap-off points of the D5$^\prime$ branes. Compare the region carved out by the blue curves to fig.~\ref{fig:braneworld}.
		Right: Corresponding Wilson loop coordinates on $\Sigma$ for 3d dual (\ref{eq:h1h2-intermediate-max-2}) with $N_5=4K$. The coordinates now cover $\Sigma$ entirely.
		\label{fig:BCFT-3d-Sigma-coords}}
\end{figure}
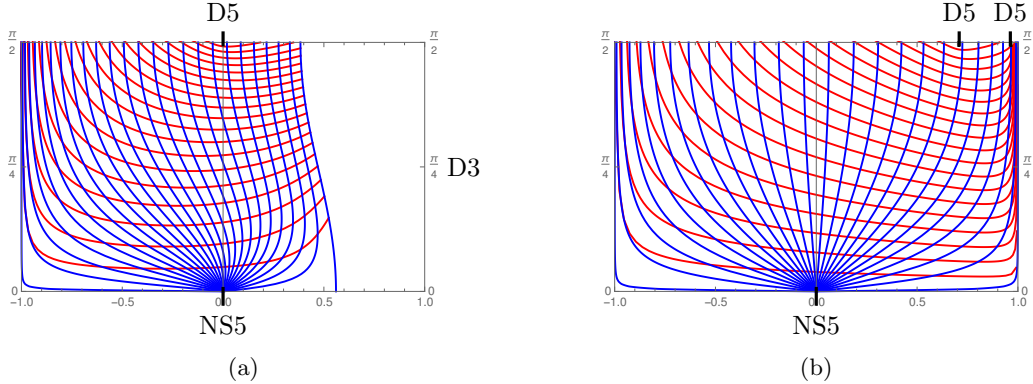

Wilson loops in antisymmetric representations associated with individual 3d gauge nodes can be realized in the brane construction by D5$^\prime$ branes \cite{Assel:2015oxa}, as shown in fig.~\ref{fig:Wilson}. The D5$^\prime$ can be characterized by a horizontal coordinate in the brane diagram, specifying between which pair of NS5-branes the D5$^\prime$ is inserted, and a vertical coordinate specifying where along the corresponding D3-brane stack the D5$^\prime$ is inserted.
In the supergravity solutions the Wilson loops are realized by probe D5$^\prime$-branes wrapping AdS$_2$ in AdS$_4$, a curve in $\Sigma$, the entire sphere $S_1^2$, and an $S^1$ in $S_2^2$ \cite{Coccia:2021lpp}.
The curve in $\Sigma$ wrapped by the D5$^\prime$ corresponds to the gauge node which the Wilson loop is associated with, and how far the D5$^\prime$ extends along this curve encodes the rank of the representation.
With that information one can transfer the aforementioned coordinate system on the brane diagram onto $\Sigma$.
This is shown in fig.~\ref{fig:BCFT-3d-Sigma-coords}. Through this connection between the geometry on $\Sigma$ and the brane diagram, the 3d quiver diagram is directly encoded in the supergravity dual.

To make things concrete  we start with the BCFT dual, with $h_{1/2}$  in (\ref{eq:h1h2-BCFT}). The dual harmonic function to $h_2$ (using $h_2=\cA_2+\bar \cA_2$, $h_2^D=i(\cA_2-\bar \cA_2)$ with holomorphic $\cA_2$) is given by
\begin{align}\label{eq:h1h2-BCFT-dual}
	h_2^D&=\frac{i\pi \alpha^\prime}{4} K e^z-\frac{i\alpha^\prime}{4}N_5\ln\tanh\left(\frac{z}{2}\right)+\rm{c.c.}
\end{align}
The Wilson loop D5$^\prime$ branes extend along curves of constant $h_2^D$ (the blue curves in fig.~\ref{fig:BCFT-coords}), and we denote the start/end points by $z_0$/$z_1$. They carry D3 and F1 charges given by
\begin{align}\label{eq:ND3-NF1}
	N_{\rm D3}&=\frac{2}{\pi\alpha'}h_2^D~,
	&
	N_{\rm F1}&=\frac{4}{\pi^2{\alpha'}^2}\left[\Im\left(\cA_1\cA_2+\cC\right)\right]_{z_0}^{z_1}~,
\end{align}
where $\cC$ is defined by $\partial\cC=\cA_1\partial\cA_2-\cA_2\partial\cA_1$.
The gauge node which the Wilson loop is associated with is set by the D3 charge, with $0\leq h_2^D\leq \frac{\pi \alpha^\prime}{2}N_5$. The rank of the representation and the point along the curve on $\Sigma$ where the D5$^\prime$ caps off are set by the F1 charge. For each point on $\Sigma$ on which a Wilson loop D5$^\prime$ ends, we can now use  $(N_{\rm D3},N_{\rm F1})$ as coordinates. This defines points on $\Sigma$ in terms of field theory data, and in turn in terms of coordinates on the brane diagram.

The resulting coordinate system in fig.~\ref{fig:BCFT-coords} covers the region marked as 3d in fig.~\ref{fig:braneworld} and one of the transition regions.
The blue curve which does not start at the NS5 pole is the Wilson loop associated with the 4d gauge node (localized on the defect to preserve defect conformal symmetry).
In summary, one can use the D3 and F1 charges of a Wilson loop D5$^\prime$ ending at a given point on $\Sigma$ as coordinates. $N_{\rm D3}$  specifies the gauge node which the Wilson loop is associated with and $N_{\rm F1}$ the representation. In the brane diagram in fig.~\ref{fig:BCFT-brane}, $N_{\rm D3}$ determines between which pair of NS5 branes the D5$^\prime$ is located and $N_{\rm F1}$ specifies the vertical position.

We now turn to the dual of only the 3d degrees of freedom (\ref{eq:h1h2-intermediate-max}).
Since the gauge structure of the 3d quiver is identical to the 3d theory appearing at the boundary of the 4d BCFT, the set of antisymmetric 3d Wilson loops is identical.
This means the entire set of points on $\Sigma$ in the BCFT dual which can be identified using the Wilson loops can be translated to the intermediate description.
The dual harmonic function $h_2$ corresponding to (\ref{eq:h1h2-intermediate-max}) is
\begin{align}\label{eq:h1h2-3d-dual}
	h_2^D&=-\frac{i\alpha^\prime}{4}N_5\ln\tanh\left(\frac{z}{2}\right)+\rm{c.c.}
\end{align}
A coordinate system on $\Sigma$ can be defined in terms of field theory input in the same way as for the BCFT dual. The result is illustrated in fig.~\ref{fig:3d-coords}.

The identification of points on $\Sigma$ between the full BCFT dual and the 3d defect dual amounts to identifying points with the same Wilson loop coordinates. This literally identifies the curves shown in fig.~\ref{fig:3d-coords} with those in fig.~\ref{fig:BCFT-coords}.
Since the Wilson loop coordinates translate to coordinates in the brane diagram, this identification is very natural.
In the 3d dual $\Sigma$ is covered entirely by the Wilson loop coordinates, and mapped to a subset of $\Sigma$ in the full BCFT dual.
In that sense the 3d dual of the intermediate description corresponds to a `3d region' in the full BCFT dual.

The Wilson loop construction identifies AdS$_2$ slices (rather than individual points) between the BCFT dual and the 3d dual.
It is intended to illustrate the general principle -- different observables may call for different identifications. For example,
for NS5$^\prime$-branes describing vortex loops the coordinates in the BCFT dual would be obtained by a vertical reflection from those in fig.~\ref{fig:BCFT-coords}, while the coordinates in the 3d dual would differ more substantially from fig.~\ref{fig:3d-coords}.

\section{4d coupling and '\lowercase{t} Hooft limit}\label{sec:S-duality}

The solutions in (\ref{eq:h1h2-BCFT}) describe the brane configurations in fig.~\ref{fig:brane-D5NS5-D3} with the axio-dilaton $\tau=\chi+i/g_s$ approaching $\tau\rightarrow i$ in the asymptotic AdS$_5\times$S$^5$ region. This fixes the coupling for the 4d $\mathcal N=4$ SYM fields, $\tau=2\pi\theta+4\pi i/g_{\rm YM}^2$, accordingly.
Solutions with different asymptotic value of $\tau$ can be generated from (\ref{eq:h1h2-BCFT}) using $SL(2,\RR)$ transformations.
This way results obtained from the Einstein-frame metric associated with (\ref{eq:h1h2-BCFT}), such as the geodesics to be discussed in sec.~\ref{sec:shortcuts} (or the black hole discussions of \cite{Uhlemann:2021nhu}), are not affected, keeping the number of parameters for these discussions minimal.
One could introduce the asymptotic string coupling as independent parameter, but we will not do so here.
The intermediate descriptions for the transformed solutions can be obtained following the strategy of sec.~\ref{sec:intermediate}, and we will discuss their precise form below.

The $SL(2,\RR)$ transformations of Type IIB supergravity transform the 5-brane charges, and the transformation has to be chosen such that the resulting charges are properly quantized.
We use
\begin{align}\label{eq:Sg}
	S_g&=\begin{pmatrix} 1/\sqrt{g} & 0 \\ 0 & \sqrt{g}\end{pmatrix}~,
	&
	S_g: \quad \tau&\rightarrow \frac{\tau}{g}~, \quad B_{(2)}+iC_{(2)}\rightarrow \sqrt{g}B_{(2)}+\frac{i}{\sqrt{g}}C_{(2)}~,
\end{align}
which preserves the D5/NS5 nature of the 5-branes.
The Einstein-frame metric and 5-form are invariant.
In terms of the harmonic functions $h_{1/2}$ the transformation (\ref{eq:Sg}) is realized by
\begin{align}\label{eq:h1h2-BCFT-g}
	(h_1,h_2)&\rightarrow \left(\frac{1}{\sqrt{g}}h_1,\sqrt{g} h_2\right).
\end{align}
It rescales the string coupling to $g_s=g$, or $g_{\rm YM}^2=4\pi g$ in 4d $\mathcal N=4$ SYM, and rescales the D5 and NS5 charges by a factor $1/\sqrt{g}$ and $\sqrt{g}$, respectively.

\begin{figure}
	\centering
	\begin{tikzpicture}[y={(0cm,1cm)}, x={(0.707cm,0.707cm)}, z={(1cm,0cm)}, scale=1.4]
		\draw[gray,fill=gray!100,rotate around={-45:(0,0,1.8)}] (0,0,1.8) ellipse (1.8pt and 3.5pt);
		\draw[gray,fill=gray!100] (0,0,0) circle (1.5pt);
		
		\foreach \i in {-0.05,0,0.05}{ \draw[thick] (0,-1,\i) -- (0,1,\i);}

		\foreach \i in {-0.075,-0.025,0.025,0.075}{ \draw (-1.1,\i,1.8) -- (1.1,\i,1.8);}
		
		\foreach \i in {-0.045,-0.015,0.015,0.045}{ \draw (0,1.4*\i,0) -- (0,1.4*\i,1.8+\i);}
		\foreach \i in  {-0.075,-0.045,-0.015,0.015,0.045,0.075}{ \draw (0,1.4*\i,1.8+\i) -- (0,1.4*\i,4);}
		
		\node at (-0.18,-0.18,3.4) {\scriptsize $2N_5 K$ D3};
		\node at (1.0,0.3,1.8) {\scriptsize $N_5/\sqrt{g}$ D5};
		\node at (0,-1.25) {\footnotesize $\sqrt{g}N_5$ NS5};
		\node at (0.18,0.18,0.8) {{\scriptsize $N_5 K+\tfrac{N_5^2}{2}$ D3}};
	\end{tikzpicture}
	\caption{Brane configuration described by the supergravity  solution (\ref{eq:h1h2-BCFT}) after the $SL(2,\RR)$ transformation (\ref{eq:h1h2-BCFT-g}).
		The parameter $g$ should be chosen such that the 5-brane numbers are integer.
	\label{fig:brane-D5NS5-D3-g}}
\end{figure}
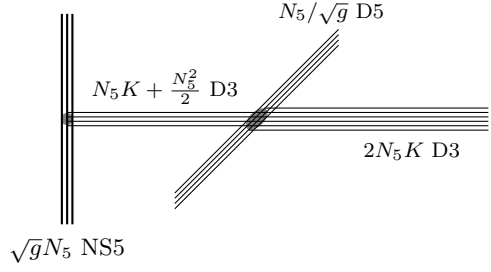

The brane configuration described by $h_{1/2}$ in (\ref{eq:h1h2-BCFT}) with the rescaling (\ref{eq:h1h2-BCFT-g}) can be obtained from fig.~\ref{fig:brane-D5NS5-D3} by rescaling the 5-brane charges, leading to fig.~\ref{fig:brane-D5NS5-D3-g}.  The numbers of D3-branes ending on each D5 and NS5 brane change accordingly.
The number of net D3-branes per D5-brane still changes sign at $N_5=2K$. The combined 3d/4d quiver for $N_5>2K$ now reads
\begin{align}\label{eq:D5NS5K-quiver-2-g}
	U(R/\sqrt{g})-U(2R/\sqrt{g})-\ldots &- U(R^2) - U(R^2-S/\sqrt{g})-\ldots - U(2N_5K+S/\sqrt{g}) - \widehat{U(2N_5K)}
	\nonumber\\
	&\ \ \ \ \ \ \,\vert\\
	\nonumber & \ \ [N_5/\sqrt{g}]	
\end{align}
Along the first ellipsis the rank increases in steps of $R/\sqrt{g}$, along the second it decreases in steps of $S/\sqrt{g}$. The total number of 3d nodes is $\sqrt{g}N_5-1$.
For $N_5<2K$ the combined 3d/4d quiver is
\begin{align}\label{eq:D5NS5K-quiver-1-g}
	U(R/\sqrt{g})-U(2R/\sqrt{g})-\ldots - U(N_5R-R/\sqrt{g}) - \widehat{U(2N_5K)}
\end{align}
The quivers are squeezed in length for small $g$, combined with an increase in the ranks of the gauge groups and the numbers of flavors at interior nodes, if present.

The 3d-only quivers and holographic duals can be obtained, as before, by introducing D5-branes terminating the 3d part of the BCFT quiver with flavors.
The 3d quiver for $N_5>2K$ is
\begin{align}\label{eq:3d-intermediate-quiver-g}
	U(R/\sqrt{g})-U(2R/\sqrt{g})-\ldots - &U(R^2) - U(R^2-S/\sqrt{g})-\ldots - U(2N_5K+S/\sqrt{g}) - [2N_5 K]
	\nonumber\\
	&\ \ \ \vert\\
	\nonumber & \!\!\![N_5/\sqrt{g}]	
\end{align}
For small $g$ the number of boundary flavors is decreased compared to the number of flavors at the interior node.
The supergravity dual for the 3d defect fields is given by the harmonic functions
\begin{align}\label{eq:h1h2-intermediate-max-g}
	h_1&=-\frac{\alpha^\prime}{4}\left[\frac{1}{\sqrt{g}}N_5\ln\tanh\left(\frac{i\pi}{4}-\frac{z-\delta_1}{2}\right)+2N_5K\ln\tanh\left(\frac{i\pi}{4}-\frac{z-\delta_2}{2}\right)\right]+\rm{c.c.}
	\nonumber\\
	h_2&=-\frac{\alpha^\prime}{4}\sqrt{g}N_5\ln\tanh\left(\frac{z}{2}\right)+\rm{c.c.}
\end{align}
with
\begin{align}
	\delta_1&=-\ln \tan \frac{\pi S}{2N_5}~,
	&
	\delta_2&=-\ln \tan \frac{\pi}{2\sqrt{g}N_5}~.
\end{align}
This 3d dual is not related to the one for $g=1$ in (\ref{eq:h1h2-intermediate-max}) by the duality transformation (\ref{eq:h1h2-BCFT-g}), since the terms in $h_1$ are not scaled uniformly and $\delta_2$ also changes; the split into 3d and 4d degrees of freedom does not commute with $SL(2,\RR)$.
For $N_5<2K$ the 3d quiver is
\begin{align}\label{eq:3d-intermediate-quiver-2-g}
	U(R/\sqrt{g})-U(2R/\sqrt{g})-\ldots - U(RN_5-R/\sqrt{g}) - [N_5R]
\end{align}
and the defect supergravity dual is defined by
\begin{align}\label{eq:h1h2-intermediate-max-2-g}
	h_1&=-\frac{\alpha^\prime}{4}N_5R\,\ln\tanh\left(\frac{i\pi}{4}-\frac{z-\delta_1}{2}\right)+\rm{c.c.}
	&
	\delta_1&=-\ln \tan \frac{\pi}{2\sqrt{g}N_5}
	\nonumber\\
	h_2&=-\frac{\alpha^\prime}{4}\sqrt{g}N_5\ln\tanh\left(\frac{z}{2}\right)+\rm{c.c.}
\end{align}
This completes the generalizations of the field theories and associated supergravity solutions.

We note that for the 3d defect SCFTs $g$ is a dimensionless parameter involved in specifying the quiver and brane diagrams, not a gauge coupling (which would be dimensionful in 3d -- the SCFTs arise as strongly-coupled IR fixed points of the gauge theories).
To make this more manifest we introduce
new parameters $N_{\rm NS5}$, $N_{\rm D5}$ and $\k$ as follows,
\begin{align}
	N_{\rm NS5}&=\sqrt{g}N_5~, & N_{\rm D5}&=N_5/\sqrt{g}~, & \k&=\frac{K}{N_5}~.
\end{align}
With these parameters the number of semi-infinite D3-branes in fig.~\ref{fig:brane-D5NS5-D3-g} is given by $2\k N_{\rm D5}N_{\rm NS5}$ and the asymptotic string coupling by $g=N_{\rm NS5}/N_{\rm D5}$.
The 't Hooft coupling of the 4d gauge node is given by $\lambda=g_{\rm YM}^2(2N_5K)=8\pi N_{\rm NS5}^2\k$.

We now discuss the large-$N$ limit.
One could keep $g$ fixed in the large-$N$ limit and realize an arbitrary fixed string coupling in the AdS$_5\times$S$^5$ region.
From the 4d $\mathcal N=4$ SYM perspective, the limit of fixed $g_{\rm YM}$ with $N_{\rm 4d}\rightarrow\infty$ would be the ``very strong coupling limit" considered  in \cite{Basu:2004dm,Binder:2019jwn,Chester:2019jas}.
A 't Hooft limit for the 4d degrees of freedom can be realized by scaling $g$ in the large-$N$ limit,
\begin{align}\label{eq:4d-tHooft}
N_{\rm D5}&\rightarrow\infty~, & N_{\rm NS5}&\gg 1~, & \lambda&=8\pi N_{\rm NS5}^2\k\gg 1~.
\end{align}
In terms of the original parameters this amounts to $N_5\rightarrow\infty$ with $\sqrt{g}N_5$ and $gN_5K$ large and fixed.
The numbers of D5-branes and D3-branes are of the same order and grow in the large-$N$ limit, while the number of NS5-branes and the 4d 't Hooft coupling are large and fixed. One could consider more general limits where $g\sim 1/N_5^{\beta}$ with $\beta>0$, but we will not discuss that here.

In gauge theory terms the number of 3d gauge nodes is large and fixed in the limit (\ref{eq:4d-tHooft}) while the ranks and flavor numbers scale.
In the BCFT dual (\ref{eq:h1h2-BCFT}) with the rescaling (\ref{eq:h1h2-BCFT-g}), $h_1$ is of order $N_{\rm D5}$ and $h_2$ of order $N_{\rm NS5}$.
The Einstein-frame metric is of order $ds^2_E\sim \sqrt{N_{\rm D5}N_{\rm NS5}}$ and the scale of the string-frame metric $ds^2_{\rm string}=e^\phi ds^2_E\sim N_{\rm NS5}$ is large and fixed.
In the 3d duals (\ref{eq:h1h2-intermediate-max-g}), (\ref{eq:h1h2-intermediate-max-2-g}), all terms in $h_{1}$ and in $h_2$ are of the same order and the positions of the 5-brane sources are fixed.

The free energies of the 3d SCFTs on $S^3$, as one measure for their central charge, can be obtained from the field theory results in \cite{Coccia:2020wtk} (another measure is the defect contribution to entanglement entropy in the BCFTs, which was analyzed in \cite{Raamsdonk:2020tin}).
For general 3d quivers with all nodes balanced, the free energy is given by \cite[eq.~(36)]{Coccia:2020wtk}.
For $N_5>2K$, when the quiver is given by  (\ref{eq:D5NS5K-quiver-2-g}), we find
\begin{align}
	F_{S^3}=\frac{N_{\rm D5}^2N_{\rm NS5}^2}{4\pi^2}\Re\Big[&\zeta(3)-\Li_3\left(-e^{2\pi i\k}\right)
	-4N_{\rm NS5}\k\left(\Li_3\left(-ie^{\pi i(\k-N_{\rm NS5}^{-1})}\right)-\Li_3\left(-ie^{\pi i(\k+N_{\rm NS5}^{-1})}\right)\right)
	\nonumber\\&
	-4N_{\rm NS5}^2\k^2 \left(\Li_3\left(e^{2\pi i N_{\rm NS5}^{-1}}\right)-\zeta(3)\right)
	\Big]~.
\end{align}
For $N_5<2K$ and the quiver in (\ref{eq:D5NS5K-quiver-1-g}), we find
\begin{align}
	F_{S^3}&=\frac{N_{\rm D5}^2N_{\rm NS5}^4}{16\pi^2}(2\k+1)^2\Re\Big[\zeta(3)-\Li_3\left(e^{2\pi i N_{\rm NS5}^{-1} } \right)
	\Big]~.
\end{align}
These field theory results exactly match the free energies obtained holographically (e.g.\ via \cite[eq.~(14)]{Coccia:2020wtk}) from the supergravity duals in (\ref{eq:h1h2-intermediate-max-g}), (\ref{eq:h1h2-intermediate-max-2-g}) as functions of $N_{\rm D5}$, $N_{\rm NS5}$ and $\k$.
The field theory derivation in \cite{Coccia:2020wtk} assumed that the number of gauge nodes, set by $N_{\rm NS5}$, is large, and that the ranks of the gauge groups, set by $N_{\rm D5}$, are large as well.
This holds for the limit (\ref{eq:4d-tHooft}) and we thus obtain a non-trivial check for the constructions.

The above discussion shows that the $SL(2,\RR)$ orbits of the BCFT duals (\ref{eq:h1h2-BCFT}) for each fixed $K/N_5$ contain solutions describing a standard 't Hooft limit for the 4d gauge node and a natural holographic regime for the 3d defect degrees of freedom. Results obtained from the Einstein-frame metric carry over immediately.
The identification of regions on $\Sigma$ according to which type of 3d loop operators are hosted in fig.~\ref{fig:braneworld} is also unaffected by the  transformation (\ref{eq:Sg}).
The form of our intermediate description depends on $g$, but the results of sec.~\ref{sec:intermediate} generalize straightforwardly:
The identification of points on $\Sigma$ between the BCFT dual and the 3d dual discussed in sec.~\ref{sec:identify-Sigma} can be carried out for general $g$. The coordinate system on $\Sigma$ in the BCFT dual is generally of the form in fig.~\ref{fig:BCFT-coords} and the coordinate system in the 3d dual generally covers $\Sigma$ entirely.
Perhaps most importantly, the conclusion of sec.~\ref{sec:compare-geometry}, that the geometry at the left end of the strip differs between the 3d dual and the BCFT dual  for non-zero $K/N_5$, extends to general $g$ and to the limit (\ref{eq:4d-tHooft}).

\section{Null geodesics in BCFT duals}\label{sec:shortcuts}

In \cite{Omiya:2021olc} it was pointed out in the context of bottom-up braneworld models that geodesics connecting a point on the ETW brane to a point on the conformal boundary of the full BCFT dual can be faster through the bulk than geodesics along the ETW brane (fig.~\ref{fig:braneworld-3d4d-1}).
In this section we analyse whether analogous geodesics exist in the full top-down duals for BCFTs. Implications for causality of the intermediate picture will be discussed in sec.~\ref{sec:causal}.

\subsection{Bottom-up braneworld models}\label{sec:bottomup-geodesics}

\definecolor{mycolor}{rgb}{0.965,0.882,0.741}
\definecolor{mycolor2}{rgb}{0.812,0.851,0.914}
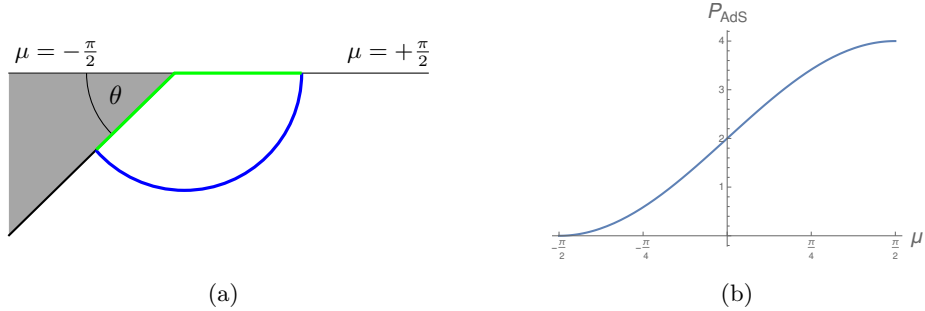
\begin{figure}
	\subfigure[][]{\label{fig:braneworld-3d4d-1}
		\begin{tikzpicture}[scale=1.3]
			\draw [white,fill=gray,opacity=0.7] (-0.8,0) -- (-2.5,0) -- (-2.5,-2/3*2.5)--(-0.8,0);
			\draw (-2.5,0) -- (1.8,0);	
			\draw[thick] (-0.8,0) -- (-2.5,-2/3*2.5);

			\node at (-0.6-0.8,-0.22) {\small $\theta$};
			\draw (-0.9-0.8,0) arc (180:225:25pt);
			\draw[white] (0,-2) circle (0.1pt);
			
			\node at (-2,0.2) {\footnotesize $\mu=-\frac{\pi}{2}$};
			\node at (1.4,0.2) {\footnotesize $\mu=+\frac{\pi}{2}$};
			
			\draw[very thick,blue] (0.5,0) arc (0:-139:1.2);
			\draw[very thick,green] (0.5,0) -- (-0.8,0) -- +(-1.7*0.47,-2/3*2.5*0.47);
		\end{tikzpicture}
	}	\hskip 10mm
	\subfigure[][]{\label{fig:braneworld-3d4d-2}
		\begin{tikzpicture}
			\node at (0,0) {\includegraphics[width=0.3\linewidth]{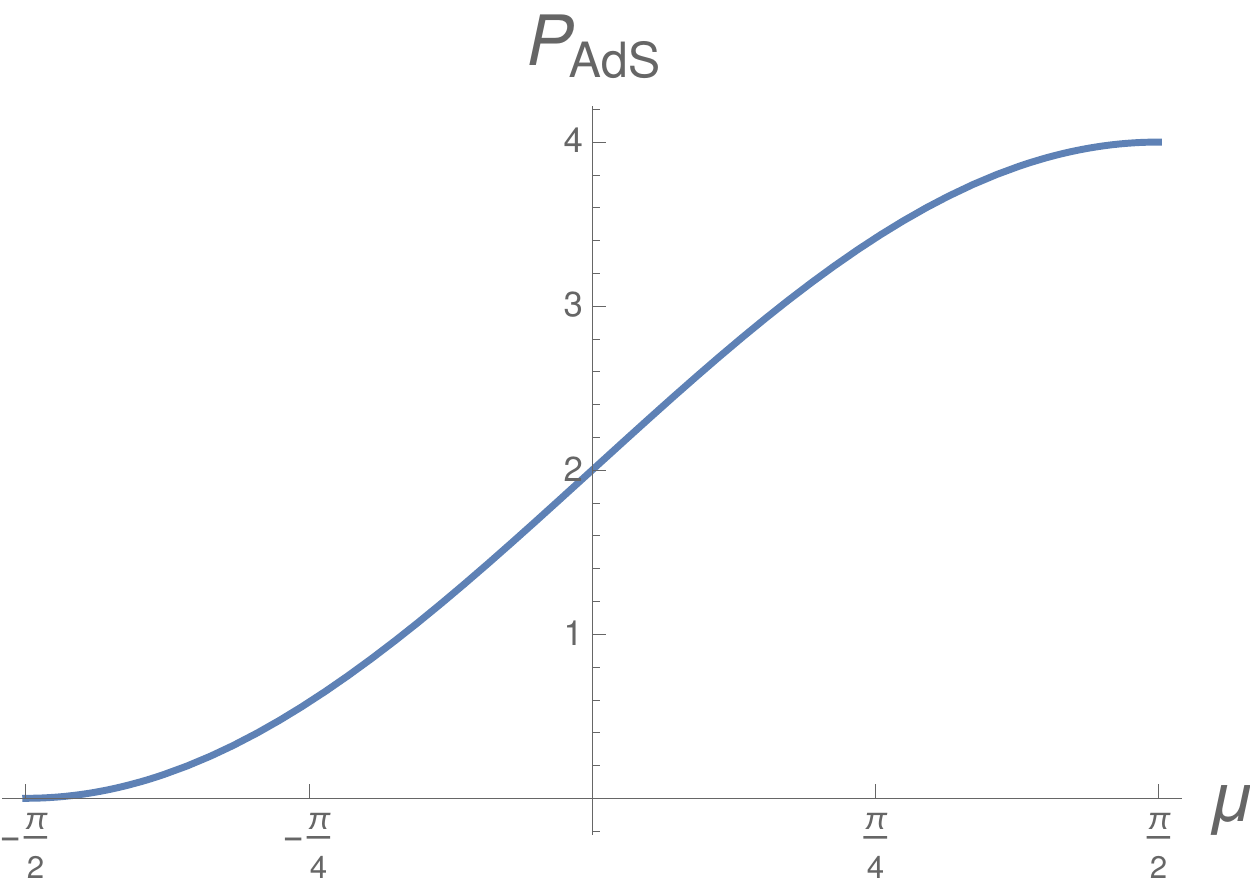}};
		\end{tikzpicture}
	}
	\caption{Left: Braneworld model with coordinates (\ref{eq:braneworld-coords}) and brane angle $\theta$. The ``along the brane" geodesic connecting a point on the ETW brane to a point in the ambient CFT is shown in green, the ``through the bulk" geodesic in blue.
	Right: $\cP_{AdS_5}(\mu)$ defined in (\ref{eq:timediff-AdS5}).\label{fig:braneworld-3d4d}}
\end{figure}

For later reference, we first review the simplest form of the ``bulk shortcut geodesics" discussed in  \cite{Omiya:2021olc} in the bottom-up braneworld models.
We start from AdS$_5$ with metric
\begin{align}\label{eq:braneworld-coords}
	ds^2&=\frac{1}{\cos^2\!\mu}\left(d\mu^2+\frac{du^2-dt^2+ds^2_{\RR^2}}{u^2}\right)~,
\end{align}
and place the ETW brane at $\mu=\mu_\star\leq 0$. The region $\mu<\mu_\star$ is cut off and the remaining part of the conformal boundary is $\mu=\frac{\pi}{2}$ (fig.~\ref{fig:braneworld-3d4d-1}).
The geodesics of interest connect a point $(\mu,u)=(\mu_\star,u_0)$ on the brane to a point $(\mu,u)=(\frac{\pi}{2},u_1)$ on the conformal boundary, at the same location on $\RR^2$.
The times taken by geodesics along the brane and through the bulk are, respectively,
\begin{align}\label{eq:bottom-up-times}
	t_{\rm brane}^2&=(u_0+u_1)^2~,
	\nonumber\\
	t_{\rm bulk}^2&=(u_0+u_1)^2-2u_0u_1(1+\sin\mu_\star)~.
\end{align}
The time difference can be expressed conveniently by introducing a function $\cP_{{\rm AdS}_5}(\mu_\star)$,
\begin{align}\label{eq:timediff-AdS5}
	t_{\rm brane}^2-t_{\rm bulk}^2&=u_0u_1 \cP_{{\rm AdS}_5}(\mu_\star)~,
	&
	\cP_{{\rm AdS}_5}(\mu_\star)&=2(1+\sin\mu_\star)~.
\end{align}
The function $\cP_{{\rm AdS}_5}(\mu_\star)$ interpolates between $4$ for $\mu_\star=\frac{\pi}{2}$ and $0$ for $\mu_\star=-\frac{\pi}{2}$ (fig.~\ref{fig:braneworld-3d4d-2}).
The qualitative features for fixed $u_0$, $u_1$ are as follows:  For blunt brane angle ($\mu_\star$ close to zero) the bulk geodesic reaches $(\frac{\pi}{2},u_1)$ significantly faster than the geodesic along the brane. As the brane angle gets sharper, the time saved through the bulk decreases. Asymptotically, for a brane approaching the conformal boundary, $\mu_\star\rightarrow -\frac{\pi}{2}$, the two paths take the same time.

For the time difference in (\ref{eq:timediff-AdS5}), $\mu_\star$ was assumed to be the location of the ETW brane.
The result, however, can be applied to geodesics starting at a generic AdS$_4$ slice with angular coordinate $\mu$, regardless of whether there is a brane or not.
The time difference then is between a geodesic connecting the bulk starting point to the 3d defect along fixed $\mu$ and from there to the end point in the ambient CFT on the one hand, and a geodesic directly connecting the two points through the bulk on the other.
We will use that perspective in the following.
For a setup with ETW brane at $\mu=\mu_\star$, the range of $\mu$ is restricted to $\mu\geq\mu_\star$ and $\cP_{{\rm AdS}_5}(\mu)$ interpolates between $4$ and $2(1+\sin\mu_\star)$.

\subsection{10d string theory BCFT duals}

In the 10d solutions we want to send a signal from the 10d analog of the ETW brane in the full BCFT dual to the 4d conformal boundary.
We can for example pick a point in the ``3d region" of the geometry (fig.~\ref{fig:braneworld})  as starting point. The 4d conformal boundary where the ambient CFT lives emerges at $\Re(z)\rightarrow\infty$. The end point should be in this region.
Since null geodesics take the same form in Einstein frame and string frame, we may choose either. The results will in particular be independent of the parameter $g$ introduced in sec.~\ref{sec:S-duality}.

The full 10d geometry in Einstein frame is given by (\ref{eq:metric-gen}), and for the AdS$_4$ part we use the coordinates
\begin{align}\label{eq:ds2-AdS4}
	ds^2_{AdS_4}=\frac{du^2-dt^2+ds^2_{\RR^2}}{u^2}~.
\end{align}
On $\Sigma$ we introduce real coordinates $x$ and $y$ such that
\begin{align}
	z&=x+i y~, & |dz|^2&=dx^2+dy^2~.
\end{align}
We consider geodesics which start and end on the same locations in $S_1^2$ and $S_2^2$ and on the $\RR^2$ in AdS$_4$, so that these parts will not play a role.
To further simplify the discussions we consider special classes of geodesics which extend along fixed $y$; we will discuss which geodesics are compatible with this constraint shortly.
The geodesics we are interested in are then parametrized by
\begin{align}\label{eq:geodesic-u-lambda}
	u&=u(\lambda)~, & t&=t(\lambda)~, & x&=x(\lambda)~.
\end{align}
The initial point should be in the 3d region of $\Sigma$, as analog of  ``on the ETW brane". The end point should be on the 4d conformal boundary of the (locally) AdS$_5\times$S$^5$ region at $x\rightarrow\infty$. That is,
\begin{align}\label{eq:geod-lambda-i}
	u(\lambda_i)&=u_0~, &	t(\lambda_i)&=0~,& x(\lambda_i)&=x_0~,\nonumber\\
	u(\lambda_f)&=u_1 &	t(\lambda_f)&=t_1~, & x(\lambda_f)&=\infty~.
\end{align}
For the supergravity solutions (\ref{eq:h1h2-BCFT}) we find 3 values of $y_0$ for which it is consistent to constrain geodesics to propagate along constant $\Im(z)=y=y_0$.
These values are
\begin{align}
	y(\lambda)=y_0\in\left\lbrace 0,\frac{\pi}{4},\frac{\pi}{2}\right\rbrace~.
\end{align}
The first and third are on the boundary of the strip.
The second can be understood from the fact that the solutions (\ref{eq:h1h2-BCFT}) are invariant under S-duality, which makes the metric invariant under the reflection $\Im(z)\rightarrow \frac{\pi}{2}-\Im(z)$, which leaves the line $y_0=\frac{\pi}{4}$ fixed.

The geodesic equations are spelled out and evaluated in app.~\ref{app:geodesics}. The result is that null geodesics along constant $y_0$ are given by $(t,u(t),x(t))$ which satisfy
\begin{align}\label{eq:null-geod-zx}
	u^2&=t^2+2u_0\dot u_0 t+u_0^2~, &
	\dot x&=\frac{f_4}{2u^2\rho}\,u_0\sqrt{1-\dot u_0^2}~.
\end{align}
The geodesics are circle segments in the $(u,t)$ plane and the geodesic equations are reduced to a single first-order equation for $x(t)$.
The initial conditions are $u(0)=u_0$, $\dot u(0)=\dot u_0$ and $x(0)=x_0$.
An integral expression for $t(x)$ is obtained in app.~\ref{app:geodesics} as
\begin{align}\label{eq:geod-t-x}
	t(x)&=u_0\sqrt{1-\dot u_0^2}\,\tan\left[\sin^{-1}\!\left(\dot u_0\right)+2\int_{x_0}^x d\hat x \frac{\rho(\hat x,y_0)}{f_4(\hat x,y_0)}\right]-u_0\dot u_0~.
\end{align}

\bigskip

\textbf{Geodesics ``along the brane":}
We first realize the 10d analog of the ``along the ETW brane" geodesics.
To this end, we note that the solution to (\ref{eq:null-geod-zx}) with $\dot u_0=\pm 1$ becomes
$u=t\pm u_0$ with constant $x=x_0$.
The ``along the brane" geodesics in the 10d solutions are realized by combining two such geodesics along constant $x$. Namely,
\begin{align}
	x&=x_0~, & u&=u_0-t~, & t&\in (0,u_0)~,
	\nonumber\\
	x&=\infty~, & u&=t-u_0~, & t&\in(u_0,u_0+u_1)~.
\end{align}
The first segment is a null geodesic connecting $u=u_0$ to $u=0$ in the AdS$_4$ space at $x=x_0$, i.e.\ at the starting point on $\Sigma$.
The second segment connects $u=0$ to $u=u_1$ along the AdS$_4$ slice at the 4d conformal boundary of the AdS$_5\times$S$^5$ region.
Combining these geodesics connects the initial point to the desired end point.
The curve arrives at the destination at
\begin{align}\label{eq:t-brane}
	t_{\rm brane}&=u_0+u_1~.
\end{align}
These geodesics do not explore the internal space and the arrival time is independent of the background data $N_5$ and $K$. This is analogous to the bottom-up models, in which the time $t_{\rm brane}$ along the brane in (\ref{eq:bottom-up-times}) is independent of the brane angle.

\bigskip

\textbf{Geodesics ``through the bulk":}
We now turn to geodesics connecting the beginning and end points ``through the bulk". We take this to mean geodesics proceeding from $x=x_0$ to $x=\infty$ without going through the conformal boundary of $AdS_4$.\footnote{One could consider ``in between" geodesics which reach the conformal boundary at some $x\in(x_0,\infty)$  and connect to a segment at $x=\infty$ from there. Since the segment at $x=\infty$ would be shared with the corresponding ``along the brane" geodesic, we do not consider this option separately.}
For given background specified by $N_5$ and $K$, the 10d geodesics depend on the initial values $x_0$, $u_0$, $\dot u_0$. The form of the geodesics only depends non-trivially on the ratio $N_5/K$, so we have one relevant parameter in the background.

\begin{figure}
	\includegraphics[width=0.31\linewidth]{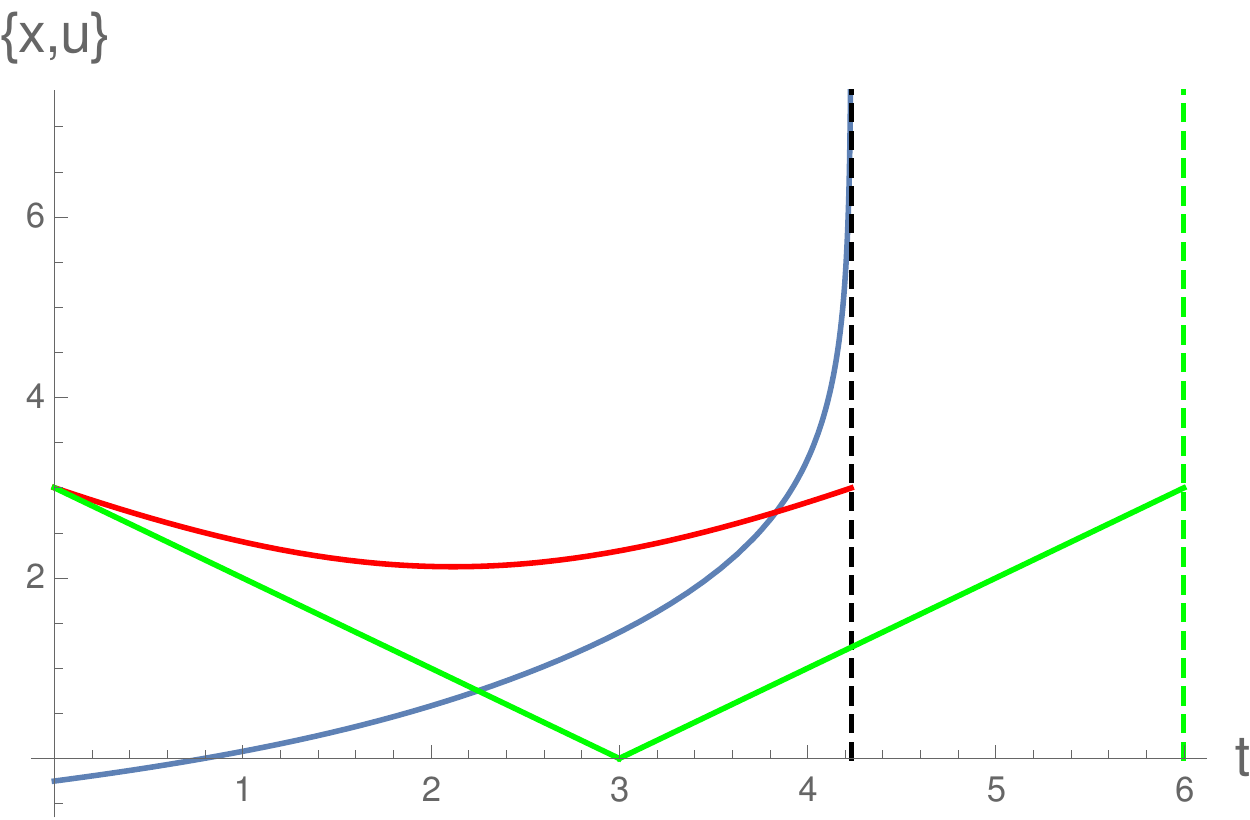}
	\includegraphics[width=0.31\linewidth]{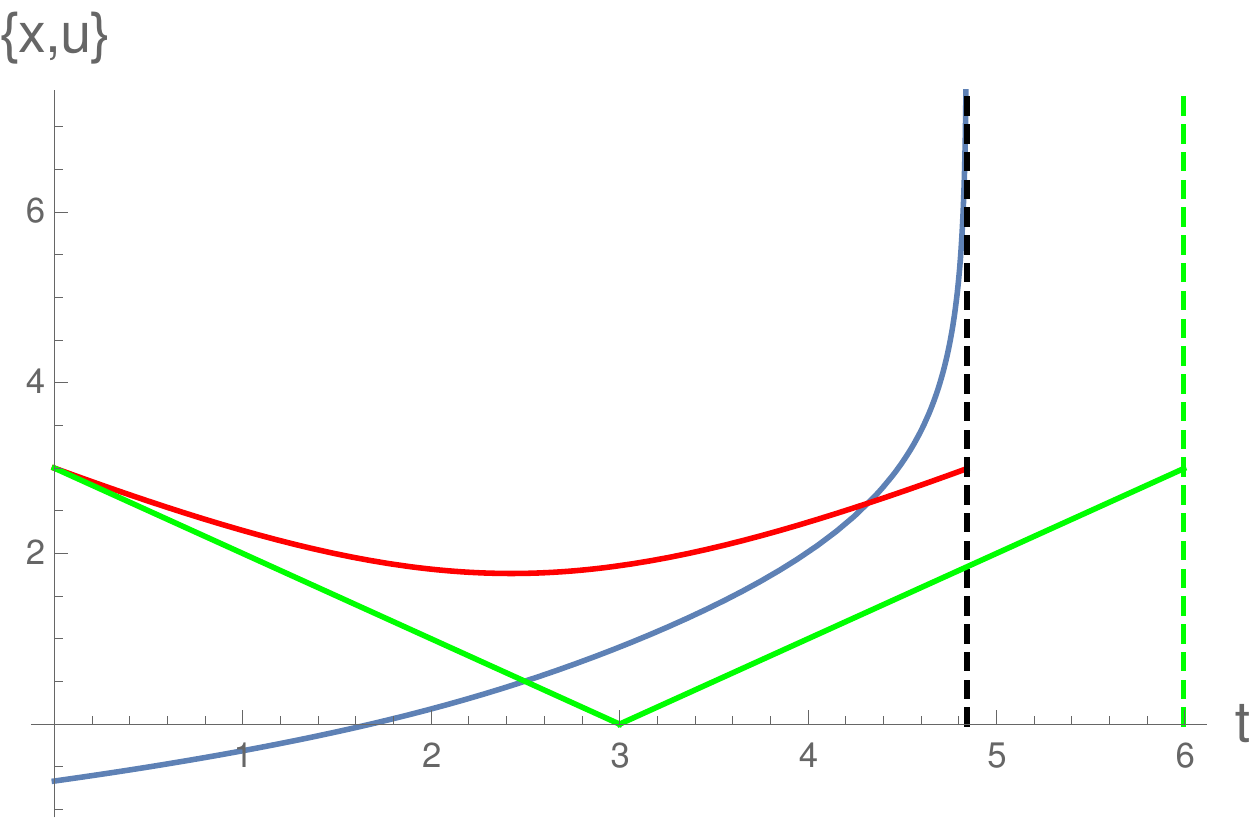}
	\includegraphics[width=0.31\linewidth]{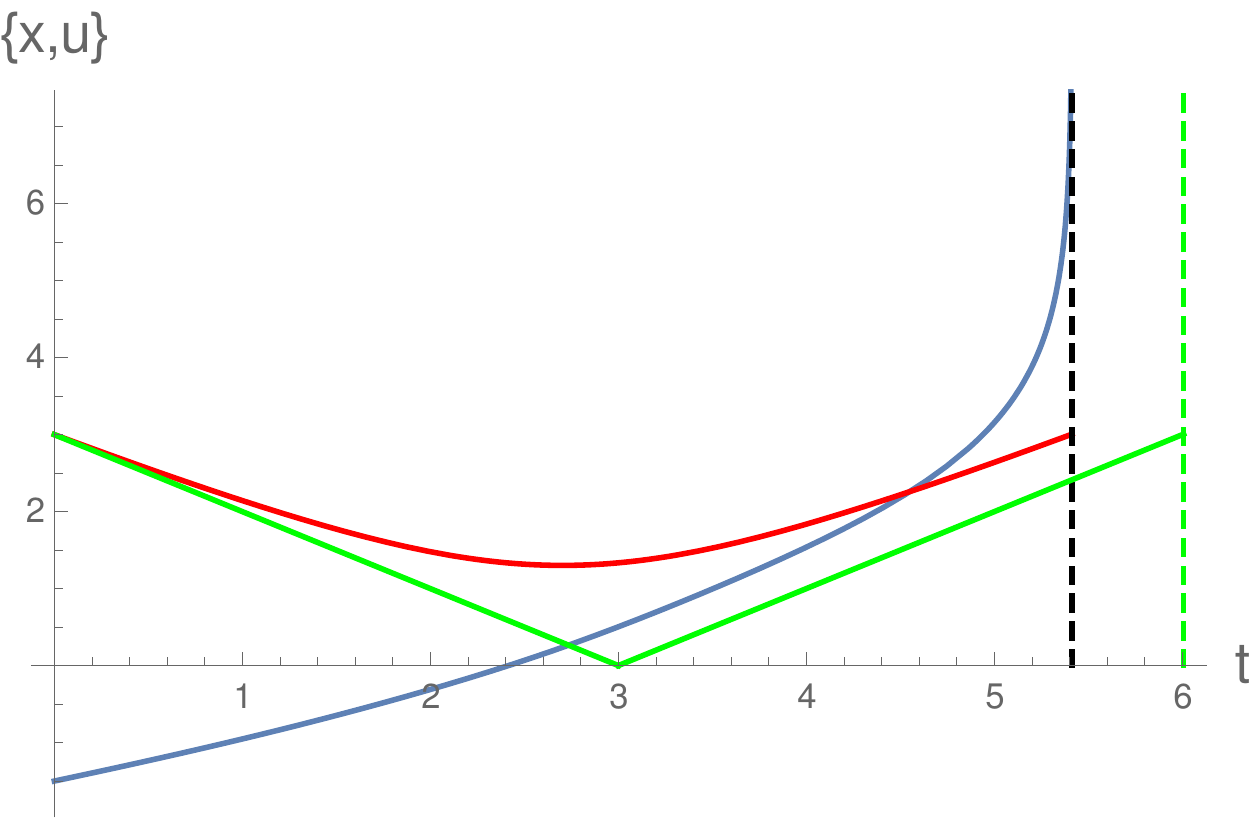}
	\\[7mm]
	\includegraphics[width=0.31\linewidth]{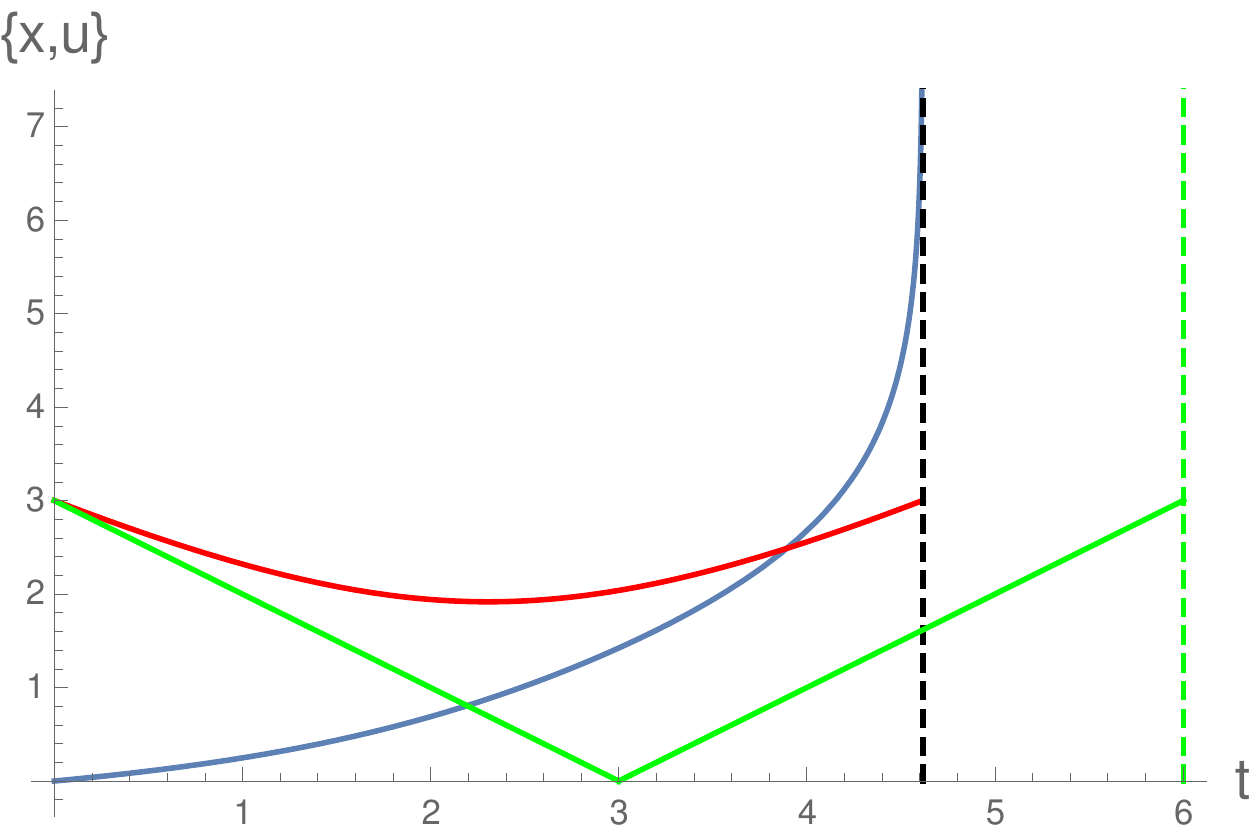}
	\includegraphics[width=0.31\linewidth]{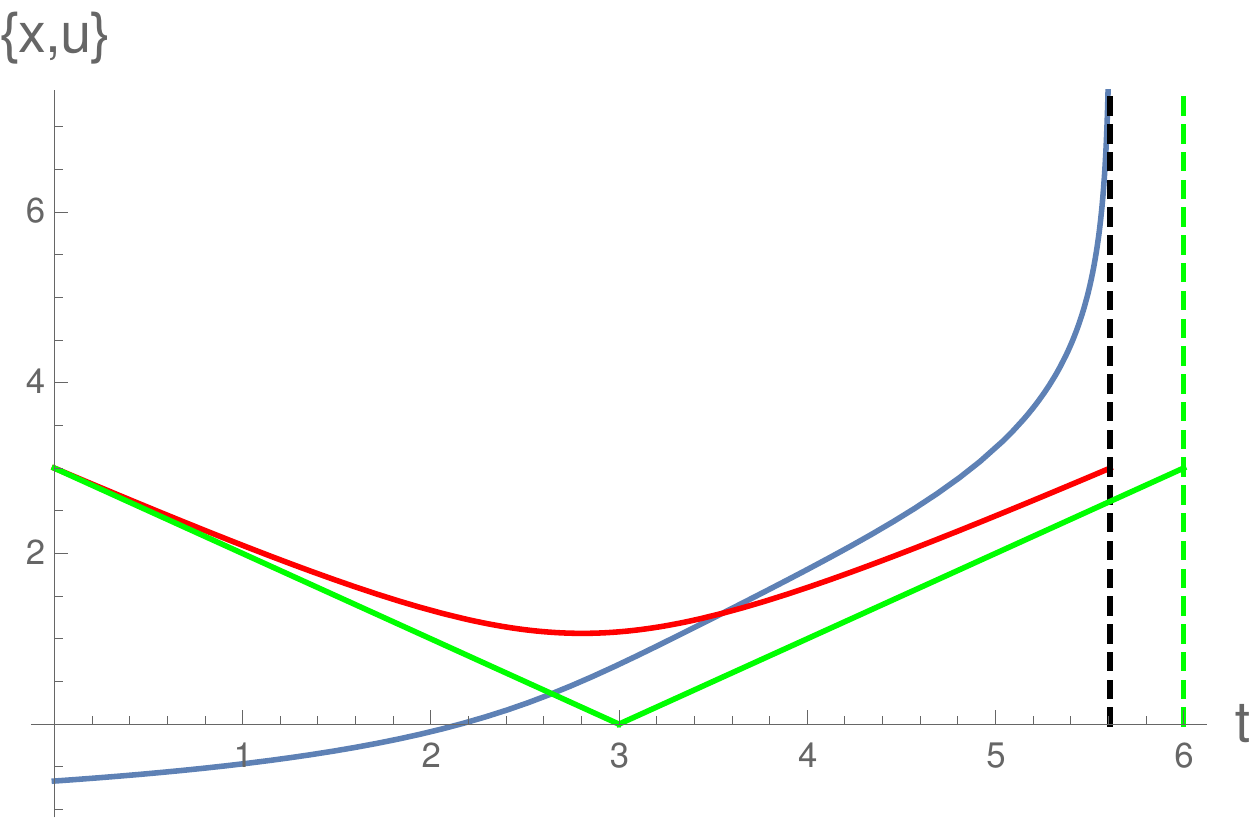}
	\includegraphics[width=0.31\linewidth]{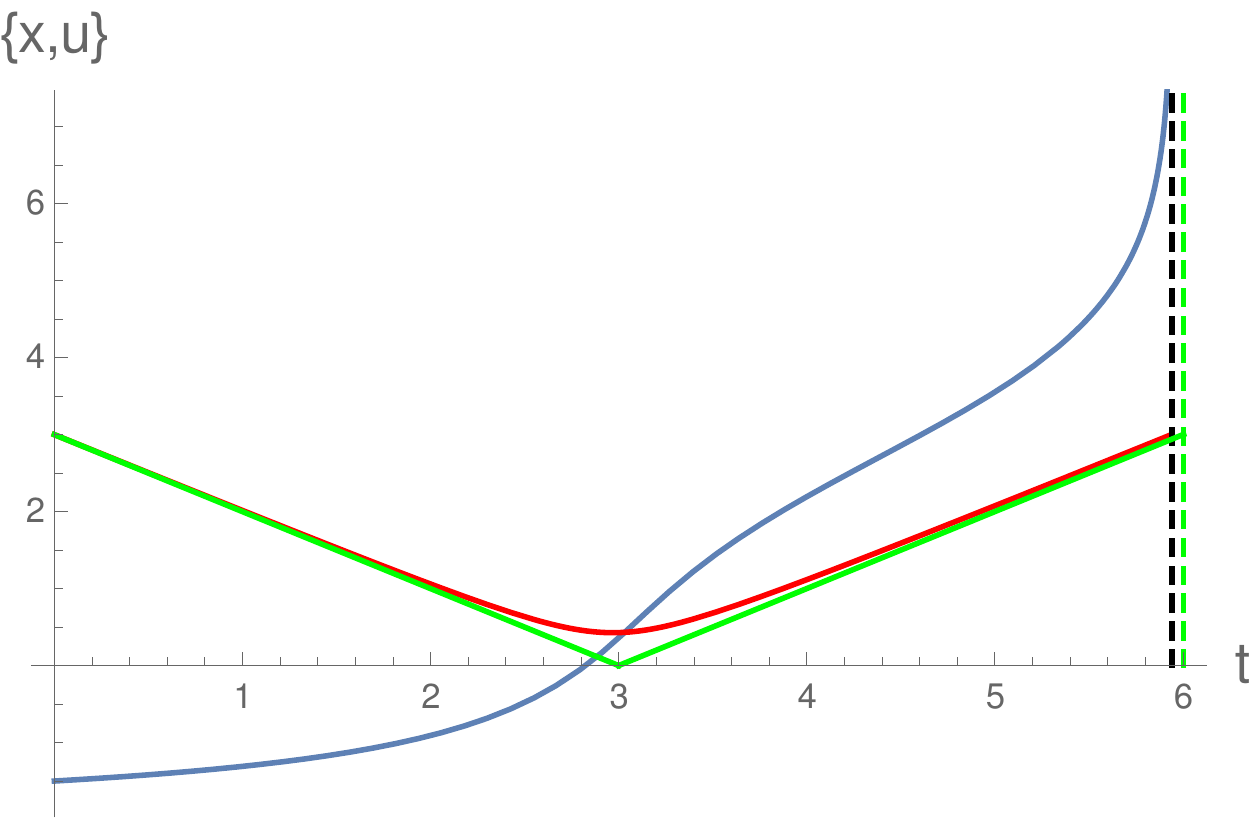}
	
	\caption{``Bulk geodesics" shown as $x(t)$ in blue, $u(t)$ in red, and ``brane geodesics" shown with $u(t)$ in green.
	Top row: $N_5=K$, $y_0=\frac{\pi}{4}$. Starting point is $u_0=3$ and, from left to right, $x_0\in\lbrace -\frac{1}{4},-\frac{2}{3},-\frac{3}{2}\rbrace$.
	$x(t)$ diverges after a finite time, which is when the geodesics approach the conformal boundary of the $AdS_5\times S^5$ region. The vertical dashed lines show the total time taken to reach the end point. Bottom row: Analogous plots for $N_5=2K$, $y_0=\frac{\pi}{4}$; geodesics starting at $u_0=3$ and, from left to right, $x_0\in\lbrace 0,-\frac{2}{3},-\frac{3}{2}\rbrace$. \label{fig:geodesics}}
\end{figure}

A sample of ``through the bulk" geodesics along the central slice of the strip, $y_0=\frac{\pi}{4}$, obtained by solving (\ref{eq:null-geod-zx}) numerically, is shown in fig.~\ref{fig:geodesics}.
The top row is for $N_5=K$.
From fig.~\ref{fig:braneworld}, the 3d region for $y=\frac{\pi}{4}$ is $x\lesssim -0.2$.
When the starting point for the geodesic is near the transition region of fig.~\ref{fig:braneworld}, the geodesic through the bulk reaches the end point notably faster than the ``along the brane" geodesic. As the starting point moves further into the 3d region, towards negative $x$, the time discrepancy decreases.

The bottom row in fig.~\ref{fig:geodesics} shows $N_5=2K$.
Compared to the previous case this corresponds to more 3d degrees of freedom, i.e.\ a sharper brane angle in the braneworld models.
The 3d region according to fig.~\ref{fig:braneworld} occupies the entire range of $y$ for $x\leq 0$.
The starting points $x_0$ shown in fig.~\ref{fig:geodesics} are again in the 3d region.
Generally, the ``through the bulk" geodesic again is faster, with the discrepancy decreasing as $x_0$ is moved further into the 3d region.

The main result, evident already in fig.~\ref{fig:geodesics}, is that the full 10d BCFT duals have analogs of the geodesics in bottom-up models reviewed in sec.~\ref{sec:bottomup-geodesics}, which can connect points in the ETW brane region to points in the ambient CFT geometry faster through the bulk than along the brane.
Their relevance for causality and locality of the intermediate picture will be discussed in sec.~\ref{sec:causal}.

\subsection{``Bulk'' vs.\ ``brane'' travel time}

We now discuss the time differences between the geodesics along the brane and through the bulk, which can be extracted from the integral expression (\ref{eq:geod-t-x}), more quantitatively.
The time difference in principle depends on $N_5/K$, $x_0$, $y_0$, $u_0$ and either $\dot u_0$ or, equivalently, the end point $u_1$ at $x=\infty$.
The symmetries of the AdS$_4$  metric (\ref{eq:ds2-AdS4}) constrain this dependence.
We find
\begin{align}\label{eq:time-diff-10d}
	t_{\rm brane}^2-t_{\rm bulk}^2&= u_0 u_1 \cP_{y_0,N_5/K}(x_0)~.
\end{align}
That is to say, for a given background and choice of $y_0$, the time difference is parametrized by a single function $\cP_{y_0,N_5/K}(x_0)$ which depends only on the starting point on $\Sigma$.\footnote{This can be understood as follows: Invariance of the AdS$_4$ metric (\ref{eq:ds2-AdS4}) under $(u,t,ds^2_{\RR^2})\rightarrow (\alpha u,\alpha t,\alpha^2ds^2_{\RR^2})$ constrains $t_{\rm brane}^2-t_{\rm bulk}^2$ to be homogeneous of degree $2$ in $u_0$, $u_1$. For $u_0\rightarrow 0$ or $u_1\rightarrow 0$, $t_{\rm brane}^2-t_{\rm bulk}^2$ should vanish, since bulk and brane geodesics then become identical. This leaves only the dependence on $u_0$, $u_1$ in (\ref{eq:time-diff-10d}).}

\begin{figure}
	\subfigure[][]{\label{fig:P-x-1}
	\includegraphics[height=4.4cm]{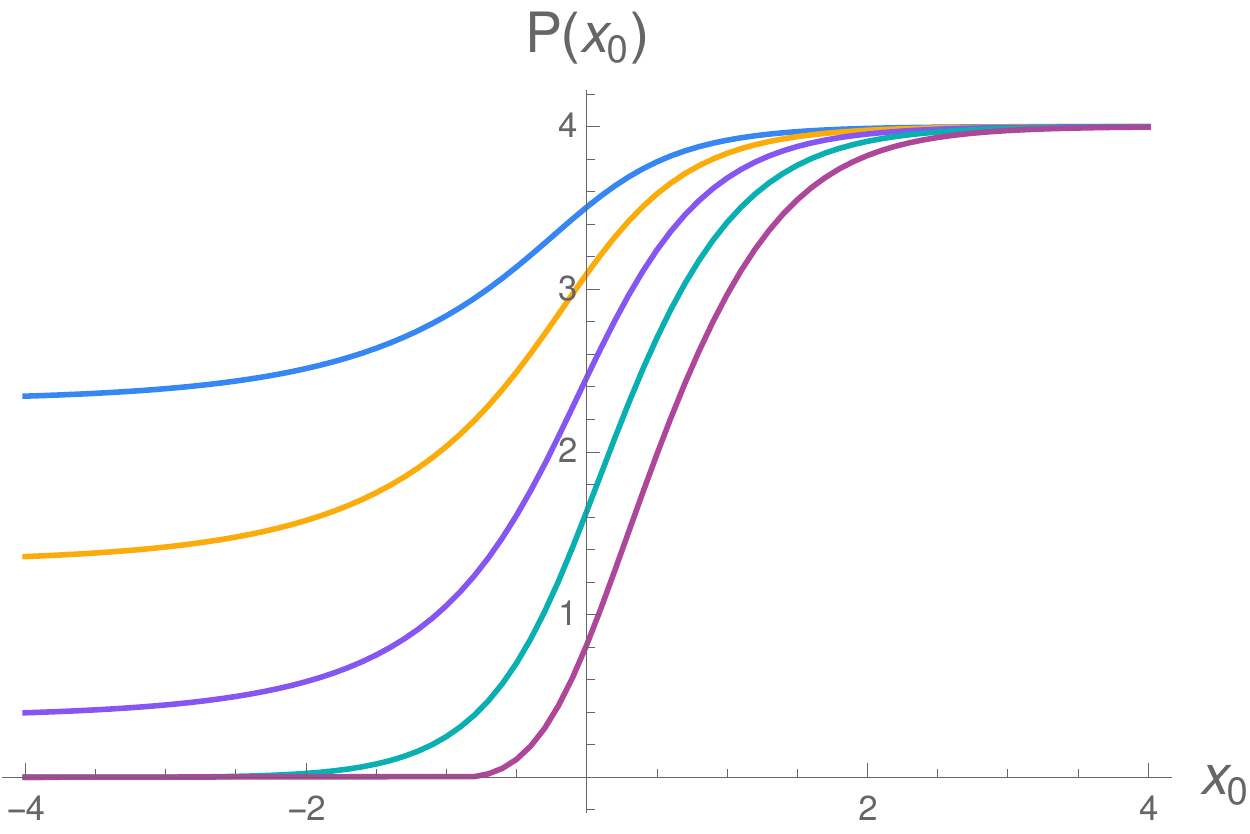}
	}
	\hskip 15mm
	\subfigure[][]{\label{fig:P-x-2}
	\includegraphics[height=4.4cm]{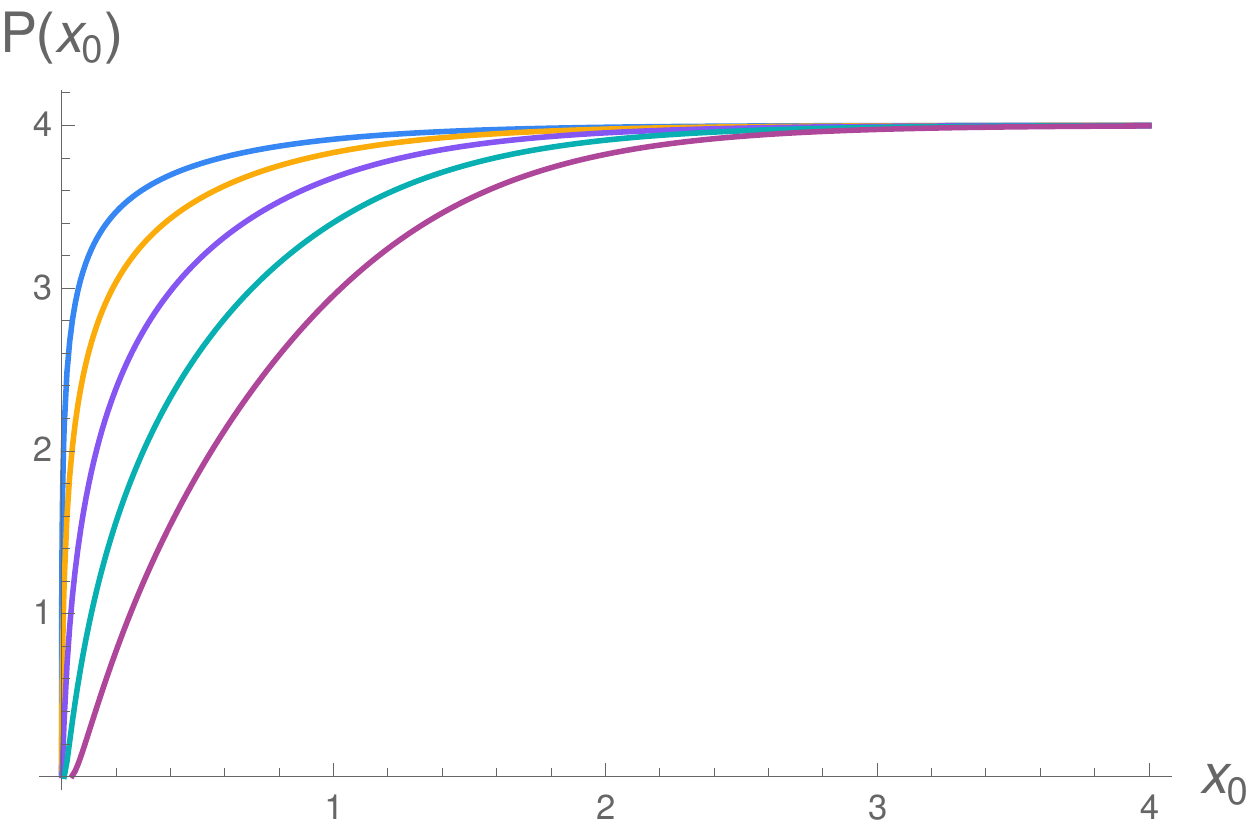}
	}
	\caption{
		On the left $\cP_{y_0,N_5/K}(x_0)$ for $y_0=\frac{\pi}{4}$, i.e.\ geodesics along the central slice of the strip, and on the right for $y_0=0$, i.e.\ geodesics along the boundary of the strip. The curves show, from top to bottom, $N_5/K\in\lbrace \frac{1}{4},\frac{1}{2},1,2,4\rbrace$.
		In the limit $N_5/K\rightarrow \infty$ at fixed $x_0$, $\cP_{y_0,N_5/K}(x_0)$ vanishes.\label{fig:P-x}}
\end{figure}

Plots of $\cP_{y_0,N_5/K}(x_0)$ for 10d solutions with different $N_5/K$ are shown in fig.~\ref{fig:P-x}.
They show that the curves universally approach $4$ as $x_0\rightarrow +\infty$.
This is consistent with $\cP_{{\rm AdS}_5}$ in (\ref{eq:timediff-AdS5}) approaching the same value $4$ for $\mu\rightarrow\frac{\pi}{2}$: for large $x_0$ the geodesic starts on an AdS$_4$ slice close to the conformal boundary of the AdS$_5\times$S$^5$ region. The entire geodesic is in the asymptotic AdS$_5\times$S$^5$ region, where the results from plain AdS$_5$ in the braneworld model should be expected to apply.

More generally, $\cP_{y_0,N_5/K}(x_0)$ for geodesics along the central slice of the strip, $y_0=\frac{\pi}{4}$ (fig.~\ref{fig:P-x-1}), interpolates between $4$ at $x_0\rightarrow +\infty$ and a value which depends on $N_5/K$ at $x_0\rightarrow -\infty$
\begin{align}
	\lim_{x_0\rightarrow -\infty}\cP_{y_0=\pi/4,N_5/K}(x_0)&=\cP_{N_5/K}^{-\infty}~.
\end{align}
This $\cP_{N_5/K}^{-\infty}$ is positive for intermediate $N_5/K$ and approaches zero for large $N_5/K$.
This is consistent with the bottom-up result (\ref{eq:timediff-AdS5}) if the AdS$_4$ slice on which the geodesic starts, $\mu$, is identified in 10d with the starting point $x_0$ of the geodesic on the strip\footnote{%
	Note that fig.~\ref{fig:P-x} shows $\cP_{y_0,N_5/K}(x_0)$ for geodesics with different starting points on $\Sigma$ without changing $N_5/K$.},
in the sense that $P_{{\rm AdS}_5}$ similarly interpolates between $4$ and a number set by the ETW brane location $\mu_\star$ which is generally positive and approaches zero when the brane angle becomes sharp.
Of course, the 10d geodesics captured by fig.~\ref{fig:P-x-1} are only a small sample of the full set of 10d geodesics connecting bulk points to points in the ambient CFT geometry.

The time differences for geodesics along the boundary of $\Sigma$ are shown in fig.~\ref{fig:P-x-2}. Due to the reflection symmetry of the geometry under $y\rightarrow \frac{\pi}{2}-y$, the geodesics for $y=0$ and $y=\frac{\pi}{2}$ behave identically. The curves $\cP_{y_0=0,N_5/K}(x_0)$ interpolate between $4$ for large $x_0$ and zero for $x_0\rightarrow 0^+$. The latter limit corresponds to the starting point of the geodesics approaching the location of one of the 5-brane sources (which are at $z=0$ and $z=\frac{i\pi}{2}$).
This in particular shows that in the 10d solutions $\cP_{y_0,N_5/K}$  reaches all the way down to zero for any value of $N_5/K$. This is different from the bottom-up models where $\cP_{\rm AdS_5}$ is bounded from below by a positive number set by the brane angle.
How large the part of $\Sigma$ is where $\cP_{y_0,N_5/K}$ is close to zero depends on $N_5/K$.
For $N_5/K\rightarrow\infty$ and fixed $x_0$ away from the 5-brane sources, we generally find $\cP_{y_0,N_5/K}(x_0)\rightarrow 0$.

In summary, we find a picture in the 10d BCFT duals which is broadly consistent with bottom-up braneworld results.
In particular, we identified analogs of the geodesics considered in \cite{Omiya:2021olc} in the 10d BCFT duals.
The 10d picture naturally provides more structure and some of the details differ, e.g.\ the lower bound on $\mathcal P$ for fixed ratio of 3d and 4d central charges. But the braneworld models provide valuable qualitative guidance for the 10d BCFT duals.

\section{Implications for intermediate picture}
\label{sec:causal}

We now discuss to what extent the geodesics of sec.~\ref{sec:shortcuts} are in tension with a local intermediate description.
The picture often advertised in the bottom-up models is that the intermediate description consists of a gravity theory on the ETW brane (or in the wedge between a tensionless brane and the ETW brane in fig.~\ref{fig:braneworld-3d4d-1}) coupled to the 4d d.o.f.\ at the conformal boundary. From that perspective the shortcuts through the bulk of sec.~\ref{sec:bottomup-geodesics} would suggest that a process can happen faster in the full BCFT dual than in the intermediate description.
This was interpreted in \cite{Omiya:2021olc} as signaling a form of non-locality at long distances in the intermediate picture.
Here we will discuss the analogous question from the 10d perspective, which will lead to a refined interpretation.

To assess whether the bulk geodesics in the 10d solutions identified in sec.~\ref{sec:shortcuts} constitute a causality violation in the intermediate picture or not, we need to compare two times:
\begin{itemize}
	\item[(a)] the time a signal takes from a point ``on the brane" in the full BCFT dual (e.g.\ in the 3d region in fig.~\ref{fig:braneworld}) to a point on the 4d conformal boundary in the AdS$_5\times$S$^5$ region at $x\rightarrow \infty$
	\item[(b)] the time taken for the same process in the intermediate holographic description
\end{itemize}
The bulk geodesics found in sec.~\ref{sec:shortcuts} would constitute a causality paradox under certain assumptions about the process in (b). Two crucial assumptions are:
\begin{itemize}
	\item[$(i)$] a part of the 10d solutions (the ETW brane region) can literally be interpreted as the geometry of the gravity theory of the intermediate description
	\item[$(ii)$] the point in the full BCFT dual from which the geodesic starts has an analog in the intermediate description and this analog is literally the point with the same coordinates
\end{itemize}
Both assumptions would be satisfied if the na\"ive intermediate picture $(I_n)$ were accurate. But both assumptions are actually problematic.
As shown in sec.~\ref{sec:intermediate}, the gravity part of the proper intermediate description $(I_p)$ is not obtained simply as part of the full BCFT dual. One aspect is that the geometry of the gravity theory in the intermediate description closes off smoothly to form a compact internal space. Qualitatively, the intermediate gravity dual may resemble a geometry obtained by taking a part of the full BCFT dual and modifying it locally so that it closes off smoothly. Quantitatively, however, the geometry changes throughout (fig.~\ref{fig:f4ratio}).
Interpreting a part of the full BCFT dual as gravitational sector of the intermediate description therefore is not accurate.
As a result, the first assumption is not justified. The second assumption is then problematic as well, as its very formulation would compare coordinates in different geometries.

We can therefore not derive non-localities from the results of sec.~\ref{sec:shortcuts}.
We found 10d analogs of the geodesics considered in the bottom-up BCFT duals in \cite{Omiya:2021olc}, and in that sense the bottom-up models capture qualitative aspects of the full 10d BCFT duals well.
But the conclusion of \cite{Omiya:2021olc} that the
BCFT causal structure is incompatible with a local intermediate description would only hold if the na\"ive intermediate description, obtained as part of the BCFT dual,  were generally accurate. We found that it is not.
Indeed, the proper intermediate description of sec.~\ref{sec:intermediate} is local by construction.
The na\"ive intermediate description only becomes accurate for $K/N_5\rightarrow 0$, and in this limit the geodesics through the bulk do not provide shortcuts and there is no tension.

\subsection{No-go for simple identifications}

Comparing causal structures on different geometries may not be immediately meaningful.
But one could attempt to construct a map from the 3d dual of the intermediate description into the full BCFT dual.
We discussed one way to relate the geometries of the 3d dual and the BCFT dual in sec.~\ref{sec:identify-Sigma}.\footnote{It identifies AdS$_2$ slices associated with identical parts of the brane constructions between the 3d dual and the full BCFT dual.
One might extend this to identify points localized in AdS$_4$, e.g.\ by including mass deformations of the Wilson loops used to relate points on $\Sigma$ in the 3d dual and in the BCFT dual, to gap them in the IR. The D5$^\prime$ would then cap off in AdS$_4$ rather than reaching all the way into the IR, and one could identify the cap-off points.}
With any such embedding one can transfer the causal structure of the 3d dual to part of the full BCFT dual, and ask if the result is compatible with the causal structure of the BCFT dual.
In this section we will address this question at a general level: can there be a map from the 3d dual into the full BCFT dual in such a way that the time needed to send a signal from a bulk point to a point in the ambient CFT generally agrees between the two descriptions?

Consider geodesics in the BCFT dual from a ``brane" point $(x,u)=(x_0,u_0)$ to a point in the ambient CFT $(x,u)=(\infty,u_1)$, where $u$ is the radial coordinate in AdS$_4$ and the complex coordinate on $\Sigma$ is $z=x+iy$.
As in sec.~\ref{sec:shortcuts}, we consider geodesics connecting the same points in the $\RR^{2}$ part of the metric (\ref{eq:ds2-AdS4}).
For the times ``along the brane" and ``through the bulk" we found
\begin{align}
	t_{\rm brane}^2&=(u_0+u_1)^2~,\nonumber\\
	t_{\rm bulk}^2&=(u_0+u_1)^2-2u_0 u_1 \mathcal P_{y_0,N/K}(x_0)~.
\end{align}
Now assume there were a simple map between points in the 3d dual and points in the BCFT dual. This would identify a point $(x,y,\tilde u)$ in the 3d dual corresponding to the starting point $(x_0,y_0,u_0)$ of the geodesic in the BCFT dual. In the intermediate description, the time for connecting  $(x,y,\tilde u)$ to the end point in the ambient 4d CFT geometry would be
\begin{align}
	t_{\rm 3d}&=t_{\rm 3d,bulk}+t_{\rm ambient}~,
\end{align}
where $t_{\rm 3d,bulk}$ is the time taken through the bulk of the 3d dual and $t_{\rm ambient}$ is the time taken in the ambient 4d CFT geometry.
One may expect the geodesic in the 3d dual to connect the bulk point $(x,y,\tilde u)$ to the boundary of AdS$_4$ in the shortest way possible, and go on from there to the end point in the ambient 4d CFT geometry. In that case the time $t_{\rm 3d,bulk}$ would be independent of the location $u_1$ of the end point in the ambient CFT, while $t_{\rm ambient}=u_1$.
The time difference between the ``bulk geodesic" in the BCFT dual and the corresponding process in the 3d dual would be
\begin{align}\label{eq:delta-t-u1}
	t_{\rm bulk}^2-t_{\rm 3d}^2&=u_0^2-t_{\rm 3d,bulk}^2+u_1\left(u_0\left(2-\mathcal P_{y_0,N/K}(x_0)\right)-2t_{\rm 3d,bulk}\right)~.
\end{align}
To claim general agreement, this would have to vanish for arbitrary $u_1$.
Since $t_{\rm 3d,bulk}$ does not depend on $u_1$, the term linear in $u_1$ in (\ref{eq:delta-t-u1}) and the term which is independent of $u_1$ would have to vanish separately. But these two conditions are incompatible unless $P_{y_0,N/K}(x_0)=0$.

We thus conclude that for such simple identifications between the 3d dual and the BCFT dual, and of the geodesics in the two pictures, the times in the BCFT dual and in the intermediate picture will not generally agree. The argument applies whenever the time in the 3d dual, $t_{\rm 3d,bulk}$, is independent of $u_1$ and the time in the ambient CFT geometry is $t_{\rm ambient}=u_1$.
The above argument also obstructs simple reinterpretations of the ``brane theory" in the bottom-up models.
We note that generally identifying bulk points in AdS/BCFT using field theory data, e.g.\ using a generalization of \cite{Hamilton:2006az}, would likely involve 3d and 4d degrees of freedom, and should presumably be treated with care in the different layers of holographic descriptions of BCFTs. From that perspective one may not expect a simple map of the form assumed above to exist.

\section{Discussion}
\label{sec:discussion}

In this work we have given an explicit recipe for how to construct a proper intermediate holographic description for a 4d BCFT from a top down perspective: the 10d string geometry dual to the 3d CFT living at the end of the BCFT ambient space is, by hand or ``semi-holographically", coupled to the ambient 4d degrees of freedom of the BCFT. One may wonder to what extent this ``proper" intermediate picture is included in the geometry dual to the full BCFT. Related to this question is the concern about locality. We have seen that the bottom-up concerns about locality \cite{Omiya:2021olc} of the intermediate picture are not resolved simply by going to the full 10d geometry dual to the BCFT: analogs of the geodesics considered in \cite{Omiya:2021olc} can be found in the 10d BCFT dual.
Nevertheless, the ``proper" intermediate picture is local by construction.
The resolution is that, in general, the intermediate picture geometry differs from the full BCFT dual throughout the entire geometry; it is not part of the full BCFT dual geometry.
The full BCFT dual is, both in the top-down and bottom-up constructions, compatible with the BCFT causal structure, as it should be. The proper intermediate description is, by construction, also compatible with the BCFT causal structure. What is not required is that the defect dual of the intermediate picture is part of the full BCFT dual geometry, and we found that it is indeed not.

What then are the implications for the recent construction of islands using double holography \cite{Almheiri:2019hni,Almheiri:2019psy,Chen:2020uac,Chen:2020hmv,Geng:2020fxl,Geng:2021mic,Uhlemann:2021nhu,Demulder:2022aij}?
The main tool of the studies in \cite{Almheiri:2019hni,Almheiri:2019psy,Chen:2020uac,Chen:2020hmv,Geng:2020fxl,Geng:2021mic,Uhlemann:2021nhu,Demulder:2022aij} was to translate the question about islands in the intermediate picture, where 4d gravity is coupled to a non-gravitating bath, into an entanglement entropy calculation in the full BCFT. As long as a coherent intermediate picture exists, and we have defined one in this work, the question about the Page curve translates into a well defined calculation in the BCFT. The latter can then be answered in the full BCFT geometry. The time evolution of the entanglement entropy of the radiation region in the BCFT is uncontentious.
The same answer would be obtained directly in the proper intermediate picture.
The observation that the proper intermediate picture differs from the na\"ive intermediate picture obtained as part of the full BCFT dual does not harm  calculations performed in the full BCFT dual.

The results presented here can further be seen as support for the computations carried out in bottom-up models. We found geodesics with qualitatively similar features as those discussed in \cite{Omiya:2021olc} in the full 10d BCFT duals. So the bottom-up models capture qualitative aspects of the full 10d string theory BCFT duals quite well.
What we argued is not accurate is the na\"ive view of the intermediate picture.
Though it may not be immediately clear how to construct a  ``proper" intermediate picture in the bottom-up models from first principles, one may expect a wedge holography dual \cite{Akal:2020wfl} for the defect degrees of freedom, where a finite region of a 5d spacetime bounded by two end-of-the-world branes gives a dual for the 3d CFT.
In any case, the calculations in the full bottom-up BCFT duals are actually supported by our discussion.

There is a limit in which the geometry of the proper intermediate picture is well approximated by (part of) the full BCFT geometry: the limit $K/N_5 \rightarrow 0$. In this limit the number of 3d degrees of freedom is much larger than the number of 4d degrees of freedom and correspondingly the AdS$_4$ curvature radius in the resolved end-of-the-world brane is large compared to the asymptotic AdS$_5$ curvature radius. The bottom-up analog is the limit of a near-critical tension brane whose location approaches the conformal boundary.
In this limit the na\"ive intermediate description becomes accurate. This is also the limit in which the localized graviton is parametrically lighter than generic modes \cite{Karch:2000ct,Bachas:2018zmb} and the bulk and brane geodesics of \cite{Omiya:2021olc} take the same time. The recent Page curve discussions include this limit in their parameter space. We emphasize, though, that the proper intermediate description presented here can be constructed for generic $K/N_5$.

\let\oldaddcontentsline\addcontentsline
\renewcommand{\addcontentsline}[3]{}
\begin{acknowledgments}
AK would like to thank Hao Geng, Suvrat Raju, and Lisa Randall for extensive discussions on the results of \cite{Omiya:2021olc}.
The work of AK was supported, in part, by the U.S.~Department of Energy under Grant DE-SC0022021 and by a grant from the Simons Foundation (Grant 651678, AK).	The work of HYS was supported by a grant from the Simons Foundation (Grant 651678, AK). CFU is supported, in part, by the US Department of Energy under Grant No.~DE-SC0007859 	and by the Leinweber Center for Theoretical Physics.
\end{acknowledgments}
\let\addcontentsline\oldaddcontentsline

\appendix

\section{Null geodesics in 10d}\label{app:geodesics}

The geodesic equation ${x^{\prime\prime}}^\mu+\Gamma_{\nu\rho}^\mu{x^\prime}^\nu {x^\prime}^\rho=0$, for curves parametrized as in (\ref{eq:geodesic-u-lambda}) with $y=y_0$, and with the primes denoting differentiation w.r.t.\ $\lambda$, becomes
\begin{align}
	\label{eq:geod}
	0&=\frac{d}{d\lambda}\left(\frac{t^\prime f_4^2(x,y_0)}{u^2}\right),
	\nonumber\\
	0&=	u^2\frac{d}{d\lambda}\left(\frac{u'f_4^2(x,y_0)}{u}\right)-{t^\prime}^2f_4^2(x,y_0)\,,
	\nonumber\\
	0&=x^{\prime\prime}+\frac{\partial_x \rho^2(x,y_0)}{2\rho^2(x,y_0)}{x^\prime}^2+({t^\prime}^2-{u^\prime}^2)\frac{\partial_x f_4^2(x,y_0)}{8u^2\rho^2(x,y_0)}\,,
	\nonumber\\
	0&=-\frac{\partial_y \rho^2(x,y_0)}{2\rho^2(x,y_0)}{x^\prime}^2+({t^\prime}^2-{u^\prime}^2)\frac{\partial_y f_4^2(x,y_0)}{8u^2\rho^2(x,y_0)}\,.
\end{align}
The last equation is the constraint resulting from the restriction to constant $y$. It is satisfied identically along $y=y_0\equiv\frac{\pi}{4}$ in the solutions (\ref{eq:h1h2-BCFT}).
It is also satisfied for $y_0=0$ and $y_0=\frac{\pi}{2}$, since the metric functions $\rho^2$ and $f_4^2$ satisfy Neumann boundary conditions.

The first equation in (\ref{eq:geod}) can be integrated straightforwardly, which leads to
\begin{align}
	t^\prime\frac{ f_4^2}{u^2}&=c_1~.
\end{align}
We will switch the parametrization to $x(\lambda)\rightarrow x(t(\lambda))$ and $u(\lambda)\rightarrow u(t(\lambda))$, and denote derivatives with respect to $t$ by a dot. The second equation in (\ref{eq:geod}) becomes
\begin{align}
	\frac{d}{dt}(u\dot u)-1&=0~.
\end{align}
This is solved by
\begin{align}\label{eq:z-sol-app}
	u^2&=t^2+2u_0\dot u_0 t+u_0^2~.
\end{align}
The only remaining non-trivial equation is the third one in (\ref{eq:geod}).
The condition for a geodesic solving (\ref{eq:geod}) to be null is
\begin{align}\label{eq:geod-null}
	\frac{f_4^2}{u^2}\left({u^\prime}^2-{t^\prime}^2\right)+4{x^\prime}^2\rho^2&=0~.
\end{align}
Using the null condition (\ref{eq:geod-null}), the third equation in (\ref{eq:geod}) can be written as
\begin{align}\label{eq:geod-x}
	0&=\frac{d}{d\lambda}\left(x^\prime f_4(x,y_0)\rho(x,y_0)\right)~.
\end{align}
After switching the parametrization to $x(t)$, this leads to
\begin{align}\label{eq:x-dot-app}
	\dot x&=c_2\frac{f_4}{u^2\rho}~.
\end{align}
The null condition becomes
\begin{align}
	\frac{f_4^2}{u^2}(\dot u^2-1)+4\dot x^2 \rho^2&=0~.
\end{align}
This fixes $\dot x_0$ (and consequently $c_2$) in terms of $x_0,u_0,\dot u_0$.
The result is $c_2=\frac{1}{2}u_0\sqrt{1-\dot u_0^2}$.

The geodesic equations have been reduced to a single first-order equation. To obtain the time taken by a geodesic connecting some $x_0$ to the conformal boundary of the AdS$_5\times$S$^5$ region at $x\rightarrow \infty$, we switch to the parametrization $t(x)$ instead of $x(t)$.
$u$ is still given by (\ref{eq:z-sol-app}). From (\ref{eq:x-dot-app})
\begin{align}
\frac{dt}{dx}&=\frac{u^2\rho}{c_2 f_4}~.
\end{align}
After bringing $u^2$ to the left and using the solution for $u$ in (\ref{eq:z-sol-app}), we find
\begin{align}
	\frac{d}{dx}\left[\tan^{-1}\left(\frac{t+u_0\dot u_0}{u_0\sqrt{1-\dot u_0^2}}\right)\right]&=\frac{2\rho}{f_4}~.
\end{align}
One may further evaluate the right hand side using $\rho^4/f_4^4=W^2/(h_1h_2)^2$ but we will not do so here.
One may now express $t$ as an explicit integral
\begin{align}
	t&=u_0\sqrt{1-\dot u_0^2}\,\tan\left[\sin^{-1}\!\left(\dot u_0\right)+2\int_{x_0}^x d\hat x \frac{\rho(\hat x,y_0)}{f_4(\hat x,y_0)}\right]-u_0\dot u_0~.
\end{align}

\bibliography{notes}

\begin{thebibliography}{40}%
\makeatletter
\providecommand \@ifxundefined [1]{%
 \@ifx{#1\undefined}
}%
\providecommand \@ifnum [1]{%
 \ifnum #1\expandafter \@firstoftwo
 \else \expandafter \@secondoftwo
 \fi
}%
\providecommand \@ifx [1]{%
 \ifx #1\expandafter \@firstoftwo
 \else \expandafter \@secondoftwo
 \fi
}%
\providecommand \natexlab [1]{#1}%
\providecommand \enquote  [1]{``#1''}%
\providecommand \bibnamefont  [1]{#1}%
\providecommand \bibfnamefont [1]{#1}%
\providecommand \citenamefont [1]{#1}%
\providecommand \href@noop [0]{\@secondoftwo}%
\providecommand \href [0]{\begingroup \@sanitize@url \@href}%
\providecommand \@href[1]{\@@startlink{#1}\@@href}%
\providecommand \@@href[1]{\endgroup#1\@@endlink}%
\providecommand \@sanitize@url [0]{\catcode `\\12\catcode `\$12\catcode
  `\&12\catcode `\#12\catcode `\^12\catcode `\_12\catcode `\%12\relax}%
\providecommand \@@startlink[1]{}%
\providecommand \@@endlink[0]{}%
\providecommand \url  [0]{\begingroup\@sanitize@url \@url }%
\providecommand \@url [1]{\endgroup\@href {#1}{\urlprefix }}%
\providecommand \urlprefix  [0]{URL }%
\providecommand \Eprint [0]{\href }%
\providecommand \doibase [0]{http://dx.doi.org/}%
\providecommand \selectlanguage [0]{\@gobble}%
\providecommand \bibinfo  [0]{\@secondoftwo}%
\providecommand \bibfield  [0]{\@secondoftwo}%
\providecommand \translation [1]{[#1]}%
\providecommand \BibitemOpen [0]{}%
\providecommand \bibitemStop [0]{}%
\providecommand \bibitemNoStop [0]{.\EOS\space}%
\providecommand \EOS [0]{\spacefactor3000\relax}%
\providecommand \BibitemShut  [1]{\csname bibitem#1\endcsname}%
\let\auto@bib@innerbib\@empty
\bibitem [{\citenamefont {Almheiri}\ \emph
  {et~al.}(2020{\natexlab{a}})\citenamefont {Almheiri}, \citenamefont
  {Mahajan}, \citenamefont {Maldacena},\ and\ \citenamefont
  {Zhao}}]{Almheiri:2019hni}%
  \BibitemOpen
  \bibfield  {author} {\bibinfo {author} {\bibfnamefont {Ahmed}\ \bibnamefont
  {Almheiri}}, \bibinfo {author} {\bibfnamefont {Raghu}\ \bibnamefont
  {Mahajan}}, \bibinfo {author} {\bibfnamefont {Juan}\ \bibnamefont
  {Maldacena}}, \ and\ \bibinfo {author} {\bibfnamefont {Ying}\ \bibnamefont
  {Zhao}},\ }\bibfield  {title} {\enquote {\bibinfo {title} {{The Page curve of
  Hawking radiation from semiclassical geometry}},}\ }\href {\doibase
  10.1007/JHEP03(2020)149} {\bibfield  {journal} {\bibinfo  {journal} {JHEP}\
  }\textbf {\bibinfo {volume} {03}},\ \bibinfo {pages} {149} (\bibinfo {year}
  {2020}{\natexlab{a}})},\ \Eprint {http://arxiv.org/abs/1908.10996}
  {arXiv:1908.10996 [hep-th]} \BibitemShut {NoStop}%
\bibitem [{\citenamefont {Almheiri}\ \emph
  {et~al.}(2020{\natexlab{b}})\citenamefont {Almheiri}, \citenamefont
  {Mahajan},\ and\ \citenamefont {Santos}}]{Almheiri:2019psy}%
  \BibitemOpen
  \bibfield  {author} {\bibinfo {author} {\bibfnamefont {Ahmed}\ \bibnamefont
  {Almheiri}}, \bibinfo {author} {\bibfnamefont {Raghu}\ \bibnamefont
  {Mahajan}}, \ and\ \bibinfo {author} {\bibfnamefont {Jorge~E.}\ \bibnamefont
  {Santos}},\ }\bibfield  {title} {\enquote {\bibinfo {title} {{Entanglement
  islands in higher dimensions}},}\ }\href {\doibase
  10.21468/SciPostPhys.9.1.001} {\bibfield  {journal} {\bibinfo  {journal}
  {SciPost Phys.}\ }\textbf {\bibinfo {volume} {9}},\ \bibinfo {pages} {001}
  (\bibinfo {year} {2020}{\natexlab{b}})},\ \Eprint
  {http://arxiv.org/abs/1911.09666} {arXiv:1911.09666 [hep-th]} \BibitemShut
  {NoStop}%
\bibitem [{\citenamefont {Chen}\ \emph
  {et~al.}(2020{\natexlab{a}})\citenamefont {Chen}, \citenamefont {Myers},
  \citenamefont {Neuenfeld}, \citenamefont {Reyes},\ and\ \citenamefont
  {Sandor}}]{Chen:2020uac}%
  \BibitemOpen
  \bibfield  {author} {\bibinfo {author} {\bibfnamefont {Hong~Zhe}\
  \bibnamefont {Chen}}, \bibinfo {author} {\bibfnamefont {Robert~C.}\
  \bibnamefont {Myers}}, \bibinfo {author} {\bibfnamefont {Dominik}\
  \bibnamefont {Neuenfeld}}, \bibinfo {author} {\bibfnamefont {Ignacio~A.}\
  \bibnamefont {Reyes}}, \ and\ \bibinfo {author} {\bibfnamefont {Joshua}\
  \bibnamefont {Sandor}},\ }\bibfield  {title} {\enquote {\bibinfo {title}
  {{Quantum Extremal Islands Made Easy, Part I: Entanglement on the Brane}},}\
  }\href {\doibase 10.1007/JHEP10(2020)166} {\bibfield  {journal} {\bibinfo
  {journal} {JHEP}\ }\textbf {\bibinfo {volume} {10}},\ \bibinfo {pages} {166}
  (\bibinfo {year} {2020}{\natexlab{a}})},\ \Eprint
  {http://arxiv.org/abs/2006.04851} {arXiv:2006.04851 [hep-th]} \BibitemShut
  {NoStop}%
\bibitem [{\citenamefont {Chen}\ \emph
  {et~al.}(2020{\natexlab{b}})\citenamefont {Chen}, \citenamefont {Myers},
  \citenamefont {Neuenfeld}, \citenamefont {Reyes},\ and\ \citenamefont
  {Sandor}}]{Chen:2020hmv}%
  \BibitemOpen
  \bibfield  {author} {\bibinfo {author} {\bibfnamefont {Hong~Zhe}\
  \bibnamefont {Chen}}, \bibinfo {author} {\bibfnamefont {Robert~C.}\
  \bibnamefont {Myers}}, \bibinfo {author} {\bibfnamefont {Dominik}\
  \bibnamefont {Neuenfeld}}, \bibinfo {author} {\bibfnamefont {Ignacio~A.}\
  \bibnamefont {Reyes}}, \ and\ \bibinfo {author} {\bibfnamefont {Joshua}\
  \bibnamefont {Sandor}},\ }\bibfield  {title} {\enquote {\bibinfo {title}
  {{Quantum Extremal Islands Made Easy, Part II: Black Holes on the Brane}},}\
  }\href {\doibase 10.1007/JHEP12(2020)025} {\bibfield  {journal} {\bibinfo
  {journal} {JHEP}\ }\textbf {\bibinfo {volume} {12}},\ \bibinfo {pages} {025}
  (\bibinfo {year} {2020}{\natexlab{b}})},\ \Eprint
  {http://arxiv.org/abs/2010.00018} {arXiv:2010.00018 [hep-th]} \BibitemShut
  {NoStop}%
\bibitem [{\citenamefont {Geng}\ \emph
  {et~al.}(2021{\natexlab{a}})\citenamefont {Geng}, \citenamefont {Karch},
  \citenamefont {Perez-Pardavila}, \citenamefont {Raju}, \citenamefont
  {Randall}, \citenamefont {Riojas},\ and\ \citenamefont
  {Shashi}}]{Geng:2020fxl}%
  \BibitemOpen
  \bibfield  {author} {\bibinfo {author} {\bibfnamefont {Hao}\ \bibnamefont
  {Geng}}, \bibinfo {author} {\bibfnamefont {Andreas}\ \bibnamefont {Karch}},
  \bibinfo {author} {\bibfnamefont {Carlos}\ \bibnamefont {Perez-Pardavila}},
  \bibinfo {author} {\bibfnamefont {Suvrat}\ \bibnamefont {Raju}}, \bibinfo
  {author} {\bibfnamefont {Lisa}\ \bibnamefont {Randall}}, \bibinfo {author}
  {\bibfnamefont {Marcos}\ \bibnamefont {Riojas}}, \ and\ \bibinfo {author}
  {\bibfnamefont {Sanjit}\ \bibnamefont {Shashi}},\ }\bibfield  {title}
  {\enquote {\bibinfo {title} {{Information Transfer with a Gravitating
  Bath}},}\ }\href {\doibase 10.21468/SciPostPhys.10.5.103} {\bibfield
  {journal} {\bibinfo  {journal} {SciPost Phys.}\ }\textbf {\bibinfo {volume}
  {10}},\ \bibinfo {pages} {103} (\bibinfo {year} {2021}{\natexlab{a}})},\
  \Eprint {http://arxiv.org/abs/2012.04671} {arXiv:2012.04671 [hep-th]}
  \BibitemShut {NoStop}%
\bibitem [{\citenamefont {Geng}\ \emph
  {et~al.}(2021{\natexlab{b}})\citenamefont {Geng}, \citenamefont {Karch},
  \citenamefont {Perez-Pardavila}, \citenamefont {Raju}, \citenamefont
  {Randall}, \citenamefont {Riojas},\ and\ \citenamefont
  {Shashi}}]{Geng:2021mic}%
  \BibitemOpen
  \bibfield  {author} {\bibinfo {author} {\bibfnamefont {Hao}\ \bibnamefont
  {Geng}}, \bibinfo {author} {\bibfnamefont {Andreas}\ \bibnamefont {Karch}},
  \bibinfo {author} {\bibfnamefont {Carlos}\ \bibnamefont {Perez-Pardavila}},
  \bibinfo {author} {\bibfnamefont {Suvrat}\ \bibnamefont {Raju}}, \bibinfo
  {author} {\bibfnamefont {Lisa}\ \bibnamefont {Randall}}, \bibinfo {author}
  {\bibfnamefont {Marcos}\ \bibnamefont {Riojas}}, \ and\ \bibinfo {author}
  {\bibfnamefont {Sanjit}\ \bibnamefont {Shashi}},\ }\bibfield  {title}
  {\enquote {\bibinfo {title} {{Entanglement Phase Structure of a Holographic
  BCFT in a Black Hole Background}},}\ }\href@noop {} {\  (\bibinfo {year}
  {2021}{\natexlab{b}})},\ \Eprint {http://arxiv.org/abs/2112.09132}
  {arXiv:2112.09132 [hep-th]} \BibitemShut {NoStop}%
\bibitem [{\citenamefont {Uhlemann}(2021)}]{Uhlemann:2021nhu}%
  \BibitemOpen
  \bibfield  {author} {\bibinfo {author} {\bibfnamefont {Christoph~F.}\
  \bibnamefont {Uhlemann}},\ }\bibfield  {title} {\enquote {\bibinfo {title}
  {{Islands and Page curves in 4d from Type IIB}},}\ }\href {\doibase
  10.1007/JHEP08(2021)104} {\bibfield  {journal} {\bibinfo  {journal} {JHEP}\
  }\textbf {\bibinfo {volume} {08}},\ \bibinfo {pages} {104} (\bibinfo {year}
  {2021})},\ \Eprint {http://arxiv.org/abs/2105.00008} {arXiv:2105.00008
  [hep-th]} \BibitemShut {NoStop}%
\bibitem [{\citenamefont {Demulder}\ \emph {et~al.}(2022)\citenamefont
  {Demulder}, \citenamefont {Gnecchi}, \citenamefont {Lavdas},\ and\
  \citenamefont {Lust}}]{Demulder:2022aij}%
  \BibitemOpen
  \bibfield  {author} {\bibinfo {author} {\bibfnamefont {Saskia}\ \bibnamefont
  {Demulder}}, \bibinfo {author} {\bibfnamefont {Alessandra}\ \bibnamefont
  {Gnecchi}}, \bibinfo {author} {\bibfnamefont {Ioannis}\ \bibnamefont
  {Lavdas}}, \ and\ \bibinfo {author} {\bibfnamefont {Dieter}\ \bibnamefont
  {Lust}},\ }\bibfield  {title} {\enquote {\bibinfo {title} {{Islands and Light
  Gravitons in type IIB String Theory}},}\ }\href@noop {} {\  (\bibinfo {year}
  {2022})},\ \Eprint {http://arxiv.org/abs/2204.03669} {arXiv:2204.03669
  [hep-th]} \BibitemShut {NoStop}%
\bibitem [{\citenamefont {Karch}\ and\ \citenamefont
  {Randall}(2001{\natexlab{a}})}]{Karch:2000gx}%
  \BibitemOpen
  \bibfield  {author} {\bibinfo {author} {\bibfnamefont {Andreas}\ \bibnamefont
  {Karch}}\ and\ \bibinfo {author} {\bibfnamefont {Lisa}\ \bibnamefont
  {Randall}},\ }\bibfield  {title} {\enquote {\bibinfo {title} {{Open and
  closed string interpretation of SUSY CFT's on branes with boundaries}},}\
  }\href {\doibase 10.1088/1126-6708/2001/06/063} {\bibfield  {journal}
  {\bibinfo  {journal} {JHEP}\ }\textbf {\bibinfo {volume} {06}},\ \bibinfo
  {pages} {063} (\bibinfo {year} {2001}{\natexlab{a}})},\ \Eprint
  {http://arxiv.org/abs/hep-th/0105132} {arXiv:hep-th/0105132} \BibitemShut
  {NoStop}%
\bibitem [{\citenamefont {Karch}\ and\ \citenamefont
  {Randall}(2001{\natexlab{b}})}]{Karch:2000ct}%
  \BibitemOpen
  \bibfield  {author} {\bibinfo {author} {\bibfnamefont {Andreas}\ \bibnamefont
  {Karch}}\ and\ \bibinfo {author} {\bibfnamefont {Lisa}\ \bibnamefont
  {Randall}},\ }\bibfield  {title} {\enquote {\bibinfo {title} {{Locally
  localized gravity}},}\ }\href {\doibase 10.1088/1126-6708/2001/05/008}
  {\bibfield  {journal} {\bibinfo  {journal} {JHEP}\ }\textbf {\bibinfo
  {volume} {05}},\ \bibinfo {pages} {008} (\bibinfo {year}
  {2001}{\natexlab{b}})},\ \Eprint {http://arxiv.org/abs/hep-th/0011156}
  {arXiv:hep-th/0011156} \BibitemShut {NoStop}%
\bibitem [{\citenamefont {Takayanagi}(2011)}]{Takayanagi:2011zk}%
  \BibitemOpen
  \bibfield  {author} {\bibinfo {author} {\bibfnamefont {Tadashi}\ \bibnamefont
  {Takayanagi}},\ }\bibfield  {title} {\enquote {\bibinfo {title} {{Holographic
  Dual of BCFT}},}\ }\href {\doibase 10.1103/PhysRevLett.107.101602} {\bibfield
   {journal} {\bibinfo  {journal} {Phys. Rev. Lett.}\ }\textbf {\bibinfo
  {volume} {107}},\ \bibinfo {pages} {101602} (\bibinfo {year} {2011})},\
  \Eprint {http://arxiv.org/abs/1105.5165} {arXiv:1105.5165 [hep-th]}
  \BibitemShut {NoStop}%
\bibitem [{\citenamefont {Fujita}\ \emph {et~al.}(2011)\citenamefont {Fujita},
  \citenamefont {Takayanagi},\ and\ \citenamefont {Tonni}}]{Fujita:2011fp}%
  \BibitemOpen
  \bibfield  {author} {\bibinfo {author} {\bibfnamefont {Mitsutoshi}\
  \bibnamefont {Fujita}}, \bibinfo {author} {\bibfnamefont {Tadashi}\
  \bibnamefont {Takayanagi}}, \ and\ \bibinfo {author} {\bibfnamefont {Erik}\
  \bibnamefont {Tonni}},\ }\bibfield  {title} {\enquote {\bibinfo {title}
  {{Aspects of AdS/BCFT}},}\ }\href {\doibase 10.1007/JHEP11(2011)043}
  {\bibfield  {journal} {\bibinfo  {journal} {JHEP}\ }\textbf {\bibinfo
  {volume} {11}},\ \bibinfo {pages} {043} (\bibinfo {year} {2011})},\ \Eprint
  {http://arxiv.org/abs/1108.5152} {arXiv:1108.5152 [hep-th]} \BibitemShut
  {NoStop}%
\bibitem [{\citenamefont {Randall}\ and\ \citenamefont
  {Sundrum}(1999)}]{Randall:1999vf}%
  \BibitemOpen
  \bibfield  {author} {\bibinfo {author} {\bibfnamefont {Lisa}\ \bibnamefont
  {Randall}}\ and\ \bibinfo {author} {\bibfnamefont {Raman}\ \bibnamefont
  {Sundrum}},\ }\bibfield  {title} {\enquote {\bibinfo {title} {{An Alternative
  to compactification}},}\ }\href {\doibase 10.1103/PhysRevLett.83.4690}
  {\bibfield  {journal} {\bibinfo  {journal} {Phys. Rev. Lett.}\ }\textbf
  {\bibinfo {volume} {83}},\ \bibinfo {pages} {4690--4693} (\bibinfo {year}
  {1999})},\ \Eprint {http://arxiv.org/abs/hep-th/9906064}
  {arXiv:hep-th/9906064} \BibitemShut {NoStop}%
\bibitem [{\citenamefont {D'Hoker}\ \emph
  {et~al.}(2007{\natexlab{a}})\citenamefont {D'Hoker}, \citenamefont {Estes},\
  and\ \citenamefont {Gutperle}}]{DHoker:2007zhm}%
  \BibitemOpen
  \bibfield  {author} {\bibinfo {author} {\bibfnamefont {Eric}\ \bibnamefont
  {D'Hoker}}, \bibinfo {author} {\bibfnamefont {John}\ \bibnamefont {Estes}}, \
  and\ \bibinfo {author} {\bibfnamefont {Michael}\ \bibnamefont {Gutperle}},\
  }\bibfield  {title} {\enquote {\bibinfo {title} {{Exact half-BPS Type IIB
  interface solutions. I. Local solution and supersymmetric Janus}},}\ }\href
  {\doibase 10.1088/1126-6708/2007/06/021} {\bibfield  {journal} {\bibinfo
  {journal} {JHEP}\ }\textbf {\bibinfo {volume} {06}},\ \bibinfo {pages} {021}
  (\bibinfo {year} {2007}{\natexlab{a}})},\ \Eprint
  {http://arxiv.org/abs/0705.0022} {arXiv:0705.0022 [hep-th]} \BibitemShut
  {NoStop}%
\bibitem [{\citenamefont {D'Hoker}\ \emph
  {et~al.}(2007{\natexlab{b}})\citenamefont {D'Hoker}, \citenamefont {Estes},\
  and\ \citenamefont {Gutperle}}]{DHoker:2007hhe}%
  \BibitemOpen
  \bibfield  {author} {\bibinfo {author} {\bibfnamefont {Eric}\ \bibnamefont
  {D'Hoker}}, \bibinfo {author} {\bibfnamefont {John}\ \bibnamefont {Estes}}, \
  and\ \bibinfo {author} {\bibfnamefont {Michael}\ \bibnamefont {Gutperle}},\
  }\bibfield  {title} {\enquote {\bibinfo {title} {{Exact half-BPS Type IIB
  interface solutions. II. Flux solutions and multi-Janus}},}\ }\href {\doibase
  10.1088/1126-6708/2007/06/022} {\bibfield  {journal} {\bibinfo  {journal}
  {JHEP}\ }\textbf {\bibinfo {volume} {06}},\ \bibinfo {pages} {022} (\bibinfo
  {year} {2007}{\natexlab{b}})},\ \Eprint {http://arxiv.org/abs/0705.0024}
  {arXiv:0705.0024 [hep-th]} \BibitemShut {NoStop}%
\bibitem [{\citenamefont {Aharony}\ \emph {et~al.}(2011)\citenamefont
  {Aharony}, \citenamefont {Berdichevsky}, \citenamefont {Berkooz},\ and\
  \citenamefont {Shamir}}]{Aharony:2011yc}%
  \BibitemOpen
  \bibfield  {author} {\bibinfo {author} {\bibfnamefont {Ofer}\ \bibnamefont
  {Aharony}}, \bibinfo {author} {\bibfnamefont {Leon}\ \bibnamefont
  {Berdichevsky}}, \bibinfo {author} {\bibfnamefont {Micha}\ \bibnamefont
  {Berkooz}}, \ and\ \bibinfo {author} {\bibfnamefont {Itamar}\ \bibnamefont
  {Shamir}},\ }\bibfield  {title} {\enquote {\bibinfo {title} {{Near-horizon
  solutions for D3-branes ending on 5-branes}},}\ }\href {\doibase
  10.1103/PhysRevD.84.126003} {\bibfield  {journal} {\bibinfo  {journal} {Phys.
  Rev. D}\ }\textbf {\bibinfo {volume} {84}},\ \bibinfo {pages} {126003}
  (\bibinfo {year} {2011})},\ \Eprint {http://arxiv.org/abs/1106.1870}
  {arXiv:1106.1870 [hep-th]} \BibitemShut {NoStop}%
\bibitem [{\citenamefont {Assel}\ \emph {et~al.}(2011)\citenamefont {Assel},
  \citenamefont {Bachas}, \citenamefont {Estes},\ and\ \citenamefont
  {Gomis}}]{Assel:2011xz}%
  \BibitemOpen
  \bibfield  {author} {\bibinfo {author} {\bibfnamefont {Benjamin}\
  \bibnamefont {Assel}}, \bibinfo {author} {\bibfnamefont {Costas}\
  \bibnamefont {Bachas}}, \bibinfo {author} {\bibfnamefont {John}\ \bibnamefont
  {Estes}}, \ and\ \bibinfo {author} {\bibfnamefont {Jaume}\ \bibnamefont
  {Gomis}},\ }\bibfield  {title} {\enquote {\bibinfo {title} {{Holographic
  Duals of D=3 N=4 Superconformal Field Theories}},}\ }\href {\doibase
  10.1007/JHEP08(2011)087} {\bibfield  {journal} {\bibinfo  {journal} {JHEP}\
  }\textbf {\bibinfo {volume} {08}},\ \bibinfo {pages} {087} (\bibinfo {year}
  {2011})},\ \Eprint {http://arxiv.org/abs/1106.4253} {arXiv:1106.4253
  [hep-th]} \BibitemShut {NoStop}%
\bibitem [{\citenamefont {Karch}\ and\ \citenamefont
  {Randall}(2001{\natexlab{c}})}]{Karch:2001cw}%
  \BibitemOpen
  \bibfield  {author} {\bibinfo {author} {\bibfnamefont {Andreas}\ \bibnamefont
  {Karch}}\ and\ \bibinfo {author} {\bibfnamefont {Lisa}\ \bibnamefont
  {Randall}},\ }\bibfield  {title} {\enquote {\bibinfo {title} {{Localized
  gravity in string theory}},}\ }\href {\doibase 10.1103/PhysRevLett.87.061601}
  {\bibfield  {journal} {\bibinfo  {journal} {Phys. Rev. Lett.}\ }\textbf
  {\bibinfo {volume} {87}},\ \bibinfo {pages} {061601} (\bibinfo {year}
  {2001}{\natexlab{c}})},\ \Eprint {http://arxiv.org/abs/hep-th/0105108}
  {arXiv:hep-th/0105108} \BibitemShut {NoStop}%
\bibitem [{\citenamefont {Penington}(2020)}]{Penington:2019npb}%
  \BibitemOpen
  \bibfield  {author} {\bibinfo {author} {\bibfnamefont {Geoffrey}\
  \bibnamefont {Penington}},\ }\bibfield  {title} {\enquote {\bibinfo {title}
  {{Entanglement Wedge Reconstruction and the Information Paradox}},}\ }\href
  {\doibase 10.1007/JHEP09(2020)002} {\bibfield  {journal} {\bibinfo  {journal}
  {JHEP}\ }\textbf {\bibinfo {volume} {09}},\ \bibinfo {pages} {002} (\bibinfo
  {year} {2020})},\ \Eprint {http://arxiv.org/abs/1905.08255} {arXiv:1905.08255
  [hep-th]} \BibitemShut {NoStop}%
\bibitem [{\citenamefont {Almheiri}\ \emph {et~al.}(2019)\citenamefont
  {Almheiri}, \citenamefont {Engelhardt}, \citenamefont {Marolf},\ and\
  \citenamefont {Maxfield}}]{Almheiri:2019psf}%
  \BibitemOpen
  \bibfield  {author} {\bibinfo {author} {\bibfnamefont {Ahmed}\ \bibnamefont
  {Almheiri}}, \bibinfo {author} {\bibfnamefont {Netta}\ \bibnamefont
  {Engelhardt}}, \bibinfo {author} {\bibfnamefont {Donald}\ \bibnamefont
  {Marolf}}, \ and\ \bibinfo {author} {\bibfnamefont {Henry}\ \bibnamefont
  {Maxfield}},\ }\bibfield  {title} {\enquote {\bibinfo {title} {{The entropy
  of bulk quantum fields and the entanglement wedge of an evaporating black
  hole}},}\ }\href {\doibase 10.1007/JHEP12(2019)063} {\bibfield  {journal}
  {\bibinfo  {journal} {JHEP}\ }\textbf {\bibinfo {volume} {12}},\ \bibinfo
  {pages} {063} (\bibinfo {year} {2019})},\ \Eprint
  {http://arxiv.org/abs/1905.08762} {arXiv:1905.08762 [hep-th]} \BibitemShut
  {NoStop}%
\bibitem [{\citenamefont {Almheiri}\ \emph {et~al.}(2021)\citenamefont
  {Almheiri}, \citenamefont {Hartman}, \citenamefont {Maldacena}, \citenamefont
  {Shaghoulian},\ and\ \citenamefont {Tajdini}}]{Almheiri:2020cfm}%
  \BibitemOpen
  \bibfield  {author} {\bibinfo {author} {\bibfnamefont {Ahmed}\ \bibnamefont
  {Almheiri}}, \bibinfo {author} {\bibfnamefont {Thomas}\ \bibnamefont
  {Hartman}}, \bibinfo {author} {\bibfnamefont {Juan}\ \bibnamefont
  {Maldacena}}, \bibinfo {author} {\bibfnamefont {Edgar}\ \bibnamefont
  {Shaghoulian}}, \ and\ \bibinfo {author} {\bibfnamefont {Amirhossein}\
  \bibnamefont {Tajdini}},\ }\bibfield  {title} {\enquote {\bibinfo {title}
  {{The entropy of Hawking radiation}},}\ }\href {\doibase
  10.1103/RevModPhys.93.035002} {\bibfield  {journal} {\bibinfo  {journal}
  {Rev. Mod. Phys.}\ }\textbf {\bibinfo {volume} {93}},\ \bibinfo {pages}
  {035002} (\bibinfo {year} {2021})},\ \Eprint
  {http://arxiv.org/abs/2006.06872} {arXiv:2006.06872 [hep-th]} \BibitemShut
  {NoStop}%
\bibitem [{\citenamefont {Bachas}\ and\ \citenamefont
  {Lavdas}(2018)}]{Bachas:2018zmb}%
  \BibitemOpen
  \bibfield  {author} {\bibinfo {author} {\bibfnamefont {Constantin}\
  \bibnamefont {Bachas}}\ and\ \bibinfo {author} {\bibfnamefont {Ioannis}\
  \bibnamefont {Lavdas}},\ }\bibfield  {title} {\enquote {\bibinfo {title}
  {{Massive Anti-de Sitter Gravity from String Theory}},}\ }\href {\doibase
  10.1007/JHEP11(2018)003} {\bibfield  {journal} {\bibinfo  {journal} {JHEP}\
  }\textbf {\bibinfo {volume} {11}},\ \bibinfo {pages} {003} (\bibinfo {year}
  {2018})},\ \Eprint {http://arxiv.org/abs/1807.00591} {arXiv:1807.00591
  [hep-th]} \BibitemShut {NoStop}%
\bibitem [{\citenamefont {Aharony}\ \emph {et~al.}(2006)\citenamefont
  {Aharony}, \citenamefont {Clark},\ and\ \citenamefont
  {Karch}}]{Aharony:2006hz}%
  \BibitemOpen
  \bibfield  {author} {\bibinfo {author} {\bibfnamefont {Ofer}\ \bibnamefont
  {Aharony}}, \bibinfo {author} {\bibfnamefont {Adam~B.}\ \bibnamefont
  {Clark}}, \ and\ \bibinfo {author} {\bibfnamefont {Andreas}\ \bibnamefont
  {Karch}},\ }\bibfield  {title} {\enquote {\bibinfo {title} {{The CFT/AdS
  correspondence, massive gravitons and a connectivity index conjecture}},}\
  }\href {\doibase 10.1103/PhysRevD.74.086006} {\bibfield  {journal} {\bibinfo
  {journal} {Phys. Rev. D}\ }\textbf {\bibinfo {volume} {74}},\ \bibinfo
  {pages} {086006} (\bibinfo {year} {2006})},\ \Eprint
  {http://arxiv.org/abs/hep-th/0608089} {arXiv:hep-th/0608089} \BibitemShut
  {NoStop}%
\bibitem [{\citenamefont {Geng}\ \emph {et~al.}(2022)\citenamefont {Geng},
  \citenamefont {Karch}, \citenamefont {Perez-Pardavila}, \citenamefont {Raju},
  \citenamefont {Randall}, \citenamefont {Riojas},\ and\ \citenamefont
  {Shashi}}]{Geng:2021hlu}%
  \BibitemOpen
  \bibfield  {author} {\bibinfo {author} {\bibfnamefont {Hao}\ \bibnamefont
  {Geng}}, \bibinfo {author} {\bibfnamefont {Andreas}\ \bibnamefont {Karch}},
  \bibinfo {author} {\bibfnamefont {Carlos}\ \bibnamefont {Perez-Pardavila}},
  \bibinfo {author} {\bibfnamefont {Suvrat}\ \bibnamefont {Raju}}, \bibinfo
  {author} {\bibfnamefont {Lisa}\ \bibnamefont {Randall}}, \bibinfo {author}
  {\bibfnamefont {Marcos}\ \bibnamefont {Riojas}}, \ and\ \bibinfo {author}
  {\bibfnamefont {Sanjit}\ \bibnamefont {Shashi}},\ }\bibfield  {title}
  {\enquote {\bibinfo {title} {{Inconsistency of islands in theories with
  long-range gravity}},}\ }\href {\doibase 10.1007/JHEP01(2022)182} {\bibfield
  {journal} {\bibinfo  {journal} {JHEP}\ }\textbf {\bibinfo {volume} {01}},\
  \bibinfo {pages} {182} (\bibinfo {year} {2022})},\ \Eprint
  {http://arxiv.org/abs/2107.03390} {arXiv:2107.03390 [hep-th]} \BibitemShut
  {NoStop}%
\bibitem [{\citenamefont {Neuenfeld}(2021)}]{Neuenfeld:2021wbl}%
  \BibitemOpen
  \bibfield  {author} {\bibinfo {author} {\bibfnamefont {Dominik}\ \bibnamefont
  {Neuenfeld}},\ }\bibfield  {title} {\enquote {\bibinfo {title} {{The
  Dictionary for Double Holography and Graviton Masses in d Dimensions}},}\
  }\href@noop {} {\  (\bibinfo {year} {2021})},\ \Eprint
  {http://arxiv.org/abs/2104.02801} {arXiv:2104.02801 [hep-th]} \BibitemShut
  {NoStop}%
\bibitem [{\citenamefont {Omiya}\ and\ \citenamefont
  {Wei}(2021)}]{Omiya:2021olc}%
  \BibitemOpen
  \bibfield  {author} {\bibinfo {author} {\bibfnamefont {Hidetoshi}\
  \bibnamefont {Omiya}}\ and\ \bibinfo {author} {\bibfnamefont {Zixia}\
  \bibnamefont {Wei}},\ }\bibfield  {title} {\enquote {\bibinfo {title}
  {{Causal Structures and Nonlocality in Double Holography}},}\ }\href@noop {}
  {\  (\bibinfo {year} {2021})},\ \Eprint {http://arxiv.org/abs/2107.01219}
  {arXiv:2107.01219 [hep-th]} \BibitemShut {NoStop}%
\bibitem [{\citenamefont {Hanany}\ and\ \citenamefont
  {Witten}(1997)}]{Hanany:1996ie}%
  \BibitemOpen
  \bibfield  {author} {\bibinfo {author} {\bibfnamefont {Amihay}\ \bibnamefont
  {Hanany}}\ and\ \bibinfo {author} {\bibfnamefont {Edward}\ \bibnamefont
  {Witten}},\ }\bibfield  {title} {\enquote {\bibinfo {title} {{Type IIB
  superstrings, BPS monopoles, and three-dimensional gauge dynamics}},}\ }\href
  {\doibase 10.1016/S0550-3213(97)00157-0} {\bibfield  {journal} {\bibinfo
  {journal} {Nucl. Phys. B}\ }\textbf {\bibinfo {volume} {492}},\ \bibinfo
  {pages} {152--190} (\bibinfo {year} {1997})},\ \Eprint
  {http://arxiv.org/abs/hep-th/9611230} {arXiv:hep-th/9611230} \BibitemShut
  {NoStop}%
\bibitem [{\citenamefont {Gaiotto}\ and\ \citenamefont
  {Witten}(2009{\natexlab{a}})}]{Gaiotto:2008sa}%
  \BibitemOpen
  \bibfield  {author} {\bibinfo {author} {\bibfnamefont {Davide}\ \bibnamefont
  {Gaiotto}}\ and\ \bibinfo {author} {\bibfnamefont {Edward}\ \bibnamefont
  {Witten}},\ }\bibfield  {title} {\enquote {\bibinfo {title} {{Supersymmetric
  Boundary Conditions in N=4 Super Yang-Mills Theory}},}\ }\href {\doibase
  10.1007/s10955-009-9687-3} {\bibfield  {journal} {\bibinfo  {journal} {J.
  Statist. Phys.}\ }\textbf {\bibinfo {volume} {135}},\ \bibinfo {pages}
  {789--855} (\bibinfo {year} {2009}{\natexlab{a}})},\ \Eprint
  {http://arxiv.org/abs/0804.2902} {arXiv:0804.2902 [hep-th]} \BibitemShut
  {NoStop}%
\bibitem [{\citenamefont {Gaiotto}\ and\ \citenamefont
  {Witten}(2009{\natexlab{b}})}]{Gaiotto:2008ak}%
  \BibitemOpen
  \bibfield  {author} {\bibinfo {author} {\bibfnamefont {Davide}\ \bibnamefont
  {Gaiotto}}\ and\ \bibinfo {author} {\bibfnamefont {Edward}\ \bibnamefont
  {Witten}},\ }\bibfield  {title} {\enquote {\bibinfo {title} {{S-Duality of
  Boundary Conditions In N=4 Super Yang-Mills Theory}},}\ }\href {\doibase
  10.4310/ATMP.2009.v13.n3.a5} {\bibfield  {journal} {\bibinfo  {journal} {Adv.
  Theor. Math. Phys.}\ }\textbf {\bibinfo {volume} {13}},\ \bibinfo {pages}
  {721--896} (\bibinfo {year} {2009}{\natexlab{b}})},\ \Eprint
  {http://arxiv.org/abs/0807.3720} {arXiv:0807.3720 [hep-th]} \BibitemShut
  {NoStop}%
\bibitem [{\citenamefont {Raamsdonk}\ and\ \citenamefont
  {Waddell}(2021)}]{Raamsdonk:2020tin}%
  \BibitemOpen
  \bibfield  {author} {\bibinfo {author} {\bibfnamefont {Mark~Van}\
  \bibnamefont {Raamsdonk}}\ and\ \bibinfo {author} {\bibfnamefont {Chris}\
  \bibnamefont {Waddell}},\ }\bibfield  {title} {\enquote {\bibinfo {title}
  {{Holographic and localization calculations of boundary F for $ \mathcal{N} $
  = 4 SUSY Yang-Mills theory}},}\ }\href {\doibase 10.1007/JHEP02(2021)222}
  {\bibfield  {journal} {\bibinfo  {journal} {JHEP}\ }\textbf {\bibinfo
  {volume} {02}},\ \bibinfo {pages} {222} (\bibinfo {year} {2021})},\ \Eprint
  {http://arxiv.org/abs/2010.14520} {arXiv:2010.14520 [hep-th]} \BibitemShut
  {NoStop}%
\bibitem [{\citenamefont {Coccia}\ and\ \citenamefont
  {Uhlemann}(2021{\natexlab{a}})}]{Coccia:2021lpp}%
  \BibitemOpen
  \bibfield  {author} {\bibinfo {author} {\bibfnamefont {Lorenzo}\ \bibnamefont
  {Coccia}}\ and\ \bibinfo {author} {\bibfnamefont {Christoph~F.}\ \bibnamefont
  {Uhlemann}},\ }\bibfield  {title} {\enquote {\bibinfo {title} {{Mapping out
  the internal space in AdS/BCFT with Wilson loops}},}\ }\href@noop {} {\
  (\bibinfo {year} {2021}{\natexlab{a}})},\ \Eprint
  {http://arxiv.org/abs/2112.14648} {arXiv:2112.14648 [hep-th]} \BibitemShut
  {NoStop}%
\bibitem [{\citenamefont {Coccia}\ and\ \citenamefont
  {Uhlemann}(2021{\natexlab{b}})}]{Coccia:2020wtk}%
  \BibitemOpen
  \bibfield  {author} {\bibinfo {author} {\bibfnamefont {Lorenzo}\ \bibnamefont
  {Coccia}}\ and\ \bibinfo {author} {\bibfnamefont {Christoph~F.}\ \bibnamefont
  {Uhlemann}},\ }\bibfield  {title} {\enquote {\bibinfo {title} {{On the planar
  limit of 3d $
  {\mathrm{T}}_{\rho}^{\sigma}\left[\mathrm{SU}\left(\mathrm{N}\right)\right]
  $}},}\ }\href {\doibase 10.1007/JHEP06(2021)038} {\bibfield  {journal}
  {\bibinfo  {journal} {JHEP}\ }\textbf {\bibinfo {volume} {06}},\ \bibinfo
  {pages} {038} (\bibinfo {year} {2021}{\natexlab{b}})},\ \Eprint
  {http://arxiv.org/abs/2011.10050} {arXiv:2011.10050 [hep-th]} \BibitemShut
  {NoStop}%
\bibitem [{\citenamefont {Assel}\ \emph {et~al.}(2012)\citenamefont {Assel},
  \citenamefont {Estes},\ and\ \citenamefont {Yamazaki}}]{Assel:2012cp}%
  \BibitemOpen
  \bibfield  {author} {\bibinfo {author} {\bibfnamefont {Benjamin}\
  \bibnamefont {Assel}}, \bibinfo {author} {\bibfnamefont {John}\ \bibnamefont
  {Estes}}, \ and\ \bibinfo {author} {\bibfnamefont {Masahito}\ \bibnamefont
  {Yamazaki}},\ }\bibfield  {title} {\enquote {\bibinfo {title} {{Large N Free
  Energy of 3d N=4 SCFTs and $AdS_4/CFT_3$}},}\ }\href {\doibase
  10.1007/JHEP09(2012)074} {\bibfield  {journal} {\bibinfo  {journal} {JHEP}\
  }\textbf {\bibinfo {volume} {09}},\ \bibinfo {pages} {074} (\bibinfo {year}
  {2012})},\ \Eprint {http://arxiv.org/abs/1206.2920} {arXiv:1206.2920
  [hep-th]} \BibitemShut {NoStop}%
\bibitem [{\citenamefont {Bachas}\ \emph {et~al.}(2018)\citenamefont {Bachas},
  \citenamefont {Bianchi},\ and\ \citenamefont {Hanany}}]{Bachas:2017wva}%
  \BibitemOpen
  \bibfield  {author} {\bibinfo {author} {\bibfnamefont {Constantin}\
  \bibnamefont {Bachas}}, \bibinfo {author} {\bibfnamefont {Massimo}\
  \bibnamefont {Bianchi}}, \ and\ \bibinfo {author} {\bibfnamefont {Amihay}\
  \bibnamefont {Hanany}},\ }\bibfield  {title} {\enquote {\bibinfo {title} {{$
  \mathcal{N}=2 $ moduli of AdS$_{4}$ vacua: a fine-print study}},}\ }\href
  {\doibase 10.1007/JHEP08(2018)100} {\bibfield  {journal} {\bibinfo  {journal}
  {JHEP}\ }\textbf {\bibinfo {volume} {08}},\ \bibinfo {pages} {100} (\bibinfo
  {year} {2018})},\ \bibinfo {note} {[Erratum: JHEP 10, 032 (2018)]},\ \Eprint
  {http://arxiv.org/abs/1711.06722} {arXiv:1711.06722 [hep-th]} \BibitemShut
  {NoStop}%
\bibitem [{\citenamefont {Assel}\ and\ \citenamefont
  {Gomis}(2015)}]{Assel:2015oxa}%
  \BibitemOpen
  \bibfield  {author} {\bibinfo {author} {\bibfnamefont {Benjamin}\
  \bibnamefont {Assel}}\ and\ \bibinfo {author} {\bibfnamefont {Jaume}\
  \bibnamefont {Gomis}},\ }\bibfield  {title} {\enquote {\bibinfo {title}
  {{Mirror Symmetry And Loop Operators}},}\ }\href {\doibase
  10.1007/JHEP11(2015)055} {\bibfield  {journal} {\bibinfo  {journal} {JHEP}\
  }\textbf {\bibinfo {volume} {11}},\ \bibinfo {pages} {055} (\bibinfo {year}
  {2015})},\ \Eprint {http://arxiv.org/abs/1506.01718} {arXiv:1506.01718
  [hep-th]} \BibitemShut {NoStop}%
\bibitem [{\citenamefont {Basu}\ \emph {et~al.}(2004)\citenamefont {Basu},
  \citenamefont {Green},\ and\ \citenamefont {Sethi}}]{Basu:2004dm}%
  \BibitemOpen
  \bibfield  {author} {\bibinfo {author} {\bibfnamefont {Anirban}\ \bibnamefont
  {Basu}}, \bibinfo {author} {\bibfnamefont {Michael~B.}\ \bibnamefont
  {Green}}, \ and\ \bibinfo {author} {\bibfnamefont {Savdeep}\ \bibnamefont
  {Sethi}},\ }\bibfield  {title} {\enquote {\bibinfo {title} {{A Curious
  truncation of N=4 Yang-Mills}},}\ }\href {\doibase
  10.1103/PhysRevLett.93.261601} {\bibfield  {journal} {\bibinfo  {journal}
  {Phys. Rev. Lett.}\ }\textbf {\bibinfo {volume} {93}},\ \bibinfo {pages}
  {261601} (\bibinfo {year} {2004})},\ \Eprint
  {http://arxiv.org/abs/hep-th/0406267} {arXiv:hep-th/0406267} \BibitemShut
  {NoStop}%
\bibitem [{\citenamefont {Binder}\ \emph {et~al.}(2019)\citenamefont {Binder},
  \citenamefont {Chester}, \citenamefont {Pufu},\ and\ \citenamefont
  {Wang}}]{Binder:2019jwn}%
  \BibitemOpen
  \bibfield  {author} {\bibinfo {author} {\bibfnamefont {Damon~J.}\
  \bibnamefont {Binder}}, \bibinfo {author} {\bibfnamefont {Shai~M.}\
  \bibnamefont {Chester}}, \bibinfo {author} {\bibfnamefont {Silviu~S.}\
  \bibnamefont {Pufu}}, \ and\ \bibinfo {author} {\bibfnamefont {Yifan}\
  \bibnamefont {Wang}},\ }\bibfield  {title} {\enquote {\bibinfo {title} {{$
  \mathcal{N} $ = 4 Super-Yang-Mills correlators at strong coupling from string
  theory and localization}},}\ }\href {\doibase 10.1007/JHEP12(2019)119}
  {\bibfield  {journal} {\bibinfo  {journal} {JHEP}\ }\textbf {\bibinfo
  {volume} {12}},\ \bibinfo {pages} {119} (\bibinfo {year} {2019})},\ \Eprint
  {http://arxiv.org/abs/1902.06263} {arXiv:1902.06263 [hep-th]} \BibitemShut
  {NoStop}%
\bibitem [{\citenamefont {Chester}\ \emph {et~al.}(2020)\citenamefont
  {Chester}, \citenamefont {Green}, \citenamefont {Pufu}, \citenamefont
  {Wang},\ and\ \citenamefont {Wen}}]{Chester:2019jas}%
  \BibitemOpen
  \bibfield  {author} {\bibinfo {author} {\bibfnamefont {Shai~M.}\ \bibnamefont
  {Chester}}, \bibinfo {author} {\bibfnamefont {Michael~B.}\ \bibnamefont
  {Green}}, \bibinfo {author} {\bibfnamefont {Silviu~S.}\ \bibnamefont {Pufu}},
  \bibinfo {author} {\bibfnamefont {Yifan}\ \bibnamefont {Wang}}, \ and\
  \bibinfo {author} {\bibfnamefont {Congkao}\ \bibnamefont {Wen}},\ }\bibfield
  {title} {\enquote {\bibinfo {title} {{Modular invariance in superstring
  theory from $ \mathcal{N} $ = 4 super-Yang-Mills}},}\ }\href {\doibase
  10.1007/JHEP11(2020)016} {\bibfield  {journal} {\bibinfo  {journal} {JHEP}\
  }\textbf {\bibinfo {volume} {11}},\ \bibinfo {pages} {016} (\bibinfo {year}
  {2020})},\ \Eprint {http://arxiv.org/abs/1912.13365} {arXiv:1912.13365
  [hep-th]} \BibitemShut {NoStop}%
\bibitem [{\citenamefont {Hamilton}\ \emph {et~al.}(2006)\citenamefont
  {Hamilton}, \citenamefont {Kabat}, \citenamefont {Lifschytz},\ and\
  \citenamefont {Lowe}}]{Hamilton:2006az}%
  \BibitemOpen
  \bibfield  {author} {\bibinfo {author} {\bibfnamefont {Alex}\ \bibnamefont
  {Hamilton}}, \bibinfo {author} {\bibfnamefont {Daniel~N.}\ \bibnamefont
  {Kabat}}, \bibinfo {author} {\bibfnamefont {Gilad}\ \bibnamefont
  {Lifschytz}}, \ and\ \bibinfo {author} {\bibfnamefont {David~A.}\
  \bibnamefont {Lowe}},\ }\bibfield  {title} {\enquote {\bibinfo {title}
  {{Holographic representation of local bulk operators}},}\ }\href {\doibase
  10.1103/PhysRevD.74.066009} {\bibfield  {journal} {\bibinfo  {journal} {Phys.
  Rev. D}\ }\textbf {\bibinfo {volume} {74}},\ \bibinfo {pages} {066009}
  (\bibinfo {year} {2006})},\ \Eprint {http://arxiv.org/abs/hep-th/0606141}
  {arXiv:hep-th/0606141} \BibitemShut {NoStop}%
\bibitem [{\citenamefont {Akal}\ \emph {et~al.}(2020)\citenamefont {Akal},
  \citenamefont {Kusuki}, \citenamefont {Takayanagi},\ and\ \citenamefont
  {Wei}}]{Akal:2020wfl}%
  \BibitemOpen
  \bibfield  {author} {\bibinfo {author} {\bibfnamefont {Ibrahim}\ \bibnamefont
  {Akal}}, \bibinfo {author} {\bibfnamefont {Yuya}\ \bibnamefont {Kusuki}},
  \bibinfo {author} {\bibfnamefont {Tadashi}\ \bibnamefont {Takayanagi}}, \
  and\ \bibinfo {author} {\bibfnamefont {Zixia}\ \bibnamefont {Wei}},\
  }\bibfield  {title} {\enquote {\bibinfo {title} {{Codimension two holography
  for wedges}},}\ }\href {\doibase 10.1103/PhysRevD.102.126007} {\bibfield
  {journal} {\bibinfo  {journal} {Phys. Rev. D}\ }\textbf {\bibinfo {volume}
  {102}},\ \bibinfo {pages} {126007} (\bibinfo {year} {2020})},\ \Eprint
  {http://arxiv.org/abs/2007.06800} {arXiv:2007.06800 [hep-th]} \BibitemShut
  {NoStop}%
\end{thebibliography}%
\end{document}